\documentclass[12pt]{article}
\usepackage{amsmath,amssymb,bm,epsf,epsfig,graphicx,hyperref,physics, cleveref, subfig, slashed, stmaryrd, tensor}
\usepackage[backend=bibtex,natbib,style=numeric-comp,sorting=none, doi=false, isbn=false,url=false]{biblatex}
\input epsf.sty
\usepackage[]{todonotes}

\usepackage[weather, alpine]{ifsym}

\definecolor{bisque}{rgb}{1.0, 0.89, 0.77}
\definecolor{forestgreen(web)}{rgb}{0.13, 0.55, 0.13}

\def\be{\begin{equation}}
\def\ee{\end{equation}}
\def\bea{\begin{eqnarray}}
\def\eea{\end{eqnarray}}
\def\ie{\begin{equation}\begin{aligned}}
\def\fe{\end{aligned}\end{equation}}
\numberwithin{equation}{section}

\renewcommand{\title}[1]{\vbox{\center\LARGE{#1}}\vspace{5mm}}
\renewcommand{\author}[1]{\vbox{\center#1}\vspace{5mm}}
\newcommand{\address}[1]{\vbox{\center\em#1}}
\newcommand{\email}[1]{\vbox{\center\tt#1}\vspace{5mm}}

\newcommand{\A}{{\alpha}}

\newcommand{\B}[1]{\bar{#1}}

\begin{document}

\unitlength = .8mm

\begin{titlepage}

\begin{center}

\hfill \\
\hfill \\
\vskip 1cm

\title{On the $AdS_5\times S^5$ Solution of Superstring Field Theory}

\author{Minjae Cho$^{\text{\Snow}}$, Jo\~ao Gomide$^{\text{\Sun}\text{\FilledWeakRainCloud}}$, Jaroslav Scheinpflug${}^{\text{\FilledWeakRainCloud}}$, Xi Yin${}^{\text{\FilledWeakRainCloud}}$}

\address{$^{\text{\Snow}}$Leinweber Institute for Theoretical Physics \& Enrico Fermi Institute, University of Chicago,\\ Chicago, IL 60637, USA
			\\
		$^{\text{\Sun}}$ICTP South American Institute for Fundamental Research
		\\
Instituto de Física Teórica, UNESP - Univ. Estadual Paulista
\\
Rua Dr. Bento Teobaldo Ferraz 271, 01140-070, São Paulo, SP, Brazil
\\
${}^{\text{\FilledWeakRainCloud}}$Jefferson Physical Laboratory, Harvard University,
Cambridge, MA 02138 USA
}

\email{cho7@uchicago.edu, jpr.gomide@unesp.br, jscheinpflug@g.harvard.edu, xiyin@fas.harvard.edu}

\end{center}

\abstract{We determine the $AdS_5\times S^5$ solution of type IIB superstring field theory (SFT) to the third order in the expansion with respect to Ramond-Ramond (RR) flux, demonstrate its supersymmetry from the SFT gauge transformations, and identify the massless RR axion in the spectrum of linearized fluctuations. We present an all-order solution in the pp-wave limit and comment on potential obstructions to, and the existence of, the all-order $AdS_5\times S^5$ solution.}
\vfill

\end{titlepage}

\eject

\begingroup
\hypersetup{linkcolor=black}

\tableofcontents

\endgroup

\section{Introduction}

The formulation of superstrings in Ramond-Ramond (RR) flux backgrounds, where a local worldsheet superconformal field theory description is unavailable in the Ramond-Neveu-Schwarz (RNS) formalism \cite{Friedan:1985ge, Polchinski:1998rr, Berenstein:1999ip, Berenstein:1999jq}, has been a longstanding challenge. One possible approach is the Green-Schwarz formalism of the superstring \cite{Green:1983wt, Grisaru:1985fv, Duff:1987bx, Metsaev:1998it}, which a priori amounts to an effective string theory \cite{Grisaru:1988sa, Roiban:2007jf, Giombi:2009gd} whose full quantization requires additional symmetry assumptions such as integrability \cite{Metsaev:2002re, Bena:2003wd}. Another promising approach is the pure spinor formalism \cite{Berkovits:2000fe, Berkovits:2001ue, Berkovits:2002zk, Berkovits:2004xu}, which is based on a worldsheet CFT with rather complicated ghost structure, whose full quantum consistency remains to be understood \cite{Berkovits:2006vi}. A third alternative is the closed superstring field theory (SFT) based on the RNS formalism \cite{deLacroix:2017lif, Sen:2024nfd}, in which the RR flux can be described through a nontrivial string field solution starting from a purely NSNS background \cite{Cho:2018nfn, Cho:2023mhw, Kim:2024dnw}. Physical observables, such as the string spectrum and scattering amplitudes in the RR flux background, can then be extracted from the fluctuations of the string fields around a given solution.

In this paper, we carry out the first steps in analyzing the type IIB superstring theory in the $AdS_5\times S^5$ spacetime in the SFT framework. In particular, we identify the classical solution of type IIB SFT that describes the $AdS_5 \times S^5$ background, to the leading nontrivial orders in the $\frac{1}{R}$-expansion around the ten-dimensional Minkowski background, where $R$ stands for the radius of $AdS_5$ and $S^5$. The SFT equation of motion takes the form \cite{Zwiebach:1992ie, Sen:2024nfd}
\be
Q \Psi + \sum\limits_{n=2}^\infty \frac{1}{n!}[\Psi^{\otimes n}] = 0,
\label{generalEOM}
\ee
where the string field $\Psi$ is a state in the worldsheet CFT associated with the Minkowskian spacetime background, and $Q$ is the BRST charge. The precise definition of the space of string fields and the construction of the string bracket $[\Psi^{\otimes n}]$ will be reviewed in section \ref{sftSec}. In particular, the classical string brackets together with the BRST charge comprise the structure of an $L_\infty$ algebra. A consistent choice of the string brackets specifies a string field frame. For explicit perturbative computations, we will employ the string brackets defined through the flat vertices introduced in \cite{Mazel:2024alu} and extended to the superstring in appendix \ref{sec:flatvert}. The corresponding string field frame will be referred to as the ``flat-vertex frame''.

When identifying the $AdS_5 \cross S^5$ background, we expect it to have $PSU(2,2|4)$ super-isometry. From the point of view of SFT, these are represented as $L_\infty$ gauge transformations which preserve (\ref{generalEOM}) and leave the background solution invariant. For simplicity, in the present work, we demand the background to be annihilated by the subset of generators consisting of $SO(5) \cross SO(5)$ rotations and 32 fermionic super-isometries, so that
\ie{}
& Q \Lambda^A + \sum_{n=1}^\infty {1\over n!} [\Psi^{\otimes n}\otimes \Lambda^A] = 0,
\label{generalGauge}
\fe
where $A$ labels both the bosonic and fermionic super-isometry generators $\Lambda^A$. The $L_\infty$ commutators of the above super-isometries are expected to yield the other generators of the full $PSU(2,2|4)$ algebra on general grounds, so that the full $PSU(2,2|4)$ is expected to preserve the solution. The existence of the $AdS_5 \cross S^5$ solution is then a $Q$-cohomology problem, which demands the sources in (\ref{generalEOM}) and (\ref{generalGauge}) to be simultaneously exact.

A perturbative solution to (\ref{generalEOM}) and (\ref{generalGauge}) takes the form
\be \label{pertExp}
\Psi = \sum_{n=1}^\infty \mu^n \Psi^{(n)}, \,\, \Lambda^A = \sum_{n=0}^\infty \mu^n (\Lambda^A)^{(n)},
\ee
where in the case of the $AdS_5 \cross S^5$ solution of interest, the parameter $\mu$ is proportional to the inverse radius $\frac{1}{R}$. Order by order, we compute the ghost number 2 and 3 $Q$-cohomologies on the space of string fields respecting the worldsheet symmetries manifestly preserved by $\Psi$. We find that a priori, the cohomology has nontrivial support, which we identify with higher-derivative deformations of IIB SUGRA supersymmetry variations. The expectation is that once we make use of the $L_\infty$ commutation relations of the full nonlinearly realized $PSU(2,2|4)$ algebra up to a given order, in addition to the symmetries that can be made manifest at the level of the worldsheet, the cohomology will become empty.

Despite the above potential cohomological obstructions, we still proceed and construct the $AdS_5 \cross S^5$ solution up to third order in the large radius expansion, explicitly verifying its maximal super-isometry up to second order.
For explicit perturbative computations, it will be useful to split the string field $\Psi$ into its massless and massive components \cite{Erbin:2020eyc},
\ie{}
\Psi = \psi + (1-\mathbb{P})\Psi,~~~~ \psi \equiv \mathbb{P} \Psi,
\fe
where $\mathbb{P}$ is the projector onto the subspace of string fields on which $L_0^+$ is nilpotent. Upon imposing Siegel gauge condition on the massive component, namely
\ie\label{masssiegel}
b_0^+(1-\mathbb{P})\Psi=0,
\fe
where $b_0^+\equiv b_0 + \bar b_0$, one can then solve $(1-\mathbb{P})\Psi$ unambiguously in terms of $\psi$, thereby reducing the SFT equation to that of the massless string field.

The $AdS_5\times S^5$ solution, up to third order in perturbation theory, is determined to be described by the massless string field $\psi$
\ie\label{masslessansatzint}
    \psi = F_{\alpha \beta}(X) V^{\alpha\beta}_{RR}+ h_{\mu\nu}(X) V^{\mu \nu}_{NSNS} + \phi(X) D + A_{\mu}(X) V^\mu_{aux.},
\fe
with $F_{\alpha \beta}$ self-dual 5-form flux, $h_{\mu \nu}$ the metric fluctuation, $\phi$ the ghost dilaton and $A_\mu$ an auxiliary field,
where the vertex operators $V^{\alpha\beta}_{RR},V^{\mu \nu}_{NSNS},D, V^\mu_{aux.}$ are defined as
\ie\label{vertdefint}
    V^{\alpha \beta}_{RR} &\equiv c\bar{c} e^{-\frac{1}{2}\phi} S^\alpha e^{-\frac{1}{2}\bar{\phi}}\bar{S}^\beta ,
    \\
    V^{\mu \nu}_{NSNS} &\equiv 4c \bar{c} e^{-\phi} \psi^\mu e^{-\bar{\phi}}\bar{\psi}^\nu ,
    \\
    D & \equiv c\bar{c}\big(\eta e^{-2\bar{\phi}} \bar{\partial} \bar{\xi} - e^{-2\phi}\partial \xi \bar{\eta}\big) ,
    \\
    V^\mu_{aux.} &\equiv c_0^+\big(c \bar{c} e^{-\phi} \psi^\mu e^{-2\bar{\phi}}\bar{\partial}\bar{\xi} + c \bar{c} e^{-2\phi} \partial \xi e^{-\bar{\phi}} \bar{\psi}^\mu\big),
\fe
and the profiles are given by
\ie\label{covsol}
    F_{\alpha \beta}(X) &= {i\over R} \left( 1- {X^2\over 4R^2} \right) (\gamma^{01234})_{\alpha \beta} + {\cal O}(R^{-5}) ,
    \\
    h_{\mu \nu}(X) &=  -\frac{1}{R^2} \left(\delta^a_\mu \delta^b_\nu X_a X_b - \delta^i_\mu \delta^j_\nu X_i X_j \right) + {\cal O}(R^{-4}),
    \\
    \phi(X) &=  -\frac{X^2}{R^2} + {\cal O}(R^{-4}),
    \\
    A_\mu(X) &= \frac{10}{R^2} \partial_\mu X^2 + {\cal O}(R^{-4}),
\fe
where $\mu,\nu=0,\cdots,9$ are the 10-dimensional Minkowski indices which we divide into $a,b=0,\cdots,4$ for $AdS_5$ and $i,j=5,\cdots,9$ for $S^5$, and $X^2\equiv X_a X^a-X_i X^i$. Note that the Siegel gauge condition has not been imposed in this solution, which allows for spacetime diffeomorphism to be represented as residual gauge transformations on $\psi$. 
Nonetheless, it can be further transformed into the Siegel gauge $b_0^+\psi=0$, where $A_{\mu}(X)=0$ and the remaining components are
\ie\label{siegelsolution}
    F_{\alpha \beta}(X) &= {i\over R} \left( 1 + {X^2\over 4R^2} \right) (\gamma^{01234})_{\alpha \beta} + {\cal O}(R^{-5}),
    \\
   h_{\mu \nu}(X) &= {1\over R^2} \left[ \frac{3}{7} \big(\delta^a_\mu \delta^b_\nu X_a X_b - \delta^i_\mu \delta^j_\nu X_i X_j\big) +\frac{5}{7}\big(\eta_{ab}\delta_{\mu}^a\delta_{\nu}^bX_cX^c-\delta_{ij}\delta_{\mu}^i\delta_{\nu}^jX_kX^k\big) \right] + {\cal O}(R^{-4}),
   \\
    \phi(X)&= {4 X^2\over R^2} + {\cal O}(R^{-4}) .
\fe
The above solution of SFT provides the starting point of a first-principle definition of type IIB string theory in $AdS_5\times S^5$. In particular, the physical string spectrum and amplitudes can in principle be extracted by analyzing string field fluctuations around the background solution.

We also note that the above solution drastically simplifies in the pp-wave limit, allowing us to find an all-order solution. The  massless part of the solution takes the form
\be
\begin{aligned}
F_{\alpha \beta}(X)= i \mu (\gamma^{+1234})_{\alpha \beta}, \quad h_{\mu \nu}(X) =  - \mu^2 X_\perp^2 \delta_\mu^+ \delta_\nu^+ , \quad \phi(X) =  0, \quad A_\mu(X) = 0.
\end{aligned}
\ee
Checking the maximal super-isometry of the pp-wave solution to all orders is left for the future.

The rest of the paper is organized as follows. We begin with an overview of the relevant SFT formalism in section \ref{sftSec}. We then proceed to discuss the general features of the extended string field equations for our background in section \ref{susySec} and continue by constructing the explicit solution in section \ref{solext} up to third order in the large radius expansion. We also present a simple all-order solution to the string field equations of motion that describes the pp-wave limit of $AdS_5\times S^5$ in section \ref{sec:pp}. The linearized fluctuations around the $AdS_5\times S^5$ background will be considered in section \ref{sec:axion}, where the example of the massless RR axion mode is analyzed explicitly. Potential obstructions to constructing the higher order solution are discussed in section \ref{obstructionSec} and finally we conclude with future prospects in section \ref{discussionSec}.

\section{Classical closed superstring field theory}
\label{sftSec}


\subsection{The space of string fields}

The closed superstring field theory in the RNS formalism\footnote{For reviews on the subject see \cite{Sen:2024nfd, deLacroix:2017lif, Maccaferri:2023vns, Erbin:2021smf, Erler:2019loq, Erler:2019vhl}.} is based on a worldsheet superconformal field theory (SCFT) that consists of a matter system, say that of free bosons $X^\mu$ and free fermions $(\psi^\mu, \bar\psi^\mu)$ that describe the 10-dimensional Minkowskian spacetime background, and the $(b,c,\bar b, \bar c, \beta,\gamma, \bar \beta, \bar \gamma)$ ghost system, of total central charge zero and a BRST charge $Q$ that satisfies
\ie
\{ Q, b(z)\} = T(z),~~~~ [Q, \beta(z)] = G(z),
\fe
where $T$ and $G$ are the matter+ghost stress-energy tensor and supercurrent respectively. The space ${\cal H}$ of string fields is of the form
\ie
\mathcal{H} = \mathcal{H}_{-1, -1} \oplus \mathcal{H}_{-1, -\frac{1}{2}} \oplus \mathcal{H}_{-\frac{1}{2}, -1} \oplus \mathcal{H}_{-\frac{1}{2}, -\frac{1}{2}},
\fe
where ${\cal H}_{m,n}$ stands for the space of states $|\Phi\rangle$ of the worldsheet SCFT in the $(m,n)$ picture, subject to type IIA or IIB GSO projection, and the constraint
\ie
b_0^- \ket{\Phi} = L_0^- \ket{\Phi} = 0,
\fe
where $b_0^\pm \equiv b_0 \pm \bar{b}_0$ and $L_0^\pm \equiv L_0 \pm \bar{L}_0$. In particular, the Neveu-Schwarz (NS) sector string fields are taken to be in the -1 picture and the Ramond (R) sector string fields are in the $-\frac{1}{2}$ picture.
The space $\mathcal{H}$ is moreover graded by the ghost number $N_{\rm gh}$, which assigns charge $+1$ to $c, \bar c, \gamma, \bar\gamma$, $-1$ to $b, \bar b, \beta, \bar\beta$, and 0 to $e^{\A\phi}, e^{\A\bar\phi}$ (where $\phi$ is the linear dilaton field in the $(\phi, \eta, \xi)$ representation of the $\beta\gamma$ system). The physical string field $\Psi$ carries ghost number 2, whereas a gauge transformation is generated by a string field $\Lambda$ of ghost number 1.

To construct the SFT action, it is useful to also consider an auxiliary space of string fields for which the R sector fields are taken to be in the $-{3\over 2}$ picture, namely
\ie
\mathcal{H}^{\rm aux} = \mathcal{H}_{-1, -1} \oplus \mathcal{H}_{-1, -\frac{3}{2}} \oplus \mathcal{H}_{-\frac{3}{2}, -1} \oplus \mathcal{H}_{-\frac{3}{2}, -\frac{3}{2}}.
\fe
At the level of equation of motion, however, the auxiliary string fields can be eliminated, and it suffices to work with string fields in ${\cal H}$.



\subsection{String vertices and brackets}
\label{sec:stringfieldbracket}

Let ${\cal P}_{n}\to {\cal M}_{0,n}$ be a fiber bundle over the moduli space ${\cal M}_{0,n}$ of $n$-punctured Riemann sphere, whose fiber is parameterized by the choice of coordinate maps
\ie\label{zwimap}
z = f_i(w_i)
\fe
on the disc $D_i$ containing the $i$-th puncture at $z=z_i\equiv f_i(0)$. We further promote ${\cal P}_{n}$ to a bundle ${\cal Q}_{n}\to {\cal M}_{0,n}$, whose fiber includes the additional data of the choice of spin structure associated with each puncture, labeled by its type (NS,NS), (NS,R), (R,NS), or (R,R), and the location $x_1,\cdots, x_{d_o}$ of $d_o= n_{\rm NS} + {n_{\rm R}\over 2} - 2$
holomorphic picture-changing operators (PCOs) and $\bar x_1,\cdots,\bar x_{\bar d_o}$ of $\bar d_o= \bar n_{\rm NS} + {\bar n_{\rm R}\over 2}-2$ holomorphic PCOs. Here $n_{\rm NS/R}$ and $\bar n_{\rm NS/R}$ denote the total number of holomorphic and anti-holomorphic punctures of NS/R type, with $n=n_{\rm NS} + n_{\rm R}$ and $\bar n=\bar n_{\rm NS} + \bar n_{\rm R}$.

Given a set of $n$ superstring fields $\Psi_1,\cdots,\Psi_n\in{\cal H}$ (of definite NS/R types), one defines a differential form $\Omega[\Psi_1\otimes\cdots\otimes \Psi_n]$ on ${\cal Q}_{n}$,
\ie\label{omegassft}
\Omega[\Psi_1\otimes\cdots\otimes \Psi_n] = \left\langle e^{\cal B} \prod_{a=1}^{d_o} \big[ {\cal X}(x_a) - d\xi(x_a) \big] \prod_{\bar a=1}^{\bar d_o} \big[ \bar{\cal X}(\bar {x}_{\bar a}) - d\bar\xi(\bar{x}_{\bar a})\big] \prod_{i=1}^n \left[ \Psi_i(0) \right]^{f_i} \right\rangle,
\fe
where $\langle\cdots\rangle$ stands for the worldsheet CFT correlator, ${\cal X}$ and $\bar{\cal X}$ are the holomorphic and anti-holomorphic PCOs (related to $\xi$ and $\bar\xi$ by BRST transformation), and $\left[ \Psi_i(0) \right]^{f_i}$ is the conformal transformation of the string field $\Psi_i$, viewed as an operator inserted at the origin of the $w_i$-disc, by the
map (\ref{zwimap}) to an operator inserted at $z = z_i$ on the Riemann sphere. ${\cal B}$ is an operator-valued 1-form built out of the $b$ ghost,
\ie
{\cal B} = -\sum_{i=1}^n \oint_{\partial D_i} {dz\over 2\pi i} \delta f_i(w_i) b(z) + c.c.
\fe
The form (\ref{omegassft}) is invariant under a constant phase rotation of each $w_i$ coordinate due to the condition $L_0^- \Psi_i =0$ satisfied by the closed string fields, and therefore induces a well-defined differential form on the quotient space
\ie
\widehat{\cal Q}_{n} = {\cal Q}_{n}/\{f_i(w_i) \sim f_i(e^{i\A_i} w_i), ~\forall \A_i\}.
\fe
The classical $n$-string vertex $\{\cdot\}:{\cal H}^{\otimes n}\to \mathbb{C}$ is a graded symmetric $n$-linear function in the string fields, constructed as
\ie\label{supersftvert}
\{\Psi^{\otimes n}\} = - {1\over (-2\pi i)^{n-3}} \int_{\Upsilon_{n}}  \left\langle e^{\cal B} \prod_{a=1}^{d_o} \big[ {\cal X}(x_a) - d\xi(x_a) \big] \prod_{\bar a=1}^{\bar d_o} \big[ \bar{\cal X}(\bar x_{\bar a}) - d\bar\xi(\bar x_{\bar a})\big] \prod_{i=1}^n \left[ \Psi(0) \right]^{f_i} \right\rangle,
\fe
for $n\geq 3$.
$\Upsilon_{n}$ is a $(2n-6)$-dimensional chain in $\widehat{\cal Q}_{n}$ that is symmetric with respect to exchange of any pair of punctures of the same (NS/R) type, and is subject to the compatibility condition (classical geometric master equation)
\ie\label{geommassuper}
- \partial \Upsilon_{n} &= {1\over 2} \sum_{\A\sqcup\beta = \{1,\cdots,n\}} \varrho_{\A,\beta} \left( \widetilde\Upsilon_{|\A|+1}\times \widetilde\Upsilon_{|\beta|+1}\times \{q:|q|=1\} \right),
\fe
where $\widetilde\Upsilon_m$ stands for a lift of $\Upsilon_m\subset\widehat{\cal Q}_m$ to a chain in ${\cal Q}_m$, and
\ie\label{plumbingmapab}
\varrho_{\A,\beta}: {\cal Q}_{|\A|+1} \times  {\cal Q}_{|\beta|+1} \times S^1  \to  {\cal Q}_{n}
\fe
is the plumbing map defined by joining a pair of punctured Riemann spheres $\Sigma_1$ and $\Sigma_2$ with PCO insertions, via the identification
\ie
w' = q/w
\fe
between the local coordinate $w$ around a puncture on $\Sigma_1$ and $w'$ around a puncture on $\Sigma_2$. Here the plumbing parameter $q$ is taken to be a phase, i.e. $|q|=1$, and the remaining $n=|\A|+|\beta|$ punctures are ordered according to the partition of $\{1,\cdots,n\}$ into the subsets $\A$ and $\beta$.

The string bracket $[\Psi^{\otimes n}]$ is a graded $n$-linear map ${\cal H}^{\otimes n}\to {\cal H}$ defined in terms of the $(n+1)$-string vertex by 
\ie\label{superstringbrack}
\langle\langle \Phi| c_0^- |[\Psi^{\otimes n}]\rangle = \sum_{h=0}^\infty \{ {\cal G} \Phi\otimes \Psi^{\otimes n} \}
\fe
for all $\Phi\in {\cal H}^{\rm aux}$. Here $\langle\langle\cdot|$ stands for the BPZ conjugate, and ${\cal G}$ is the picture-adjusting operator defined by
\be
\mathcal{G} =
\begin{cases}
    1 \, \, \, \, \, \, \hspace{0.5 cm} \text{on} \, \, \, \mathcal{H}_{-1, -1} \\
    \mathcal{X}_0  \, \, \, \hspace{0.45cm} \text{on} \, \, \, \mathcal{H}_{-\frac{3}{2}, -1} \\
    \bar{\mathcal{X}}_0 \, \, \, \hspace{0.45cm}\text{on} \, \, \, \mathcal{H}_{-1, -\frac{3}{2}} \\
    \mathcal{X}_0 \bar{\mathcal{X}}_0  \, \, \, \hspace{0.03cm}\text{on} \, \, \, \mathcal{H}_{-\frac{3}{2}, -\frac{3}{2}},
\end{cases}
\ee
where the $\mathcal{X}_0, \bar{\mathcal{X}}_0$ are related to the holomorphic PCO ${\cal X}(z)$ and the anti-holomorphic PCO $\bar{\cal X}(\bar z)$ by
\ie
{\cal X}_0 = \oint {dz\over 2\pi i} {\cal X}(z),~~~~ \bar {\cal X}_0 = -\oint {d\bar z\over 2\pi i} \bar {\cal X}(\bar z).
\fe
A key consistency condition on the string brackets, which follows from (\ref{geommassuper}), is the $L_\infty$ relation
\ie\label{linfty}
Q  [ \Psi^{\otimes n}  ] + n  [ Q \Psi \otimes \Psi^{\otimes (n-1)} ]
+ \sum_{m=2}^{n-1} {n\choose m}  \big[  [ \Psi^{\otimes m} ] \otimes \Psi^{\otimes (n-m)}  \big] = 0,
\fe
for all $n\geq 2$.

\subsection{Equations of motion, gauge transformations and isometries}
\label{gaugeTransforms}

The classical equations of motion for a closed superstring field $\Psi\in\mathcal{H}$ take the form (\ref{generalEOM}). These equations of motion are written with respect to a given field frame and are subject to gauge redundancy. The gauge redundancies act on a ghost number 2 physical string field via
\ie\label{gaugepsi}
\delta_\Lambda \Psi = Q \Lambda + \sum\limits_{n=1}^\infty [\Psi^{\otimes n}\otimes \Lambda],
\fe
where $\Lambda$ is a string field of ghost number 1 with opposite Grassmann-parity with respect to $\Psi$. As usual, both the choice of field frame \cite{Hata:1993gf} and the gauge ambiguity do not change physical observables. At the first few orders in the large radius expansion, we can choose our field frame to align with that of a given SUGRA solution, and will later do so for the sake of comparison. Note that at higher orders no such comparison is available as there are $\alpha'$ corrections to the background solution not captured by SUGRA. The physical observables in our setup are the physical spectrum and boundary correlators. Both are defined on the asymptotic boundary, but thanks the the homogeneity of $AdS_5 \cross S^5$, one is able to identify the physical spectrum by observing how the $PSU(2,2|4)$ super-isometry group acts in the bulk.  The $PSU(2,2|4)$ super-isometry of the $AdS_5 \cross S^5$ background is represented by SFT gauge transformations that preserve the background solution; they are generated by a set of ghost number 1 string fields $\Lambda^A$ that obey
\ie{}
& Q \Lambda^A + \sum_{n=1}^\infty {1\over n!} [\Psi^{\otimes n}\otimes \Lambda^A] = 0,
\\
&\sum_{n=0}^\infty {1\over n!}  [\Psi^{\otimes n}\otimes \Lambda^A\otimes \Lambda^B ] = f^{AB}{}_C \Lambda^C + Q \Omega^{AB} + \sum_{n=1}^\infty {1\over n!} [\Psi^{\otimes n}\otimes \Omega^{AB}] ,
\label{superiso}
\fe
where $A$ labels super-isometry generators, $f^{AB}{}_C$ are the $PSU(2,2|4)$ structure constants, and $\Omega^{AB}$ are auxiliary string fields that are needed due to the $L_\infty$ nature of the SFT gauge algebra \cite{Hohm:2017pnh, Mazel:2025fxj}.

By substituting in the expansion (\ref{pertExp}), we solve the equations (\ref{generalEOM}) order by order in $\mu$. Proceeding as for any polynomial equation, this formally necessitates inverting $Q$. To perform such an inversion, following \cite{Erbin:2020eyc}, we introduce the projector $\mathbb{P}$ which projects to the $L_0^+$-nilpotent sector of $\mathcal{H}$. String fields split as $\Psi = \psi + (1-\mathbb{P})\Psi$, where $\mathbb{P} \Psi = \psi$. We impose Siegel gauge on the $(1-\mathbb{P})$-projected `massive' components (\ref{masssiegel}). Then $Q$ can be explicitly inverted on them as
\be\label{massivesf}
(1-\mathbb{P})\Psi = -\frac{b_0^+}{L_0^+}\sum\limits_{n=2}^\infty \frac{1}{n!}(1-\mathbb{P})[\Psi^{\otimes n}].
\ee
The $\mathbb{P}$-projected `massless part' of the solution $\psi$ then obeys the equations
\be\label{masslesseom}
Q \psi + \sum\limits_{n=2}^\infty \frac{1}{n!}[\psi^{\otimes n}]' = 0,
\ee
where the massless bracket is defined via
\be
\begin{aligned}
	[\psi^{\otimes 2}]' &= \mathbb{P}[\psi^{\otimes 2}], \\
	[\psi^{\otimes 3}]' &= \mathbb{P}[\psi^{\otimes 3}] - 3 \mathbb{P}\left[\psi \otimes \frac{b_0^+}{L_0^+}(1-\mathbb{P})[\psi^{\otimes 2}]\right]
    \label{masslessthree}
\end{aligned}
\ee
and so on for the higher brackets. These are the massless effective SFT EOM for the massless part of the physical string field $\psi = \sum_{n=1}^\infty \mu^n \psi^{(n)}$.

The EOM of the massless effective SFT are subject to a (massless effective) gauge redunancy. Since the Siegel gauge has been fixed on the massive modes, one must make sure that Siegel gauge on the massive modes is preserved when one does a gauge transform in the massless effective SFT \cite{Mazel:2025fxj}. The condition that the Siegel gauge on massive modes is preserved is $b_0^+ (1-\mathbb{P}) \delta_\Lambda \Psi = 0$. One can again solve this iteratively, splitting $\Lambda = \lambda + \Xi$ such that $\mathbb{P}\Lambda = \lambda$ and $(1-\mathbb{P})\Lambda=\Xi$. Choosing Siegel gauge on the massive generator $b_0^+ \Xi = 0$, one obtains
\be
\Xi = - \frac{b_0^+}{L_0^+} \sum\limits_{n=1}^\infty \frac{1}{n!}(1-\mathbb{P})[\Psi^{\otimes n}\otimes \Lambda].
\ee
The gauge transformation in the massless effective SFT then takes the form
\be
\delta_\lambda \psi \equiv \mathbb{P} \delta_\Lambda \Psi = Q \lambda + \sum\limits_{n=1}^\infty \frac{1}{n!}[\psi^{\otimes n}\otimes \lambda]',
\label{variationMassless}
\ee
with $\lambda = \sum_{n=0}^\infty \mu^n \lambda^{(n)}$ and the massless brackets $[\cdots]'$ defined iteratively as before
\be
\begin{aligned}
    \label{bracket1}
    &[\psi \otimes \lambda]' = \mathbb{P}[\psi \otimes \lambda], \\
    &[\psi^{\otimes 2} \otimes \lambda]' = \mathbb{P}[\psi^{\otimes 2} \otimes \lambda] - 2\mathbb{P} \left[\psi \otimes \frac{b_0^+}{L_0^+}(1-\mathbb{P})[\psi \otimes \lambda]\right] - \mathbb{P}\left[\lambda \otimes \frac{b_0^+}{L_0^+}(1-\mathbb{P})[\psi^{\otimes 2}]\right],
\end{aligned}
\ee
and analogously for higher brackets. In the massless effective language, analogously to (\ref{superiso}), the background being preserved by a super-isometry amounts to \cite{Mazel:2025fxj}

\be
\begin{aligned}
	& Q \lambda^A + \sum_{n=1}^\infty {1\over n!} [\psi^{\otimes n}\otimes \lambda^A]'= 0,
\\
&\sum_{n=0}^\infty {1\over n!}  [\psi^{\otimes n}\otimes \lambda^A\otimes \lambda^B ]' = f^{AB}{}_C \lambda^C + Q \omega^{AB} + \sum_{n=1}^\infty {1\over n!} [\psi^{\otimes n}\otimes \omega^{AB}]',
\end{aligned}
\ee
with $\omega^{AB}$ a massless analogue of the auxiliary field $\Omega^{AB}$.

In summary, finding a perturbative SFT solution with a given set of super-isometries (in our case $PSU(2,2|4)$) amounts to simultaneously solving the following set of equations
\be
\begin{aligned}
& Q \psi + \sum\limits_{n=2}^\infty \frac{1}{n!}[\psi^{\otimes n}]' = 0, \\
& Q \lambda^A + \sum_{n=1}^\infty {1\over n!} [\psi^{\otimes n}\otimes \lambda^A]' = 0,
\\
&\sum_{n=0}^\infty {1\over n!}  [\psi^{\otimes n}\otimes \lambda^A\otimes \lambda^B ]' = f^{AB}{}_C \lambda^C + Q \omega^{AB} + \sum_{n=1}^\infty {1\over n!} [\psi^{\otimes n}\otimes \omega^{AB}]',
\label{extendedProblem}
\end{aligned}
\ee
of the massless effective SFT.

\section{General features of the extended string field equations}
\label{susySec}
In this section, we investigate the general properties of the extended string field equations (\ref{extendedProblem}) for the special case of demanding the SFT solution to be preserved by a subalgebra of $PSU(2,2|4)$, which consists of $SO(5) \cross SO(5)$ rotations and 32 super-isometries. That is, we look at the properties of the extended system
\be
\begin{aligned}
& Q \psi + \sum\limits_{n=2}^\infty \frac{1}{n!}[\psi^{\otimes n}]' = 0, \\
& Q \lambda + \sum_{n=1}^\infty {1\over n!} [\psi^{\otimes n}\otimes \lambda]' = 0, \\
& Q \rho + \sum_{n=1}^\infty {1\over n!} [\psi^{\otimes n}\otimes \rho]' = 0,
\label{extendedProblemNew}
\end{aligned}
\ee
where $\lambda$ labels the massless component of the fermionic super-isometry string field and $\rho$ the $SO(5) \cross SO(5)$ rotation one. In particular, we state the cohomology problem that has to be solved in order to find a solution of such an extended system and identify potential obstructions. The statement is that the potential obstructions are characterized by a constant and a linear term in the dilatino variation vanishing condition and a divergenceless 2-form-spinor obstructing the integrability condition for the vanishing of the gravitino variation. Both of these have to respect the symmetries of the ansatz (\ref{masslessansatz}) and are subject to the shifts (\ref{shifts}) coming from adding a $Q$-closed term to the string field solution.

It is very important to note that as a limitation of restricting to a subalgebra of $PSU(2,2|4)$, we will not exhibit the full $PSU(2,2|4)$ invariance (in particular, we do not see the full $SO(6)$ rotations). This invariance is expected to be recovered by considering the $L_\infty$ commutation relations of the above subalgebra of the $SO(5) \cross SO(5)$ rotations and the 32 super-isometries, but we do not make use of these commutation relations in this paper. The drawbacks of not realizing the full $PSU(2,2|4)$ invariance will be further discussed in section \ref{obstructionSec}.

\subsection{String field ansatz}
In this subsection, we write down the massless string field ansatz with which we try to solve the extended system (\ref{extendedProblemNew}).

A general perturbative ansatz for the massless background string field $\psi$, under the assumption of slowly-varying spacetime field profile and the appropriate ghost number and worldsheet parity assignment, is
\ie\label{masslessansatz}
    \psi = F_{\alpha \beta}(X) V^{\alpha\beta}_{RR}+ h_{\mu\nu}(X) V^{\mu \nu}_{NSNS} + \phi(X) D + A_{\mu}(X) V^\mu_{aux.},
\fe
where $V^{\alpha\beta}_{RR},V^{\mu \nu}_{NSNS},D$ and $V^\mu_{aux.}$ are defined as in (\ref{vertdefint}). By appropriate worldsheet parity assignment, we mean odd parity in the RR sector and even parity in the NSNS sector. This parity is chosen so that the $B$-field and RR 3-form flux are turned off as in the SUGRA solution of appendix \ref{sugraApp}. Furthermore, the SUGRA solution has definite charge under two other discrete transformations, symmetry under which we also impose. Consequently, the ansatz is required to preserve the following discrete symmetries:
\begin{itemize}
	\item Worldsheet parity ($\mathbb{Z}_2$): odd in the RR sector, even in the NSNS sector.
	\item Orientifold parity ($\mathbb{Z}_2$): even in both sectors so that $X^\mu$ dependence in both RR and NSNS sectors is even under total spacetime parity $X^\mu \to - X^\mu$.
	\item Swapping $AdS_5$ and $S^5$ directions ($\mathbb{Z}_4$): after Wick rotating to Euclidean signature, splitting the spacetime directions into 5 orthogonal planes $(X_0, X_5), \ldots, (X_4, X_9)$, then performing a rotation by angle $\frac{\pi}{2}$. This effectively performs the swap $X^a \to X^i, X^i \to - X^a$. We also rescale the radius $R \to i R$ and the solution is to be even in both the RR and NSNS sectors.
\end{itemize}
Moreover, we demand the profiles to be $SO(5) \cross SO(5)$ covariant. Since worldsheet parity and rotations are symmetries of the underlying worldsheet CFT and commute with the BRST charge, it follows that this ansatz is consistent with the equations of motion (\ref{masslesseom}). It is also consistent with (\ref{masslesseom}) and the SUGRA solution to turn on RR fields at odd order in perturbation theory and NSNS fields at even order in perturbation theory. Therefore, this is what we will do when searching for the perturbative solution. In appendix \ref{diffeoApp}, following the reasoning that was initially developed in the context of classical bosonic closed SFT in \cite{Mazel:2025fxj}, we use linearized diffeomorphism transformations to check that this ansatz makes good physical sense: the fields that are being turned on are the RR 5-form $F_{\alpha \beta}$ \footnote{ Note that the discrete symmetries we impose on the solution would also allow us to turn on a RR 1-form field strength of total parity even profile, but this is not allowed by $SO(5) \cross SO(5)$ invariance (the axion scalar would be total parity odd).}, metric $h_{\mu \nu}$ and the physical dilaton $\Phi = \frac{1}{4}(h-\phi)$. The 5-form flux is required to be self-dual in order for the solution to be invariant under the $AdS_5 \leftrightarrow S^5$ swap symmetry.

In order to complete the ansatz for the solution to the extended string field equations, we take
\begin{align}
    \lambda = \epsilon_{\alpha}(X) c\bar{c}e^{-\frac{1}{2}\phi} S^\alpha e^{-2\bar{\phi}}\bar{\partial}\bar{\xi} + \bar{\epsilon}_{\alpha}(X) c\bar{c} e^{-2\phi}\partial \xi e^{-\frac{\bar{\phi}}{2}}\bar{S}^\alpha,
    \label{phiAnsatz}
\end{align}
for the massless part of the 32 fermionic super-isometries (these come in two independent sets of 16, but for now we suppress the labeling of these sets). Here, $\epsilon_{\alpha}(X)$, $\bar{\epsilon}_{\alpha}(X)$ are Grassmann odd super-isometry generators.  We also choose to focus on one of the two sets of super-isometries for the remainder of the paper. This can be achieved by taking $\bar{\epsilon}^{(0)}_\alpha = 0$ at zeroth order, which means that for $n$ odd the super-isometry is in the (NS,R) sector and for $n$ even it is in the (R,NS) sector, as dictated by the structure of our solution. As it is clear from worldsheet parity, this is done without the loss of generality, i.e., if one set of super-isometries exists, the existence of the other one automatically follows.

Note that the $SO(5) \cross SO(5)$ rotations $\rho$ are trivial to implement as the rotations are preserved by the string brackets (the rotations are a global symmetry of the worldsheet CFT and commute with the BRST charge) and are manifest at the level of the ansatz (\ref{masslessansatz}), and so we do not consider them in solving for the generators and string field solution of (\ref{extendedProblemNew}).

\subsection{Existence of the solution as a cohomology problem}
In this subsection, we formulate the existence of a solution to the extended system (\ref{extendedProblemNew}) as a cohomology problem.

There exists a solution to the cohomology problem provided that the sources in (\ref{extendedProblemNew}) are $Q$-exact.  Generic sources restricted by the worldsheet parity of the ansatz (\ref{masslessansatz}) are such that the extended system (\ref{extendedProblemNew}) takes the form (the sources are also further restricted by $SO(5) \cross SO(5)$ covariance and the other discrete symmetries, but this does not change the form of the equations below)
\be
\begin{aligned}
	Q \psi = \,&s_{\alpha\beta}(X)c_0^+c\B{c}e^{-\frac{1}{2}\phi}S^{\alpha}e^{-\frac{1}{2}\B{\phi}}\B{S}^{\beta}+ t^{\alpha}_{\beta}(X)c\eta e^{\frac{1}{2}\phi}S_{\alpha}\bar{c}e^{-\frac{1}{2}\B{\phi}}\B{S}^{\beta} + (t^T)_{\alpha}^{\beta}(X)c e^{-\frac{1}{2}\phi}S^{\alpha}\bar{c}\B{\eta}e^{\frac{1}{2}\B{\phi}}\B{S}_{\beta} + \\
		   & e_{\mu\nu}(X)c_0^+ c\bar{c} e^{-\phi} \psi^\mu e^{-\bar{\phi}} \bar{\psi}^\nu +f(X)c_0^+c\bar{c}\big(\eta e^{-2\bar{\phi}} \bar{\partial} \bar{\xi} - e^{-2\phi}\partial \xi \bar{\eta}\big) +g_{\mu}(X)c\B{c}\big(\eta e^{-\B{\phi}}\B{\psi}^{\mu}+e^{-\phi}\psi^{\mu}\B{\eta}\big) \\
	Q \lambda =\,& v_{\alpha}(X)c_0^+c\B{c}e^{-\frac{1}{2}}S^{\alpha}e^{-2\B{\phi}}\B{\partial}\B{\xi}+u_{\mu\alpha}(X)c\B{c}e^{-\frac{1}{2}\phi}S^{\alpha}e^{-\B{\phi}}\B{\psi}^{\mu}+w^{\alpha}(X)c\B{c}\eta e^{\frac{1}{2}\phi}S_{\alpha}e^{-2\B{\phi}}\B{\partial}\B{\xi} + \\
		    &\bar{v}_{\alpha}(X) c_0^+ c\bar{c}e^{-2\phi}\partial \xi e^{-\frac{\bar{\phi}}{2}}\bar{S}^\alpha +\bar{u}_{\mu \alpha}(X) ce^{-\phi}\psi^\mu \bar{c}e^{-\frac{\bar{\phi}}{2}}\bar{S}^\alpha + \bar{w}^\alpha(X) c e^{-2\phi}\partial\xi \bar{c}\bar{\eta}e^{\frac{\bar{\phi}}{2}}\bar{S}_\alpha,
		    \label{genericSuperIsometry}
\end{aligned}
\ee
Since the kinetic terms are given by
\be
\begin{aligned}
	\label{kinpart}
	Q \psi = \, & -\frac{1}{2}\Box F_{\alpha \beta} c_0^+ c\bar{c}e^{-\frac{1}{2}\phi}S^\alpha e^{-\frac{1}{2}\bar{\phi}}\bar{S}^\beta + \frac{i}{4} (\slashed{\partial} F)^\alpha_\beta c\eta e^{\frac{1}{2}\phi} S_\alpha \bar{c}e^{-\frac{1}{2}\bar{\phi}}\bar{S}^\beta + \frac{i}{4} (F  \overset{\leftarrow}{\slashed{\partial}})^\beta_\alpha c e^{-\frac{1}{2}\phi}S^\alpha \bar{c}\bar{\eta}e^{\frac{1}{2}\bar{\phi}}\bar{S}_\beta + \\
		    &  \frac{1}{2} \big(-4\Box h_{\mu\nu} -\partial_\mu A_{\nu} - \partial_\nu A_\mu \big) c_0^+c\B{c}e^{-\phi}\psi^{\mu}e^{-\B{\phi}}\B{\psi}^{\nu}+\\
    &\frac{1}{4}\Big(-2\Box\phi-\partial^{\mu}A_\mu\Big)c_0^+c\B{c}\big(\eta e^{-2\B{\phi}}\B{\partial}\B{\xi}-e^{-2\phi}\partial\xi\B{\eta}\big)+\\
    &\frac{1}{4}\Big(A_\mu-2\partial_{\mu}\phi +4 \partial^{\nu}h_{\mu\nu}\Big) c\B{c}\big(\eta e^{-\B{\phi}}\B{\psi}^{\mu}+e^{-\phi}\psi^{\mu}\B{\eta}\big) \\
    Q \lambda = \,&\frac{1}{2}\Box \epsilon_{\alpha}(X) c_0^+ c\bar{c} e^{-\frac{1}{2}\phi}S^\alpha e^{-2\bar{\phi}}\bar{\partial} \bar{\xi} - \frac{1}{2}\partial_\mu \epsilon_{\alpha}(X) c e^{-\frac{1}{2}\phi}S^\alpha \bar{c}e^{-\bar{\phi}}\bar{\psi}^\mu - \frac{i}{4} (\slashed{\partial}\epsilon)^\alpha c\eta e^{\frac{1}{2}\phi}S_\alpha \bar{c}e^{-2\bar{\phi}}\bar{\partial}\bar{\xi} \\
    & \frac{1}{2}\Box \bar{\epsilon}_{\alpha}(X) c_0^+ c\bar{c}e^{-2\phi}\partial \xi e^{-\frac{\bar{\phi}}{2}}\bar{S}^\alpha - \frac{1}{2}\partial_\mu \bar{\epsilon}_{\alpha}(X) ce^{-\phi}\psi^\mu \bar{c}e^{-\frac{\bar{\phi}}{2}}\bar{S}^\alpha - \frac{i}{4}(\slashed{\partial}\bar{\epsilon})^\alpha c e^{-2\phi}\partial\xi \bar{c}\bar{\eta}e^{\frac{\bar{\phi}}{2}}\bar{S}_\alpha,
\end{aligned}
\ee
the equations (\ref{genericSuperIsometry}) lead to the following relations between the string field profiles
\be
\begin{aligned}
	&-\frac{1}{2} \Box F_{\alpha \beta} = s_{\alpha \beta}, \,\,\, \frac{i}{4} (\slashed{\partial}F)^\alpha_\beta = t^\alpha_\beta, \,\,\,  \frac{i}{4} (F  \overset{\leftarrow}{\slashed{\partial}})^\beta_\alpha = (t^T)^\beta_\alpha, \\
	& \frac{1}{2}\big(-4\Box h_{\mu\nu} -\partial_\mu A_{\nu} - \partial_\nu A_\mu\big) = e_{\mu \nu}, \,\,\,  \frac{1}{4}\big(-2\Box \phi - \partial^\mu A_\mu\big) = f, \,\,\, \frac{1}{4}\big(A_\mu - 2\partial_\mu \phi + 4\partial^\nu h_{\mu \nu}\big) = g_\mu,\\
	&\frac{1}{2}\Box\epsilon_{\alpha}=v_{\alpha}, \, \, \, \frac{1}{2}\partial_{\mu}\epsilon_{\alpha}=u_{\mu\alpha}, \,\,\, -\frac{i}{4}(\slashed{\partial}\epsilon)^{\alpha}=w^{\alpha}, \\
	&\frac{1}{2}\Box\bar{\epsilon}_{\alpha}=\bar{v}_{\alpha},  \, \, \, \frac{1}{2}\partial_{\mu}\bar{\epsilon}_{\alpha}=\bar{u}_{\mu\alpha},  \,\,\, -\frac{i}{4}(\slashed{\partial}\bar{\epsilon})^{\alpha}=\bar{w}^{\alpha}.
\end{aligned}
\ee
The cohomology problem then reduces to the above equations being compatible with each other and the source profiles being exact in the sense of differential cohomology. The various compatibility conditions (we suppress the antiholomorphic conditions on super-isometry since they are related by worldsheet parity) to be satisfied are
\begin{equation}
	\begin{aligned}
    &s_{\alpha\beta}-2i(\slashed{\partial}t)_{\alpha\beta}=0,\quad s_{\alpha\beta}-2i(t^T\overset{\leftarrow}{\slashed{\partial}})_{\alpha\beta}=0, \quad \frac{1}{2} \partial^\nu e_{\mu \nu} - \partial_\mu f + \Box g_\mu = 0, \\
    &\partial_\mu v_{\alpha}-\Box u_{\mu\alpha}=0, \quad v_{\alpha}-\partial^{\mu}u_{\mu\alpha}=0,\quad \frac{i}{2}(\gamma^{\mu}u_{\mu})^{\alpha}+w^{\alpha}=0,\quad v_{\alpha}-2i(\slashed{\partial}w)_{\alpha}=0. \\
\end{aligned}
\end{equation}
Moreover, we need the gravitino variation
\be
\delta \psi_{\mu \alpha} \equiv \frac{1}{2} \partial_\mu \epsilon_\alpha - u_{\mu \alpha}
\ee
to vanish. This can be arranged by tuning the Killing spinor $\epsilon$ provided that the integrability condition (a differential cohomology condition)
\be
\partial_{[\mu} u_{\nu] \alpha} = 0
\ee
is satisfied.
These constraints are to be satisfied on the space of sources compatible with the self-consistent ansatz (\ref{masslessansatz}), (\ref{phiAnsatz}). This space is explored in detail in section \ref{obstructionSec}.

\subsection{Potential obstructions}
In this subsection, we characterize the potential obstructions to finding a solution to the extended system (\ref{extendedProblemNew}). Then we identify their shifts under the addition of a $Q$-closed term to the string field solution, which further refines the cohomology problem.

\subsubsection{Form of the potential obstructions}

The fact that the source must be $Q$-closed implies that
\begin{equation}
	\begin{aligned}
    &s_{\alpha\beta}-2i(\slashed{\partial}t)_{\alpha\beta}=0,\quad s_{\alpha\beta}-2i(t^T\overset{\leftarrow}{\slashed{\partial}})_{\alpha\beta}=0, \quad \frac{1}{2} \partial^\nu e_{\mu \nu} - \partial_\mu f + \Box g_\mu = 0,\\
    &\partial_\mu v_{\alpha}-\Box u_{\mu\alpha}=0, \quad v_{\alpha}-\partial^{\mu}u_{\mu\alpha}=0, \quad \frac{i}{2}(\slashed{\partial}u_{\mu})^{\alpha}+\partial_{\mu}w^{\alpha}=0,\quad v_{\alpha}-2i(\slashed{\partial}w)_{\alpha}=0,
	\end{aligned}
\end{equation}
which suffices to establish that some physical background string field with the ansatz (\ref{masslessansatz}) can be found, but is weaker than the consistency conditions required for the existence of 32 super-isometries. As a consequence, maximal super-isometry is potentially obstructed. To make this precise, the above relations only guarantee that
\be
\begin{aligned}
	\partial^\mu \partial_{[\mu} u_{\nu] \alpha} = 0, \quad \frac{i}{2}(\slashed{\partial}u_{\mu})^{\alpha}+\partial_{\mu}w^{\alpha} = \partial_{\mu}\Big(\frac{i}{2}(\gamma^{\nu}u_{\nu})^{\alpha}+w^{\alpha}\Big)+\frac{i}{2}(\gamma^{\nu} \partial_{[\mu} u_{\nu]})^{\alpha} = 0,
\label{eqWithTwoForm}
\end{aligned}
\ee
so that the obstructions correspond to the following combinations of source profiles in (\ref{genericSuperIsometry})
\be
\delta \lambda^\alpha \equiv \frac{i}{2}(\gamma^{\mu}u_{\mu})^{\alpha}+w^{\alpha}=\chi^\alpha - \frac{i}{2} \int \dd X^\mu (\gamma^\nu \Omega_{\mu \nu})^\alpha, \,\,\, \partial_{[\mu} u_{\nu] \alpha} = \Omega_{\mu \nu \alpha}
\label{obstructions}
\ee
and their conjugates, with $\chi^\alpha$ a constant spinor and $\Omega_{\mu \nu \alpha}$ a divergenceless 2-form-spinor (compatible with the symmetries of the ansatz). The above potential obstructions (\ref{obstructions}) correspond to having nonzero dilatino variation $\delta \lambda^\alpha$, and an obstruction to the gravitino variation integrability equation. The constant term $\chi^\alpha$ and the divergenceless $\Omega_{\mu \nu \alpha}$ are sufficient to fully characterize the potential obstructions and will be chosen to parametrize them in the following discussion. In subsection \ref{interpret}, the physical interpretation of these obstructions will be related to modifying the dilatino variation by a constant term not present in two-derivative IIB SUGRA and to turning on a nontrivial quadratic profile for the physical dilaton.

\subsubsection{Shifting the potential obstructions}

Of course, finding the form (\ref{obstructions}) of the obstructions is not the full story. Indeed, the obstructions can potentially be shown to be exact by considering shifts of the background solution which preserve the equations of motion.

To see how this works, note that we can shift the $n$-th order background solution by $Q$-closed terms (respecting the ansatz (\ref{masslessansatz})) without affecting its equations of motion. However, this procedure potentially shifts the super-isometry sources at $n$-th order. Due to the general structure of our solution explained below (\ref{masslessansatz}),  at odd order $n$ we can only shift the (R,R) component, whereas at even $n$ only the (NS,NS) part can be shifted.

After computing the relevant 2-brackets with zeroth order super-isometry generators (those are computed in appendix \ref{detailsEOM} provided one trivially relabels the string fields), one can show that, at this stage, the dilatino variation and the gravitino variation integrability conditions get shifted as (we remind the reader about the superscript notation we introduce along the lines of (\ref{pertExp}))
\be
\begin{aligned}
&\delta \lambda^{(2n)\alpha} \to \delta \lambda^{(2n)\alpha} + \frac{i}{16}(\epsilon^{(0)}\slashed{\partial})^{\alpha}\Big(\delta h^{(2n)}-\delta \phi^{(2n)}\Big), \\
& \Omega^{(2n)}_{\mu \nu \alpha} \to \Omega^{(2n)}_{\mu \nu \alpha} -\frac{1}{8}(\epsilon^{(0)}\gamma^{\rho}\slashed{\partial})_{\alpha}\partial_{[\mu}\delta h^{(2n)}_{\nu]\rho}+\frac{1}{8}\epsilon^{(0)}_{\alpha}\partial^{\rho}\partial_{[\mu}\delta h^{(2n)}_{\nu]\rho}, \\
& \delta \bar{\lambda}^{(2n-1)\alpha} \to \delta \bar{\lambda}^{(2n-1)\alpha}  + \frac{1}{8}(\gamma^{\mu}\delta F^{(2n-1)}\gamma_{\mu}\epsilon^{(0)})^{\alpha}, \\
&\bar{\Omega}^{(2n-1)}_{\mu \nu \alpha} \to \bar{\Omega}^{(2n-1)}_{\mu \nu \alpha} - \frac{i}{4}\partial_{[\mu}\delta F^{(2n-1)}\gamma_{\nu]}{\epsilon}^{(0)}_{\alpha},
\end{aligned}
\ee
where the fluctuations are on-shell, that is (the auxiliary field shift $\delta A_\mu^{(2n)}$ is integrated out)
\be
\begin{aligned}
    &\slashed{\partial} \delta F^{(2n-1)} = \delta F^{(2n-1)}  \overset{\leftarrow}{\slashed{\partial}} = 0, \\
    &\Box \delta h^{(2n)}_{\mu\nu} - \partial_\mu \partial^\rho \delta h_{\nu \rho}^{(2n)} - \partial_\nu \partial^\rho \delta h_{\mu \rho}^{(2n)} +  \partial_\mu \partial_\nu \delta \phi^{(2n)} = 0, \\
    &\Box\big(\delta h^{(2n)} - \delta \phi^{(2n)}) = 0.
\end{aligned}
\ee

It turns out that there is only a single on-shell fluctuation that shifts the obstructions and respects the ansatz (\ref{masslessansatz}) i.e. that preserves $SO(5) \cross SO(5)$ invariance and has the correct discrete symmetry charges (one has to keep in mind that the radius transforms, making this shift possible only at orders $2(2n-1)$). It takes the form
\be
\begin{aligned}
&\delta h_{\mu\nu}^{(2[2n-1])}(X)=\frac{1}{4}(\delta_{\mu}^a\delta_{\nu}^bX_aX_b-\delta_{\mu}^i\delta_{\nu}^jX_iX_j),\\
&\delta \phi^{(2[2n-1])}(X)=\frac{3}{4} (X_a X^a - X_i X^i).
\label{osshift}
\end{aligned}
\ee
The physical interpretation of this fluctuation is that it turns on a radial profile for the physical dilaton $\Phi$ (see appendix \ref{diffeoApp}), shifting both the dilatino and gravitino variations via (\ref{obstructions}) since it turns on a nontrivial $\Omega_{\mu \nu \alpha}^{(2[2n-1])}$.
This leads to the refined form of the cohomology problem
\be
\begin{aligned}
\label{shifts}
&\partial_\mu \chi^{{(2n)}\alpha} = 0, \, \chi^{(2n)\alpha} \sim \chi^{{(2n)}\alpha} ,\\
& \partial^\mu \Omega_{[\mu \nu] \alpha}^{(2[2n-1])} = 0, \, \Omega_{\mu \nu \alpha}^{(2[2n-1])} \sim \Omega_{\mu \nu \alpha} ^{(2[2n-1])}+ (\delta_{\mu}^a\delta_{\nu}^b\gamma_{ab}-\delta_{\mu}^i\delta_{\nu}^j\gamma_{ij})_\alpha^\beta\epsilon^{(0)}_\beta, \\
& \partial^\mu \Omega_{[\mu \nu] \alpha}^{(2[2n+1])} = 0, \, \Omega_{\mu \nu \alpha}^{(2[2n+1])} \sim \Omega_{\mu \nu \alpha} ^{(2[2n+1])},\\
& \partial_\mu \bar{\chi}^{{(2n-1)}\alpha} = 0, \, \bar{\chi}^{(2n-1)\alpha} \sim \bar{\chi}^{{(2n-1)}\alpha},\\
&\partial^\mu \bar{\Omega}_{[\mu \nu] \alpha} ^{(2n-1)}= 0, \, \bar{\Omega}_{\mu \nu \alpha} ^{(2n-1)}\sim \bar{\Omega}_{\mu \nu \alpha}^{(2n-1)},
\end{aligned}
\ee
on the space of obstructions compatible with the symmetries of (\ref{masslessansatz}). We see that the shifts one can make are very limited and so to prove any kind of existence of our solution, the space of obstructions must be largely restricted by symmetry. We return to the study of what the space of obstructions is in full generality in the later section \ref{obstructionSec}. For now, it suffices to say that the second order in perturbation theory $AdS_5 \cross S^5$ solution found in the next section is not obstructed. This means that one can tune the background string field so that the second-order solution to (\ref{extendedProblemNew}) is found. This requires turning on a constant self-dual 5-form flux and keeping the physical axiodilaton constant, which is in line with the SUGRA expectation.

\section{The $AdS_5 \cross S^5$ solution to the extended string field equations}
\label{solext}
In this section, we find the $AdS_5 \cross S^5$ solution of the extended string field equations (\ref{extendedProblemNew}). Recall that this corresponds to having an $SO(5) \cross SO(5)$ invariant background solution, which supports 32 super-isometries and appropriately transforms under the various discrete symmetries introduced in the last section. The solution of this extended system is found up to second order in the large radius expansion $\psi = \sum_{n=1}^\infty \mu^n \psi^{(n)}, \, \lambda = \sum_{n=0}^\infty \mu^n \lambda^{(n)}$ and we also give a candidate solution to the massless equations of motion (\ref{masslesseom}) at third order.

\subsection{Zeroth order}
At zeroth order, the extended equations are simply $Q \lambda^{(0)} = 0$, which using the kinetic terms (\ref{kinpart}) gives
\be
\partial_\mu \epsilon^{(0)}_\alpha = 0, \quad \partial_\mu \bar{\epsilon}^{(0)}_\alpha = 0.
\ee
This simply means that the Killing spinor is constant and, as remarked in the last section, we take $\epsilon^{(0)}_\alpha \neq 0$ with $\bar{\epsilon}^{(0)}_\alpha = 0$ to focus on a single set of supercharges (without loss of generality).

\subsection{First order}
\label{firstOrd}
To proceed to first order, we have to solve
\be
\begin{aligned}
	&Q \psi^{(1)} = 0, \\
	&Q \lambda^{(1)} = -[\psi^{(1)} \otimes \lambda^{(0)}]'
\end{aligned}
\ee
so that we need to compute the massless string bracket $[\psi^{(1)} \otimes \lambda^{(0)}]'$. This is done in appendix \ref{firstSUSY} and the result is
\begin{align}
    [\psi^{(1)} \otimes \lambda^{(0)} ]' = \frac{i}{4} (F^{(1)}\gamma_{\mu}\epsilon^{(0)})_{\alpha}ce^{-\phi}\psi^{\mu}\bar{c}e^{-\frac{1}{2}\B{\phi}}\B{S}^{\alpha}.
\end{align}
This gives the equations
\be
\begin{aligned}
	&\Box F^{(1)}_{\alpha\beta}=0,\,\,\,(\slashed{\partial}F^{(1)})^\alpha_\beta = 0, \,\,\, (F^{(1)} \overset{\leftarrow}{\slashed{\partial}})^\alpha_\beta = 0,\\
	& \Box \epsilon^{(1)}_\alpha = 0, \,\,\, (\slashed{\partial} \epsilon^{(1)})^\alpha = 0, \,\,\, \partial_\mu \epsilon^{(1)}_\alpha = 0 ,\\
	& \Box \bar{\epsilon}^{(1)}_\alpha = 0, \,\,\, (\slashed{\partial} \bar{\epsilon}^{(1)})^\alpha = 0, \,\,\, \partial_\mu \bar{\epsilon}^{(1)}_\alpha = \frac{i}{2} (F^{(1)} \gamma_\mu \epsilon^{(0)})_\alpha.
	\label{eqFirst}
\end{aligned}
\ee
It is clear that a solution to these equations is a constant self-dual 5-form flux, which we take to be of the form $F^{(1)} = N_{\text{R}} \gamma^{01234}$ \footnote{One could also write $F^{(1)} = \frac{1}{2} N_{\text{R}} (\gamma^{01234} - \gamma^{56789})$, where we used that $\gamma^{56789} = -\gamma^{11} \gamma^{01234}$ with $\gamma^{11} = \gamma^{0\ldots9}$ and the $\gamma^{11}-$eigenvalue assignment +1 of the chiral spinors $S^\alpha, \bar{S}^\alpha$ multiplying the flux.} ($N_{\text{R}}$ is a normalization to be determined later), together with $\epsilon^{(1)} = 0,\, \bar{\epsilon}^{(1)} = \frac{i}{2} F^{(1)} \slashed{X} \epsilon^{(0)}$. To show that the dilatino variation vanishes and that the condition for vanishing of the gravitino variation is integrable (for both refer to (\ref{obstructions})), one simply uses $\gamma^\mu F^{(1)} \gamma_\mu = 0$ and the fact that $F^{(1)}$ is constant.

\subsection{Second order}
\label{secondOrd}
At second order, we solve the extended system
\be
\begin{aligned}
 &Q \psi^{(2)} = -\frac{1}{2!} [(\psi^{(1)})^{\otimes 2}]', \\
 &Q \lambda^{(1)} = -[\psi^{(1)} \otimes \lambda^{(1)}]' - [\psi^{(2)} \otimes \lambda^{(0)}]' - \frac{1}{2!} [(\psi^{(1)})^{\otimes 2} \otimes \lambda^{(0)}]'.
\end{aligned}
\ee

Using the first-order results, we can immediately write down
\begin{align}
    [\psi^{(1)} \otimes \lambda^{(1)}]' =&-\frac{i}{4} F^{(1)}_{\alpha \beta} (\gamma_\mu)^{\beta \gamma} \bar{\epsilon}^{(1)}_{\gamma}  c e^{-\frac{1}{2}\phi}S^\alpha \bar{c}e^{-\bar{\phi}}\bar{\psi}^\mu
\end{align}
and so we are only left with computing the brackets $[(\psi^{(1)})^{\otimes 2}]',\, [\psi^{(2)} \otimes \lambda^{(0)}]'$ and $[(\psi^{(1)})^{\otimes 2} \otimes \lambda^{(0)}]'$. The source for the background string field equation of motion is computed in appendix \ref{secondEOMdetails}, and takes the form
\be
[(\psi^{(1)})^{\otimes 2}]' = 2 \Tr (F^{(1)} \gamma_\mu F^{(1)} \gamma_\nu) c_0^+ c\bar{c}e^{-\phi}\psi^\mu e^{-\bar{\phi}}\bar{\psi}^\nu.
\ee
As far as the super-isometry goes, the bracket $[\psi^{(2)} \otimes \lambda^{(0)}]'$ is computed in Appendix \ref{secondSUSY} and gives
\be
\begin{aligned}
[\psi^{(2)} \otimes \lambda^{(0)}]' =
&+ \big(\frac{3i}{16} \epsilon^{(0)}_{\beta}(\gamma^\mu)^{ \beta\alpha} \partial^\nu h^{(2)}_{\mu\nu} + \frac{2 i}{64} \slashed{A}^{(2)\alpha \beta} \epsilon^{(0)}_{\beta} \big) c \eta e^{\frac{1}{2}\phi} S_\alpha \bar{c} e^{-2\bar{\phi}}\bar{\partial}\bar{\xi}  \\
& + \big( \frac{1}{8} \epsilon^{(0)}_{\beta}(\gamma^\nu)^{\beta \gamma} \slashed{\partial}_{\gamma \alpha} h^{(2)}_{\mu\nu} + \frac{3}{16} \partial_\mu \phi^{(2)} \epsilon^{(0)}_{\alpha} + \frac{1}{32} \epsilon^{(0)}_{\alpha} A^{(2)}_\mu\big) c e^{-\frac{1}{2}\phi} S^\alpha \bar{c} e^{-\bar{\phi}}\bar{\psi}^\mu \\
& -\big(\frac{3}{8}\epsilon^{(0)}_{\beta}(\gamma^\nu)^{\beta \gamma} \partial^\mu \slashed{\partial}_{ \gamma\alpha} h^{(2)}_{\mu\nu} + \frac{2}{32} \epsilon^{(0)}_{ \beta} \slashed{A}^{(2)\beta\gamma}  \overset{\leftarrow}{\slashed{\partial}}_{ \gamma \alpha} \big) c_0^+ c e^{-\frac{1}{2}\phi} S^\alpha \bar{c} e^{-2\bar{\phi}}\bar{\partial} \bar{\xi}.
\end{aligned}
\ee
Finally, the three-bracket $[(\psi^{(1)})^{\otimes2}\otimes\lambda^{(0)}]'$ is also computed in Appendix \ref{secondSUSY} and shown to be zero.

This means that the second-order extended string field equations are
\begin{align}
	\label{curvatureEq}
	&4\Box h^{(2)}_{\mu\nu} + \partial_\mu A^{(2)}_{\nu} + \partial_\nu A^{(2)}_\mu = 2 \Tr (F^{(1)} \gamma_\mu F^{(1)} \gamma_\nu), \\        \label{boxEq}
	&2\Box\phi^{(2)} + \partial^\mu A^{(2)}_\mu = 0, \\
	\label{auxEOMeq}
	&A^{(2)}_\mu - 2\partial_\mu\phi^{(2)} + 4\partial^\nu h^{(2)}_{\mu\nu} = 0, \\
	\label{firstSUSYeq}
	&\frac{i}{4} (\slashed{\partial}\epsilon^{(2)})^\alpha = \frac{3i}{16} \epsilon^{(0)}_{\beta}(\gamma^\mu)^{\beta \alpha} \partial^\nu h^{(2)}_{\mu\nu}  + \frac{2 i}{64} \slashed{A}^{(2)\alpha \beta} \epsilon^{(0)}_{ \beta}, \\
	\label{secondSUSYeq}
	&\frac{1}{2}\partial_\mu \epsilon^{(2)}_{\alpha} = \frac{1}{8} \epsilon^{(0)}_{\beta}(\gamma^\nu)^{\beta \gamma} \slashed{\partial}_{\gamma \alpha} h^{(2)}_{\mu\nu} + \frac{3}{16} \partial_\mu \phi^{(2)} \epsilon^{(0)}_{\alpha} + \frac{1}{32} \epsilon^{(0)}_{\alpha} A^{(2)}_\mu -\frac{i}{4} F^{(1)}_{\alpha \beta}(\gamma_{\mu})^{\beta\gamma}\B{\epsilon}^{(1)}_{\gamma},\\
	&\frac{1}{2} \Box \epsilon^{(2)}_{\alpha} = \frac{3}{8}\epsilon^{(0)}_{\beta}(\gamma^\nu)^{\beta \gamma} \partial^\mu \slashed{\partial}_{ \gamma\alpha} h^{(2)}_{\mu\nu} + \frac{2}{32} \epsilon^{(0)}_{ \beta} \slashed{A}^{(2)\beta\gamma}  \overset{\leftarrow}{\slashed{\partial}_{ \gamma \alpha}},
\end{align}
and we continue by examining the vanishing of the dilatino and gravitino variations, i.e., under which conditions do these equations have a solution. The dilatino vanishing condition is the compatibility between (\ref{firstSUSYeq}) and (\ref{secondSUSYeq}) when the latter is hit with a gamma matrix. Using (\ref{auxEOMeq}) together with the extended equations from previous order, it is explicitly given by
\be
\epsilon^{(0)} \big(6 \partial_\mu \phi^{(2)} - A^{(2)}_\mu + 4\gamma^\nu\slashed{\partial}  h^{(2)}_{\mu \nu} - 12 \partial^\nu h^{(2)}_{\mu \nu}\big)\gamma^\mu  = 4 \epsilon^{(0)} \slashed{\partial}\big(\phi^{(2)} - h^{(2)}\big) = 0,
\label{dilatonEq}
\ee
which just means that the physical dilaton has to be kept constant (see the discussion on diffeomorphism in appendix \ref{diffeoApp}). The vanishing of the gravitino variation (\ref{secondSUSYeq}) is subject to an integrability condition, which using the super-isometry equation (\ref{eqFirst}) for $\bar{\epsilon}^{(1)}$ has the form
\be
\frac{1}{8} \epsilon^{(0)} \gamma^\nu \slashed{\partial} \partial_\rho h^{(2)}_{ \mu \nu} + \frac{1}{32} \epsilon^{(0)} \partial_\rho A^{(2)}_\mu + \frac{1}{8} F^{(1)} \gamma_\mu F^{(1)} \gamma_\rho \epsilon^{(0)} - \mu \leftrightarrow \rho = 0.
\label{integrEq}
\ee

For the sake of simplicity, we first solve these constraints in Siegel gauge $A_\mu^{(2)} = 0$ (later we will also present the solution in a different gauge, in which the metric takes the form of (\ref{sugraSol})). Note that the extended system is overdetermined and we first solve the system of equations (\ref{boxEq}), (\ref{auxEOMeq}) with (\ref{dilatonEq}), (\ref{integrEq}), and then show that (\ref{curvatureEq}) is satisfied. The ansatz for the metric that is compatible with (\ref{masslessansatz}) takes the form
\be
\begin{aligned}
    \label{hAnsatz}
    h^{(2)}_{ \mu \nu} =  \alpha_1 &\big(\delta^a_\mu \delta^b_\nu X_a X_b - \delta^i_\mu \delta^j_\nu X_i X_j\big) + \alpha_2 \big(\eta_{ab}\delta_{\mu}^a\delta_{\nu}^bX_cX^c-\delta_{ij}\delta_{\mu}^i\delta_{\nu}^jX_kX^k\big)
\end{aligned}
\ee
and in appendix \ref{integrabilityAppendix}, we show that the above extended system implies $\alpha_1 = -\frac{3}{7} N_{\text{R}}^2, \, \alpha_2 = -\frac{5}{7} N_{\text{R}}^2$. Then, the expression for the ghost dilaton is determined from (\ref{dilatonEq}). In the appendix, we also check that the linearized Einstein equation (\ref{curvatureEq}) is satisfied. This thus leads us to the unique Siegel gauge solution
\be
\begin{aligned}
    h^{(2)}_{\mu \nu} &=-\frac{3}{7} N_{\text{R}}^2 \big(\delta^a_\mu \delta^b_\nu X_a X_b - \delta^i_\mu \delta^j_\nu X_i X_j\big) -\frac{5}{7} N_{\text{R}}^2 \big(\eta_{ab}\delta_{\mu}^a\delta_{\nu}^bX_cX^c-
    \delta_{ij}\delta_{\mu}^i\delta_{\nu}^jX_kX^k\big), \\
	    \phi^{(2)} &= - 4 N_{\text{R}}^2 X^2 ,\\
    A^{(2)}_\mu &= 0.
    \label{ssol}
\end{aligned}
\ee
In appendix \ref{gaugeTransformAppendix}, we show that the Siegel gauge solution can be gauge transformed to the solution
\be
\begin{aligned}
	h^{(2)}_{\mu \nu} &=N_{\text{R}}^2\big(\delta^a_\mu \delta^b_\nu X_a X_b - \delta^i_\mu \delta^j_\nu X_i X_j\big),\\
	\phi^{(2)} &= N_{\text{R}}^2X^2 ,\\
	A^{(2)}_\mu &= -10 N_{\text{R}}^2\partial_\mu X^2,
	\label{oos}
\end{aligned}
\ee
which better manifests the isometries of $AdS_5 \times S^5$, having the form of the SUGRA solution in appendix \ref{sugraSol}. This form of the solution will enable us to pin down the normalization constant $N_R$ by demanding that up to second order, the string field frame is aligned with the field frame of the SUGRA solution.

This requires $h_{\mu \nu}^{(2)}$ to be precisely the second-order metric fluctuation of the SUGRA solution, which gives $\mu^2 N_R^2 = -\frac{1}{R^2}$. This is because the metric fluctuation $h_{\mu \nu}$ is normalized such that it corresponds to a deformation of the worldsheet action by a NSNS sector picture 0 vertex operator (this is the reason for the factor of 4 in the vertex operator $V_{NSNS}^{\mu \nu}$, which is tied to the factors of $\frac{1}{2}$ in (\ref{PCO1})) and, even though it is not an on-shell profile, the mixing of $h_{\mu \nu}^{(2)}$ is potentially only with two first-order RR fluxes, which are constant. We choose the orientation such that $\mu N_{\text{R}} = \frac{i}{R}$.  Note that the factor of $i$ matches the normalization of \cite{Alexandrov:2021dyl} (we use the same worldsheet CFT conventions). The normalization will be further checked in the next subsection, as well as in subsection \ref{theAxion} by yielding a canonically normalized massless wave equation for the RR axion.

Note that we should make sure that the above solutions have the correct $r_0$ dependence (see the appendix \ref{sec:flatvert}) as dictated by the Hata-Zwiebach \cite{Hata:1993gf} double 2-bracket (\ref{hata2}). This prescription requires that we shift the second-order metric fluctuation by the on-shell fluctuation $-4 N_{\text{R}}^2 \ln r_0 \widetilde{R}_{\mu \nu}$ with $\widetilde{R}_{\mu \nu}$ the rescaled Ricci tensor (\ref{ricciT}).

\subsection{Third order}
At third order in the radius expansion, we only find a solution to the equations of motion
\be
Q \psi^{(3)} = -[\psi^{(2)} \otimes \psi^{(1)}]' - \frac{1}{3!} [(\psi^{(1)})^{\otimes 3}]',
\ee
leaving the determination of whether it is maximally super-isometric to the future. We note that since there is only one profile for the RR component as opposed to the three profiles we had to solve for in the NSNS one, we expect the solution to the equations of motion to be essentially unique without imposing further constraints, and thus very likely maximally super-isometric. One purpose of writing down this solution at present is that it provides an extra check on the normalization constant $N_{\text{R}}$, as advertised above.

The computation of the bracket $[\psi^{(2)} \otimes \psi^{(1)}]'$ can be found in appendix \ref{AppEOM3rd} and leads to
\begin{equation}
\begin{split}
    [\psi^{(2)} \otimes \psi^{(1)}]'=&+\frac{1}{8}(\slashed{\partial}\gamma^{\mu} F^{(1)}\gamma^{\nu}\overset{\leftarrow}{\slashed{\partial}})_{\alpha \beta}h^{(2)}_{\mu\nu}c_0^+c\B{c}e^{-\frac{1}{2}\phi}S^{\alpha}e^{-\frac{1}{2}\B{\phi}}\B{S}^{\beta},\\
    &-\frac{i}{16}(\slashed{\partial}\gamma^{\mu}F^{(1)}\gamma^{\nu})_{\alpha}^{\beta}h^{(2)}_{\mu\nu}ce^{-\frac{1}{2}\phi}S^{\alpha}\B{c}\B{\eta}e^{\frac{1}{2}\B{\phi}}\B{S}_{\beta},\\
    &-\frac{i}{16}(\gamma^{\mu} F^{(1)}\gamma^{\nu}\overset{\leftarrow}{\slashed{\partial}})^\alpha_{
    \beta} h^{(2)}_{\mu\nu}c\eta e^{\frac{1}{2}\phi}S_{\alpha}\B{c}e^{-\frac{1}{2}\B{\phi}}\B{S}^{\beta}.
\end{split}
\end{equation}
One can also check that the 3-bracket $[(\psi^{(1)})^{\otimes 3}]'$ vanishes as is also shown in appendix \ref{AppEOM3rd}. The equations of motion for $F^{(3)}$ can therefore be written as
\begin{align}
    \frac{1}{2}\Box F^{(3)}&=\frac{1}{8}\slashed{\partial}\gamma^{\mu}F^{(1)}\gamma^{\nu}\slashed{\partial}h^{(2)}_{\mu\nu},\\
    \label{f31}
    \frac{i}{4}\slashed{\partial}F^{(3)}&=\frac{i}{16}\gamma^{\mu}F^{(1)}\gamma^{\nu}\slashed{\partial}h^{(2)}_{\mu\nu},\\
    \label{f32}
    \frac{i}{4}F^{(3)}\overset{\leftarrow}{\slashed{\partial}}&=\frac{i}{16}\slashed{\partial}\gamma^{\mu}F^{(1)}\gamma^{\nu}h^{(2)}_{\mu\nu}.
\end{align}
Note that it is sufficient to solve (\ref{f31}) as the other two equations follow.

We wish to solve the Dirac equation (\ref{f31}) for the third order flux $F^{(3)}$
\begin{equation}
	\frac{i}{4}\slashed{\partial}F^{(3)}=\frac{i}{16} \gamma^{\mu}F^{(1)}\gamma^{\nu}\slashed{\partial}h^{(2)}_{\mu\nu}.
\end{equation}
By plugging in the relaxed Siegel gauge solution (\ref{oos}), we can simplify
\begin{equation}
	\frac{i}{16}\gamma^{\mu}F^{(1)}\gamma^{\nu}\slashed{\partial}h^{(2)}_{\mu\nu}=\frac{i}{8} N_{\text{R}}^2 F^{(1)}\slashed{X},
\ee
so that it is straightforward to check that the equation of motion is solved for
\begin{equation}
	F^{(3)}=\frac{N_{\text{R}}^3}{4}X^2\gamma^{01234}.
\end{equation}
Consequently, the Ramond-Ramond flux solution up to third order in $\mu$ is given by
\begin{equation}
    F=\mu N_{\text{R}}\big(1 + \frac{1}{4}\mu^2 N_{\text{R}}^2 X^2 + \ldots\big)\gamma^{01234},
\end{equation}
and plugging in the previously chosen normalization $\mu N_{\text{R}} = \frac{i}{R}$ gives the flux
\ie{}
F &= \frac12(\omega_{AdS_5} + \omega_{S^5}) \gamma^{01234}, \\
\omega_{AdS_5} &= \frac{1}{R} \Big(1 + \frac{1}{R^2} X_a X^a \Big)^{-1/2} \sim \frac{1}{R} \Big(1 - \frac{1}{2R^2} X_a X^a + \ldots\Big), \\
\omega_{S^5} &= \frac{1}{R} \Big(1 - \frac{1}{R^2} X_i X^i \Big)^{-1/2} \sim \frac{1}{R} \Big(1 + \frac{1}{2R^2} X_i X^i + \ldots\Big),
\fe
as can be found in appendix \ref{sugraSol}.
We can thus finally write down the solution (\ref{covsol}).  As far as the Siegel gauge solution goes, after doing the inverse gauge transform, one obtains
\be
F = \frac{i}{R}\big(1 + \frac{X^2}{R^2} + \ldots)\gamma^{01234},
\ee
finally yielding (\ref{siegelsolution}). Note that at this order one might also turn on a constant flux coming with a power of $\alpha'$ by adding a $Q$-closed term to the solution.

\section{The pp-wave limit}
\label{sec:pp}

It is well-known that in the BMN limit \cite{Berenstein:2002jq}, the $AdS_5 \times S^5$ solution turns into a pp-wave background \cite{Blau:2001ne}. The SUGRA background solution then truncates at second order and the background is exactly solvable in the Green-Schwarz formalism since the magnons on the Green-Schwarz string become free \cite{Metsaev:2001bj, Metsaev:2002re}. We are interested in seeing whether a similar truncation happens in SFT. To see this, we first construct the solution up to second order by imposing maximal super-isometry, essentially obtaining the solution of \cite{Cho:2018nfn}. We then show that there are no higher order corrections for the massless string fields. This follows from a simple worldsheet symmetry argument.

The analysis essentially mirrors that of the last section and so we shall be brief. The first-order 5-form constant flux for this solution is $F^{(1)} = N_{\text{R}} \gamma^+ \gamma^1 \gamma^2 \gamma^3 \gamma^4$ and preserves supersymmetry as in subsection \ref{firstOrd}. At second order, the extended string field equations take the same form as in subsection \ref{secondOrd}. We solve them in Siegel gauge $A_\mu^{(2)} = 0$ by starting with a general pp-wave ansatz
\be
h^{(2)}_{\mu \nu} = H(X_\perp, X_+, X_-) \delta^+_\mu \delta^+_\nu.
\ee
Since $h^{(2)} = 0$, we take the constant physical dilaton solution $\phi^{(2)} = 0$ so that (\ref{dilatonEq}) is satisfied. The auxiliary equation of motion (\ref{auxEOMeq}) in Siegel gauge then gives $\partial^\nu h_{\mu \nu}^{(2)} = 0$, which eliminates $X_+$ dependence from $H$. The gravitino variation integrability condition (\ref{integrEq}) splits into two equations
\begin{align}
\label{firstpp}
&\gamma^+ \slashed{\partial}\partial_- h^{(2)}_{ + +} = 0, \\
\label{secondpp}
&\gamma^+ \slashed{\partial} \partial_i h^{(2)}_{ + +} - 2 N_{\text{R}}^2 \gamma^+ \gamma_i  = 0.
\end{align}
The first equation (\ref{firstpp}) is trivially satisfied since $H$ has no $X_+$ dependence
and the second equation (\ref{secondpp}) sets $\partial_i \partial_j h^{(2)}_{++} = -2 \eta_{i j}$, giving the unique Siegel gauge solution of the BMN type \cite{Berenstein:2002jq}
\be
h^{(2)}_{\mu \nu} =  N_{\text{R}}^2 (X_\perp)^2 \delta^+_\mu \delta^+_\nu,
\ee
where we used that the arbitrary $X_-$ dependence can be gauged away. The equation (\ref{curvatureEq}) is satisfied since
\be
\Box h^{(2)}_{\mu \nu} = 16 N_{\text{R}}^2 \delta^+_\mu \delta^+_\nu = \frac{1}{2}\Tr\big(F^{(1)} \gamma_\mu F^{(1)} \gamma_\nu\big).
\ee
 This uniqueness is consistent with the analysis of \cite{Blau:2001ne}, which shows that only such a pp-wave preserves maximal SUSY at the SUGRA level. Note that this SFT solution was already found just by solving the equations of motion in \cite{Cho:2018nfn}.

 We now argue that the massless string field equations of motion receive no corrections at higher orders. This follows from considering the boost symmetry with the holomorphic current
\be
J = X^+ \partial X^- - X^- \partial X^+ + 2\psi^+ \psi^-
\ee
and its conjugate. These currents define a boost charge, which commutes with the BRST charge and consequently with the PCO. Because OPE respects covariance, it is then clear that the string bracket preserves the boost symmetry.
Under this symmetry, $\psi^{(1)}$ carries charge 1, $\psi^{(2)}$ charge 2 and using lightcone kinematics, one can verify that there are no zero lightcone momentum invariant operators of charge higher than 2 in the RR sector and charge higher than 3 in the NSNS sector. Since the boost charge adds under the string bracket as noted before, this means that all brackets at higher than second order in the string field equations vanish.

Note that this does not mean that the massive part of the solution truncates. In particular, this has the effect that observables in this background generically receive corrections in the mass parameter. Similarly, the full extended system does not truncate, and the super-isometry generators get corrected to all orders in perturbation theory.

\section{Spectrum of fluctuations}
\label{sec:axion}

\subsection{Linearized fluctuations}

Given a background solution, one wants to compute observables. A natural start is to compute the spectrum of linearized fluctuations $\Phi$, which can then be used to, e.g., compute string scattering. These are solutions of the linearized equations of motion
\be
Q_\Psi \Phi = Q \Phi + \sum\limits_{n=1}^\infty \frac{1}{n!}[\Psi^{\otimes n}\otimes \Phi] = 0,
\ee
which can be again solved by splitting $\Phi$ into massless and massive modes via $\Phi = \phi + \Gamma$ with $\mathbb{P}\Phi = \phi$ and $(1-\mathbb{P})\Phi=\Gamma$ and the massive modes in Siegel gauge $b_0^+ \Gamma = 0$. One then has
\be
\Gamma = -\frac{b_0^+}{L_0^+}\sum\limits_{n=1}^\infty \frac{1}{n!}[\Psi^{\otimes n}\otimes \Phi],
\ee
which then yields the following linearized equations of motion for the massless modes
\be
Q\phi + \sum\limits_{n=1}^\infty \frac{1}{n!}[\psi^{\otimes n}\otimes \phi]' = 0,
\label{linEOM}
\ee
with the massless brackets defined as in (\ref{bracket1}). One can again solve these after taking an ansatz $\phi = \sum_{n=0}^\infty \mu^n \phi^{(n)}$.

Note that one could also demand the fluctuation to transform in a certain way under $L_\infty$ gauge transformations corresponding to super-isometries, thus leading to an extended string field system. In the next subsection, we shall examine the axion scalar fluctuation, which should be demanded to be a $\frac{1}{2}-$BPS massless scalar. However, this will be evident from its equations of motion and as such we will not consider the extended system explicitly.

\subsection{The axion}
\label{theAxion}
In this section, we identify the linearized fluctuation corresponding to the axion to second order in the large radius expansion. This will provide nontrivial checks for the background solution found in section \ref{solext}.

We will solve the linearized equations of motion (\ref{linEOM}) with the ansatz
\be
\phi = f_{\alpha \beta}(X) c\bar{c}e^{-\frac{1}{2}\phi} S^\alpha e^{-\frac{1}{2}\bar{\phi}} \bar{S}^\beta,
\ee
where we demand $\phi$ to be $SO(5) \cross SO(5)$ invariant, worldsheet parity odd and orientifold parity odd so that $f_{\alpha \beta}(X)$ is a one-form field strength with spacetime parity even axion scalar (this being even is compatible with $SO(5) \cross SO(5)$ invariance). Note that it is not necessary to turn on the NSNS string fields (self-consistently imposed to be worldsheet parity even when turning on pure flux at zeroth order). This is because the parities restrict their profiles to be incompatible with $SO(5) \cross SO(5)$ invariance (an odd profile for the metric and ghost dilaton, an even profile for the auxiliary field). By construction, the axion then receives corrections only at even orders in the large radius expansion.
Proceeding as in section \ref{susySec} for the background RR string field, we see that there are no obstructions to solving (\ref{linEOM}).

Similarly to (\ref{kinpart}), the kinetic piece of the equations of motion is given by
\begin{equation}
	Q\phi=-\frac{1}{2}\Box f_{\alpha\beta}c_0^+c\B{c}e^{-\frac{1}{2}\phi}S^{\alpha}e^{-\frac{1}{2}\B{\phi}}\B{S}^{\beta}+\frac{i}{4}\slashed{\partial}^{\gamma\alpha}f_{\alpha\beta}c\eta e^{\frac{1}{2}\phi}S_{\gamma}\B{c}e^{-\frac{1}{2}\B{\phi}}\B{S}^{\beta}+\frac{i}{4}f_{\alpha\beta}\overset{\leftarrow}{\slashed{\partial}}\,^{\beta\gamma}ce^{-\frac{1}{2}\phi}S^{\alpha}\B{c}\B{\eta}\B{S}_{\gamma}
\end{equation}
and it conveniently splits into a 0-form and 2-form part as can be seen via
\be
\begin{aligned}
	\label{diracIdentity}
f \overset{\leftarrow}{\slashed{\partial}} = \partial^\mu f_{\mu} - \partial_\mu f_{\nu}\gamma^{\mu \nu}.
\end{aligned}
\ee
Field strengths obeying the Bianchi identity do not contribute to the 2-form part of the above kinetic term and so solving the linearized equations of motion will amount to solving the 2-form part first and then fixing the resulting ambiguity (obeying the Bianchi identity) from the 0-form part of the equations.

\subsubsection{Zeroth order}
At zeroth order, we solve the linearized equations
\be
Q\phi^{(0)} = 0
\ee
which using (\ref{diracIdentity}) amounts to solving the Bianchi identity by writing $f^{(0)}_\mu = \partial_\mu c^{(0)}$ with
\be
\Box c^{(0)} = 0.
\ee

\subsubsection{First order}
At first order, we solve the linearized equations
\be
Q \phi^{(1)} = -[\psi^{(1)} \otimes \phi^{(0)}]'.
\ee
As noted before, there should be no source for the NSNS fields. To see this explicitly, compute the source as in the appendix (\ref{secondEOMdetails}). It is zero because of gamma matrix kinematics
\be
\Tr(F^{(1)} \gamma^\mu f^{(0)} \gamma^\nu) = 0
\ee
and so one can tune $f^{(1)} = 0$, giving no first-order NSNS contribution to the axion as expected.

\subsubsection{Second order}
\label{secondOrdExt}
At second order, we solve the linearized equations
\be
\label{lineom2}
Q \phi^{(2)} = -[\psi^{(2)} \otimes \phi^{(0)}]' - \frac{1}{2}[(\psi^{(1)})^{\otimes 2}\otimes \phi^{(0)}]'.
\ee
In appendix \ref{axionEOMAppendix}, we wrote out these equations of motion explicitly (working in the relaxed Siegel gauge of (\ref{ssol})) and got
\be
\begin{aligned}
	f^{(2)}  \overset{\leftarrow}{\slashed{\partial}} =  \,&\frac{1}{2}N_{\text{R}}^2\big(5 f^{(0)a} X_a + \partial^b f^{(0)a} X_a X_b - 5 f^i_0 X_i - \partial^j f^i_0 X_i X_j\big)+ \\
&\frac{1}{2}N_{\text{R}}^2 \partial_\mu (\gamma^{\mu b} f^{(0)a} X_a X_b - \gamma^{\mu j}f^{(0)i} X_i X_j - \frac{1}{2} (X_a X^a - X_i X^i) \gamma^{\mu \nu} f^{(0)}_{ \nu}) + \\
& \frac{1}{8} \ln r_0 \big[\Box (2\partial^\mu f^{(0)\nu} h^{(2)}_{\mu \nu}+ 2f^{(0)\nu} \partial^\mu h^{(2)}_{\mu \nu} - f^{(0)}_\mu \partial^\mu h^{(2)}) + \\
&\gamma^{\mu \nu} \partial_\mu \Box(2  f^{(0)\rho} h^{(2)}_{\nu \rho} - f^{(0)}_\nu h^{(2)})
\big],
\end{aligned}
\ee
with $h_{\mu \nu}^{(2)}$ the second-order solution background metric.
Using (\ref{diracIdentity}), we solve the 2-form part of this equation by
\be
\begin{aligned}
&f^{(2)}_{ a} = -\frac{1}{2}N_{\text{R}}^2\big(f^{(0)b} X_b X_a-\frac{1}{2}X^2 f^{(0)}_{ a} \big) - \frac{1}{8}\ln r_0\, \Box(2 f^{(0) b} h^{(2)}_{a b} - f^{(0)}_a h^{(2)}) + \hat{f}^{(2)}_a, \\
&f^{(2)}_{ i} = -\frac{1}{2}N_{\text{R}}^2\big(-f^{(0)j} X_j X_i-\frac{1}{2}X^2 f^{(0)}_{ i}\big) - \frac{1}{8}\ln r_0\, \Box(2 f^{(0) j} h^{(2)}_{i j} - f^{(0)}_i h^{(2)}) + \hat{f}^{(2)}_i,
\end{aligned}
\ee
with $\hat{f}^{(2)}$ obeying the Bianchi identities, which can again be solved via $\hat{f}^{(2)}_\mu = \partial_\mu \hat{c}^{(2)}$. We then use this to compute
\be
\begin{aligned}
	\partial^\mu f^{(2)}_{\mu} = -&\frac{1}{2} N_{\text{R}}^2\big(5 f^{(0)a} X_a - 5 f^{(0)i} X_i + \partial_a f^{(0)b} X_a X_b - \partial_i f^{(0)j} X_i X_j) - \\
				      &-\frac{1}{8} \ln r_0 \Box (2\partial^\mu  f^{(0)\nu} h^{(2)}_{\mu \nu}+2 f^{(0)\nu} \partial^\mu h^{(2)}_{\mu \nu} - f^{(0)}_\mu \partial^\mu h^{(2)}) + \Box \hat{c}^{(2)},
\end{aligned}
\ee
so that once more using the split (\ref{diracIdentity}), the equation for $\hat{c}^{(2)}$ becomes
\be
\begin{aligned}
&\Box \hat{c}^{(2)} - N_{\text{R}}^2(5 X_a \partial^a + X_a X_b \partial^a \partial^b - 5 X_i \partial^i - X_i X_j \partial^i \partial^j) c^{(0)} - \\
& \frac{1}{4}\ln r_0 \Box (2\partial^\mu \partial^\nu c^{(0)} h^{(2)}_{\mu \nu}+2\partial^\nu c^{(0)} \partial^\mu h^{(2)}_{\mu \nu} - \partial_\mu c^{(0)} \partial^\mu h^{(2)})  = 0,
\end{aligned}
\ee
where we plugged in $f^{(0)}_\mu = \partial_\mu c^{(0)}$.
After the field redefinition
\be
c^{(2)} = \hat{c}^{(2)} - \frac{1}{4}\ln r_0 (2\partial^\mu \partial^\nu c^{(0)} h^{(2)}_{\mu \nu}+2\partial^\nu c^{(0)} \partial^\mu h^{(2)}_{\mu \nu} - \partial_\mu c^{(0)} \partial^\mu h^{(2)}),
\ee
this equation gets rewritten as
\be
\Box c^{(2)} - N_{\text{R}}^2(5 X_a \partial^a + X_a X_b \partial^a \partial^b - 5 X_i \partial^i - X_i X_j \partial^i \partial^j) c^{(0)}  = 0,
\ee
which is $\Box_{\text{AdS}} \, \left(c^{(0)}+\mu^2 c^{(2)}+{\cal O}(\mu^3)\right) = 0$ expanded to order $\frac{1}{R^2}$ as written in (\ref{AdSBox}), provided that $\mu^2 N_{\text{R}}^2 = -\frac{1}{R^2}$. This is precisely the normalization of our solution, giving a nontrivial check. Note that in Siegel gauge, one obtains the wave equation (after doing the above field redefinition)
\be
\begin{aligned}
	\Box c^{(2)} + \frac{N_{\text{R}}^2}{7}\big(20&\big[-6 X_a \partial^a + 2 X_b X^b \partial_a \partial^a + 6 X_i \partial^i - 2 X_j X^j \partial_i \partial^i \big]+ \\
	3 &\big[5 X_a \partial^a + X_a X_b \partial^a \partial^b - 5 X_i \partial^i - X_i X_j \partial^i \partial^j\big]\big) c^{(0)}  = 0.
\end{aligned}
\ee

\section{Possible obstructions to constructing the higher-order solution}
\label{obstructionSec}

In section \ref{susySec}, we have shown that establishing the existence of a solution to the extended system (\ref{extendedProblemNew}) reduces to the cohomology problem characterized by the equations (\ref{shifts}). The obstructions to the cohomology problem are thus characterized by the constant spinors $\chi^\alpha, \bar{\chi}^\alpha$ and the divergenceless 2-form-spinors $\Omega_{\mu \nu \alpha}, \bar{\Omega}_{\mu \nu \alpha}$, whose definition (\ref{obstructions}) depends on the gravitino and dilatino variations of (\ref{genericSuperIsometry}).

In this section, we show how the space of obstructions is restricted by the symmetries preserved by the ansatz (\ref{masslessansatz}), (\ref{phiAnsatz}). Symmetry under worldsheet parity is trivially implemented by noting that the existence of 32 super-isometries (two sets of 16) is implied by the existence of a single set 16 super-isometries with $\bar{\epsilon}^{(0)}_\alpha = 0$. We thus focus on this single set of super-isometries and do not consider worldsheet parity further. As before, because of this choice, the nonzero obstructions can be further labeled only with the orders in which they can arise so that only $\chi^{(2n)\alpha}, \Omega_{\mu \nu \alpha}^{(2n)}$ and $\bar{\chi}^{(2n-1) \alpha}, \bar{\Omega}_{\mu \nu \alpha}^{(2n-1)}$ are possibly nonzero.

\subsection{Restrictions from $SO(5) \cross SO(5)$ rotations}
The $SO(5) \cross SO(5)$ rotations are the easiest to implement, as one simply transforms every tensor and spinor index with their appropriate rotation matrix and demands every expression to be covariant. It is then not difficult to see that the only possible obstructions compatible with $SO(5) \cross SO(5)$ rotation invariance are
\be
\begin{aligned}
	     &\chi^{(2n)\alpha} \sim (\xi_1 \gamma^{01234} + \xi_2 \gamma^{56789})^{\alpha \beta} \epsilon^{(0)}_\beta, \\
	     &\Omega_{\mu \nu \alpha}^{(2n)} \sim (\omega_1 \delta^a_\mu \delta^b_\nu \gamma_{ab} + \omega_2 \delta^i_\mu \delta^j_\nu \gamma_{ij})_\alpha^\beta \epsilon^{(0)}_\beta, \\
	     &\bar{\chi}^{(2n-1)\alpha} \sim (\bar{\xi}_1 \gamma^{01234} + \bar{\xi}_2 \gamma^{56789})^{\alpha \beta} \epsilon^{(0)}_\beta, \\
	     &\bar{\Omega}_{\mu \nu \alpha}^{(2n-1)} \sim (\bar{\omega}_1 \delta^a_\mu \delta^b_\nu \gamma_{ab} + \bar{\omega}_2 \delta^i_\mu \delta^j_\nu \gamma_{ij})_\alpha^\beta \epsilon^{(0)}_\beta,
	     \label{SOobs}
\end{aligned}
\ee
where $\xi_1, \xi_2, \omega_1, \omega_2$ and $\bar{\xi}_1, \bar{\xi}_2, \bar{\omega}_1, \bar{\omega}_2$ are so far arbitrary coefficients.

\subsection{Restrictions from orientifold parity}
As mentioned under (\ref{masslessansatz}), we demand our background solution to be invariant under orientifold parity. Equivalently, we can say that under total spacetime parity acting as
\be
\begin{aligned}
	&X^\mu \to -X^\mu, \\
	&\psi^\mu \to -\psi^\mu, \,\, \bar{\psi}^\mu \to - \bar{\psi}^\mu, \\
	&S^\alpha \to (\gamma^{11})^{\alpha}_{\beta} S^\beta, \,\, \bar{S}^\alpha \to (\gamma^{11})^{\alpha}_{\beta} \bar{S}^\beta, \\
	&S_\alpha \to (\gamma^{11})_{\alpha}^{\beta} S_\beta, \,\, \bar{S}_\alpha \to (\gamma^{11})_{\alpha}^{\beta} \bar{S}_\beta,
	\label{parityTransf}
\end{aligned}
\ee
where $(\gamma^{11})^\alpha_\beta \equiv (\gamma^{0123456789})^\alpha_\beta$ with $(\gamma^{11})^2 = 1$ and $\acomm{\gamma^{11}}{\gamma^\mu} = 0$, the RR component of the background is odd and the NSNS component is even (the $X^\mu$ dependence must then be even for each). Note that the SUGRA solution of appendix \ref{sugraApp} is consistent with this symmetry.

As far as the super-isometry (\ref{phiAnsatz}) is concerned, we can simply transform the generators via
\be
\epsilon_\alpha \to (\gamma^{11})_{\alpha}^{\beta} \epsilon_\beta, \,\, \bar{\epsilon}_\alpha \to (\gamma^{11})_{\alpha}^{\beta} \bar{\epsilon}_\beta,
\label{supergenparity}
\ee
so that the super-isometry string field is invariant as a whole. This means that after applying (\ref{parityTransf}) and (\ref{supergenparity}), the source of (\ref{genericSuperIsometry}) and therefore the obstructions (\ref{obstructions}) have to be odd at odd orders and even at even orders. To see the effect of applying the parity transformation (\ref{parityTransf}) on the obstructions, note that from (\ref{genericSuperIsometry}) it is not difficult to check that the parity on the vertex operators effectively transforms
\be
\begin{aligned}
	     &u_{\mu \alpha}(X) \to - (\gamma^{11})_{\alpha}^{\beta} \, u_{\mu\beta}(-X), \,\, \bar{u}_{\mu \alpha}(X) \to - (\gamma^{11})_\alpha^{\beta} \,\bar{u}_{\mu\beta}(-X), \\
	     & w^\alpha(X) \to (\gamma^{11})^{\alpha}_{\beta} \,w^\beta(-X), \,\, \bar{w}^\alpha(X) \to (\gamma^{11})^{\alpha}_{\beta} \,\bar{w}^\beta(-X).
\end{aligned}
\ee
This means that under the transformation (\ref{parityTransf}) of the vertex operators the obstructions (\ref{obstructions}) transform as (there is an extra minus sign for the integrability obstructions coming from a derivative acting on a function of opposite argument)
\be
\begin{aligned}
&\chi^{(2n) \alpha}(X) \to + (\gamma^{11})^{\alpha}_{\beta} \, \chi^{(2n)\beta}(-X), \\
&\Omega^{(2n)}_{\mu \nu \alpha}(X) \to + (\gamma^{11})_{\alpha}^{\beta} \, \Omega^{(2n)}_{\mu \nu\beta}(-X),\\
&\bar{\chi}^{(2n-1)\alpha}(X) \to + (\gamma^{11})^{\alpha}_{\beta} \, \bar{\chi}^{(2n-1)\beta}(-X), \\
&\bar{\Omega}^{(2n-1)}_{\mu \nu \alpha}(X) \to + (\gamma^{11})_{\alpha}^{\beta} \, \bar{\Omega}^{(2n-1)}_{\mu \nu \beta}(-X). \\
\end{aligned}
\ee
Note that one can drop the $X^\mu$ dependence above as the obstructions (\ref{SOobs}) are constant, but we keep it for clarity.
Since we know the obstructions as functions (\ref{SOobs}) of the super-isometry generators, we easily see that under the transformation of the generators (\ref{supergenparity}), there is the further transformation
\be
\begin{aligned}
	&\chi^{(2n) \alpha}(-X)  \to - (\gamma^{11})^{\alpha}_\beta (\xi_1 \gamma^{01234} + \xi_2 \gamma^{56789})^{\beta \gamma} \epsilon^{(0)}_\gamma \sim -(\gamma^{11})^{\alpha}_{\beta} \chi^{(2n)\beta}(X),  \\
	&\Omega^{(2n)}_{\mu\nu\alpha}(-X) \to (\gamma^{11})_{\alpha}^{\beta} (\omega_1 \delta^a_\mu \delta^b_\nu \gamma_{ab} + \omega_2 \delta^i_\mu \delta^j_\nu \gamma_{ij})_\beta^\gamma \epsilon^{(0)}_\gamma \sim + (\gamma^{11})_{\alpha}^{\beta} \Omega^{(2n)}_{\mu \nu \beta}(X), \\
	&\bar{\chi}^{(2n-1) \alpha} (-X) \to - (\gamma^{11})^{\alpha}_\beta (\bar{\xi}_1 \gamma^{01234} + \bar{\xi}_2 \gamma^{56789})^{\beta \gamma} \epsilon^{(0)}_\gamma \sim -(\gamma^{11})^{\alpha}_{\beta} \bar{\chi}^{(2n)\beta}(X), \\
	&\bar{\Omega}^{(2n-1)}_{\mu\nu\alpha} (-X) \to (\gamma^{11})_{\alpha}^{\beta} (\bar{\omega}_1\delta^a_\mu \delta^b_\nu \gamma_{ab} + \bar{\omega}_2 \delta^i_\mu \delta^j_\nu \gamma_{ij})_\beta^\gamma \epsilon^{(0)}_\gamma \sim + (\gamma^{11})_{\alpha}^{\beta} \bar{\Omega}^{(2n-1)}_{\mu \nu \beta}(X),
\end{aligned}
\ee
so that using $(\gamma^{11})^2 = 1$ the combined transformation finally yields
\be
\begin{aligned}
&\chi^{(2n) \alpha}(X) \to -\, \chi^{(2n)\alpha}(X), \\
&\Omega^{(2n)}_{\mu \nu \alpha}(X) \to + \, \Omega^{(2n)}_{\mu \nu\alpha}(X), \\
&\bar{\chi}^{(2n-1)\alpha}(X) \to - \, \bar{\chi}^{(2n-1)\alpha}(X), \\
&\bar{\Omega}^{(2n-1)}_{\mu \nu \alpha}(X) \to +\, \bar{\Omega}^{(2n-1)}_{\mu \nu \alpha}(X). \\
\end{aligned}
\ee
Since obstructions at odd order should be odd and obstructions at even order should be even, we immediately read off that $\chi^{(2n)\alpha}$ and $\bar{\Omega}^{(2n-1)}_{\mu \nu \alpha}$ are absent.

\subsection{Restrictions from swapping $AdS_5$ and $S^5$ directions}
As also mentioned under (\ref{masslessansatz}), we demand our background to be invariant under the $\mathbb{Z}_4$ transform that swaps Euclidean $AdS_5$ and $S^5$. This is done via splitting the spacetime coordinates into 5 orthogonal planes $(X^0, X^5), \ldots, (X^4, \ldots, X^9)$ and perfoming a rotation by angle $\frac{\pi}{2}$ in each so that $X^a \to X^i, X^i \to - X^a$. We also make the rescaling of the radius $R \to i R$ a part of this transformation. At the level of the worldsheet fields, the swap acts via the $\mathbb{Z}_4$ transform
\be
\begin{aligned}
&(X^0, X^5) \to (X^5, -X^0), \ldots, (X^4, X^9) \to (X^9, -X^4), \\
&(\psi^0, \psi^5) \to (\psi^5, -\psi^0), \ldots, (\psi^4, \psi^9) \to (\psi^9, -\psi^4), \\
&(\bar{\psi}^0, \bar{\psi}^5) \to (\bar{\psi}^5, -\bar{\psi}^0), \ldots, (\bar{\psi}^4, \bar{\psi}^9) \to (\bar{\psi}^9, -\bar{\psi}^4), \\
& S^\alpha \to \tensor{R}{^\alpha_\beta} S^\beta, \,\, \bar{S}^\alpha \to \tensor{R}{^\alpha_\beta} \bar{S}^\beta,\\
& S_\alpha \to  \tensor{(R^{-T})}{_\alpha^\beta} S_\beta, \,\, \bar{S}_\alpha \to  \tensor{(R^{-T})}{_\alpha^\beta} \bar{S}_\beta,
\label{rotTransf}
\end{aligned}
\ee
where the spin fields transform under the degree $\frac{\pi}{2}$ rotation as chiral and antichiral spinors. The same rotation of tensors will be distinguished by the indices. The OPEs involving the spin fields are preserved thanks to the identities
\be
\begin{aligned}
	&\tensor{R}{^\alpha_\gamma} \tensor{(R^{-T})}{_\beta^\gamma} = \tensor{R}{^\alpha_\gamma} \tensor{(R^{-1})}{^\gamma_\beta} = \delta^\alpha_\beta, \\
	&\tensor{(R^{-T})}{_\sigma^\alpha} (\gamma^\mu)^{\sigma \lambda} \tensor{(R^{-T})}{_\lambda^\beta} = \tensor{R}{^\mu_\nu} (\gamma^\nu)^{\alpha \beta}, \\
	&\tensor{R}{^\sigma_\alpha} (\gamma^\mu)_{\sigma \lambda} \tensor{R}{^\lambda_\beta} = \tensor{R}{^\mu_\nu} (\gamma^\nu)_{\alpha \beta}
\end{aligned}
\ee
that will routinely be used in the following.

In Euclidean signature, a self-dual 5-form flux takes the form $F_5(X) = f(X) (i \gamma^{01234} + \gamma^{56789})$ so that $\star F_5(X) = i F_5(X)$. Under the above worldsheet transformations, this flux effectively transforms as $F_5(X^a, X^i) \to i F_5(X^i, -X^a)$. Together with the transformation $R \to i R$ and the fact that the first-order flux has a constant profile, this implies that the all-order RR component of (\ref{masslessansatz}) is even. It is immediate that the NSNS component is even as well so that the entire string field (\ref{masslessansatz}) is even under the above transform. The SUGRA solution of appendix \ref{sugraApp} is again consistent with this symmetry.

To make the super-isometry even under this transformation, we transform the generators according to
\be
\epsilon_\alpha \to \tensor{(R^{-T})}{_\alpha^\beta} \epsilon_\beta, \,\,\, \bar{\epsilon}_\alpha \to \tensor{(R^{-T})}{_\alpha^\beta} \bar{\epsilon}_\beta.
\label{epstransf}
\ee
The source (\ref{genericSuperIsometry}) of the extended system is then even to all orders and the remaining obstructions have to be even as well. Proceeding exactly as in the last subsection, the transformations (\ref{rotTransf}) are equivalent to the following transformations on the string field profiles in (\ref{genericSuperIsometry})
\be
\begin{aligned}
	     &u_{\mu \alpha}(X) \to \tensor{(R^T)}{_\mu^\nu} \tensor{R}{^\beta_\alpha} \, u_{\nu\beta}(R \cdot X ), \,\, \bar{u}_{\mu \alpha}(X) \to \tensor{(R^T)}{_\mu^\nu} \tensor{R}{^\beta_\alpha}  \,\bar{u}_{\nu\beta}(R \cdot X), \\
	     & w^\alpha(X) \to \tensor{(R^{-T})}{_\beta^\alpha} \,w^\beta(R \cdot X), \,\,  \bar{w}^\alpha(X) \to \tensor{(R^{-T})}{_\beta^\alpha} \,\bar{w}^\beta(R \cdot X)
\end{aligned}
\ee
so that under (\ref{rotTransf}), the remaining obstructions transform as
\be
\begin{aligned}
	&\Omega^{(2n)}_{\mu \nu \alpha}(X) \to \tensor{(R^T)}{_\mu^\lambda} \tensor{(R^T)}{_\nu^\sigma} \tensor{R}{^\beta_\alpha} \, \Omega_{\lambda \sigma \beta}(R \cdot X), \\
	&\bar{\chi}^{(2n-1)\alpha}(X) \to (R^{-T})\indices{_\beta^\alpha} \, \bar{\chi}^{(2n-1)\beta}(R \cdot X).
	\end{aligned}
\ee
Note that in Euclidean signature, we have to replace the remaining obstructions with
\be
\begin{aligned}
&\Omega_{\mu \nu \alpha}^{(2n)} \sim (\omega_1 \delta^a_\mu \delta^b_\nu \gamma_{ab} + \omega_2 \delta^i_\mu \delta^j_\nu \gamma_{ij})_\alpha^\beta \epsilon^{(0)}_\beta, \\
&\bar{\chi}^{(2n-1)\alpha} \sim \big(i\bar{\xi}_1 \gamma^{01234} + \bar{\xi}_2\gamma^{56789}\big)^{\alpha \beta} \epsilon^{(0)}_\beta
\end{aligned}
\ee
so that the transformation (\ref{epstransf}) of the super-isometry generators gives
\be
\begin{aligned}
	&\Omega^{(2n)}_{\mu \nu \alpha}(R \cdot X)  \to \tensor{(R^{-T})}{_\alpha^\beta} (\omega_1 \delta^a_\mu \delta^b_\nu\, \tensor{R}{_a^\sigma} \tensor{R}{_b^\lambda} + \omega_2 \delta^i_\mu \delta^j_\nu \,\tensor{R}{_i^\sigma} \tensor{R}{_j^\lambda}) (\gamma_{\sigma \lambda})_\beta^\gamma \epsilon^{(0)}_\gamma, \\
	&\bar{\chi}^{(2n-1)\alpha}(R \cdot X) \to \tensor{R}{^\alpha_\delta} \tensor{(R^{-T})}{_\sigma^\delta} \big(i\bar{\xi}_1 \gamma^{01234} + \bar{\xi}_2\gamma^{56789}\big)^{\sigma \beta} \tensor{(R^{-T})}{_\beta^\gamma} \epsilon^{(0)}_\gamma,
\end{aligned}
\ee
which using $\tensor{(R^{-T})}{_\sigma^\delta} \big(i\bar{\xi}_1 \gamma^{01234} + \bar{\xi}_2 \gamma^{56789}\big)^{\sigma \beta} \tensor{(R^{-T})}{_\beta^\gamma} = i  \big(i \bar{\xi}_2 \gamma^{01234} + \bar{\xi}_1 \gamma^{56789}\big)^{\delta \gamma}$  together with some more rotation algebra can be simplified to
\be
\begin{aligned}
	& \Omega^{(2n)}_{\mu \nu \alpha}(X) \to  (\omega_2 \delta^a_\mu \delta^b_\nu \gamma_{ab} + \omega_1 \delta^i_\mu \delta^j_\nu \gamma_{ij})_\alpha^\beta \epsilon^{(0)}_\beta,\\
	&\bar{\chi}^{(2n-1)\alpha}(X) \to i \big(i\bar{\xi}_2 \gamma^{01234} + \bar{\xi}_1\gamma^{56789}\big)^{\alpha \beta} \epsilon^{(0)}_\beta.
\end{aligned}
\ee
Taking into account the rescaling of the radius, we then finally get
\be
\begin{aligned}
	& \frac{1}{R^{2n}} \Omega^{(2n)}_{\mu \nu \alpha}(X) \to \frac{(-1)^n}{R^{2n}} (\omega_2 \delta^a_\mu \delta^b_\nu \gamma_{ab} + \omega_1 \delta^i_\mu \delta^j_\nu \gamma_{ij})_\alpha^\beta \epsilon^{(0)}_\beta,\\
	&\frac{1}{R^{2n-1}} \bar{\chi}^{(2n-1)\alpha}(X) \to \frac{(-1)^{n+1}}{R^{2n-1}} \big(i\bar{\xi}_2 \gamma^{01234} + \bar{\xi}_1\gamma^{56789}\big)^{\alpha \beta} \epsilon^{(0)}_\beta,
\end{aligned}
\ee
which effectively switches the $AdS_5$ and $S^5$ directions with a possible extra sign. The only combinations that can remain as possible obstructions have to be even under the above transformation and it is clear that these are (rotated back to Lorentzian signature)
\be
\begin{aligned}
	&\bar{\chi}^{(n)\alpha} \sim \big(\gamma^{01234} - \gamma^{56789}\big)^{\alpha \beta} \epsilon^{(0)}_\beta, \,\,\,\,\, n = 1 \bmod 4 \\
	&\Omega_{\mu \nu \alpha}^{(n)} \sim (\delta^a_\mu \delta^b_\nu \gamma_{ab} - \delta^i_\mu \delta^j_\nu \gamma_{ij})_\alpha^\beta \epsilon^{(0)}_\beta, \,\, n = 2 \bmod 4 \\
	&\bar{\chi}^{(n)\alpha} \sim \big(\gamma^{01234} + \gamma^{56789}\big)^{\alpha \beta} \epsilon^{(0)}_\beta, \,\,\,\,\, n = 3 \bmod 4 \\
		&\Omega_{\mu \nu \alpha}^{(n)} \sim (\delta^a_\mu \delta^b_\nu \gamma_{ab} + \delta^i_\mu \delta^j_\nu \gamma_{ij})_\alpha^\beta \epsilon^{(0)}_\beta, \,\,  n = 0 \bmod 4.
\end{aligned}
\ee
Writing $\gamma^{56789} = -\gamma^{11} \gamma^{01234}$ and noting that the antichiral spinor $\bar{S}_\alpha$ has $\gamma^{11}$-eigenvalue equal to $-1$, we see that $\big(i \gamma^{01234} + \gamma^{56789}\big)^{\alpha \beta} \epsilon^{(0)}_\beta \bar{S}_\alpha = 0$ so that there is no obstruction for $n = 1 \bmod 4$. Moreover, the shifts (\ref{shifts}) are such that the possible obstruction at orders $2 \bmod 4$ can be removed. We are finally left with potential obstructions at orders $3\bmod4$ and $0\bmod4$.

\subsection{Remaining possible obstructions and their interpretation}
\label{interpret}
As a result of the previous subsections, we were able to get rid of most of the potential obstructions to the existence of the solution to the extended system (\ref{extendedProblemNew}) using worldsheet symmetry arguments. Nevertheless, the potential obstructions of (\ref{obstructions}) characterized via
\be \label{ObsRem}
\begin{aligned}
	&\bar{\chi}^{(n)\alpha} \sim (\gamma^{01234})^{\alpha \beta} \bar{\epsilon}^{(0)}_\beta, \,\, \hspace{1.76 cm} n = 3 \bmod 4 \\
	&\Omega^{(n)}_{\mu \nu \alpha} \sim (\delta^a_\mu \delta^b_\nu \gamma_{ab} + \delta^i_\mu \delta^j_\nu \gamma_{ij})^\beta_\alpha \epsilon^{(0)}_\beta, \,\, n = 0 \bmod 4
\end{aligned}
\ee
remain. In this subsection, we give physical interpretation to them and explain why they cannot be removed by any bosonic symmetry. We also discuss how they might possibly be shown to be absent.

The $\bar{\chi}^{(n)\alpha}$ obstruction was previously interpreted as a constant term in the condition for the vanishing of the dilatino variation. One may easily check that it can be rewritten as
\be
\bar{\chi}^{(n)\alpha} \sim \widetilde{R}_{\mu \nu} (\gamma^\mu \gamma^{01234} \gamma^\nu)^{\alpha \beta} \epsilon^{(0)}_\beta, \,\, n = 3 \bmod 4,
\ee
where $\widetilde{R}_{\mu \nu}$ is the Ricci tensor (\ref{ricciT}) for the $AdS_5 \cross S^5$ SUGRA solution. This suggests that such a term can come from modifying the IIB SUGRA dilatino SUSY variation by adding the higher derivative covariant term
\be
\alpha'^{k} \tensor{R}{_{\mu_1 \mu_2}} \tensor{R}{^{\mu_2}_{\mu_3}} \ldots \tensor{R}{^{\mu_{k}}_{\mu_{k+1}}} \gamma^{\mu_1} F_5 \gamma^{\mu_{k+1}} \epsilon.
\ee
Note that had we removed an odd number of the Ricci tensors, the result would be proportional to $\gamma^\mu F_5 \gamma_\mu = 0$, which is consistent with there being no obstructions at orders $n = 1 \bmod 4$.

Similarly, the $\Omega^{(n)}_{\mu \nu \alpha}$ obstruction was previously interpreted as coming from a divergenceless term in the integrability condition for the vanishing of the gravitino variation. As before, one can check that it can be rewritten as
\be
\Omega^{(n)}_{\mu \nu \alpha} \sim \tensor{\widetilde{R}}{^{\rho}_{[\mu}} \tensor{\widetilde{R}}{_{\nu] \rho \lambda \sigma}} \gamma^\lambda \gamma^\sigma \epsilon^{(0)}_\alpha, \,\, n = 0 \bmod 4,
\ee
where $\widetilde{R}_{\mu \nu \lambda \sigma}$ is the Riemann tensor (\ref{riemannT}) for the $AdS_5 \cross S^5$ SUGRA solution. This again suggests that this term can come from adding the following covariant term to the IIB SUGRA gravitino variation
\be
\alpha'^k \tensor{R}{_{\mu}^{\kappa_1}} \tensor{R}{_{\kappa_1}^{\kappa_2}} \ldots \tensor{R}{_{\kappa_{k-1}}^\rho} R_{\nu \rho \lambda \sigma} X^\nu (\gamma^\lambda \gamma^\sigma \epsilon)_\alpha.
\ee
Moreover, by looking at (\ref{obstructions}), $\Omega_{\mu \nu \alpha}^{(n)}$ being nonzero for $n = 0 \bmod 4$ means that there is the following term in the dilatino variation
\be
\alpha'^k (\gamma^\mu)^{\alpha \beta} \tensor{R}{_{\mu}^{\kappa_1}} \tensor{R}{_{\kappa_1}^{\kappa_2}} \ldots \tensor{R}{_{\kappa_{k-1}}^\rho} R_{\nu \rho \lambda \sigma} X^\nu (\gamma^\lambda \gamma^\sigma \epsilon)_\beta.
\ee
The constant part of the $\epsilon_\alpha$ gives a term proportional to $(\slashed{X} \epsilon^{(0)})^\alpha$, which is equivalent to the contribution coming from an anharmonic dilaton profile of the form $X_\mu X^\mu$. When an odd number of Ricci tensors is removed, the obstruction comes from a radial dilaton profile proportional to $X^2$, which can be shifted away by turning on (\ref{osshift}). This is consistent with there being no obstructions at orders $n = 2 \bmod 4$. The $X_\mu X^\mu$ profile for the dilaton on the other hand cannot be shifted away as there is no gauge nontrivial on-shell fluctuation that turns on such a profile.

Having demonstrated that there exists a covariant (but not necessarily exhaustive) way of writing the remaining possible obstructions, it is clear that the remaining obstructions (\ref{ObsRem}) cannot be removed by using any bosonic symmetries. That being said, it is highly plausible that the potential obstructions will be absent once the simplified extended system (\ref{extendedProblemNew}) is enhanced to the original extended system (\ref{extendedProblem}), which realizes the full $PSU(2,2|4)$ super-isometry of $AdS_5 \cross S^5$ \footnote{In fact, having access to the full $PSU(2,2|4)$ invariance was crucial in investigating the same problem in the pure spinor formalism \cite{Berkovits:2004xu}.}. At the practical level, this means that the mutual consistency of the variations of the background and the commutation relations of the super-isometries should restrict the form of the potential obstructions. In other words, the above modifications of the IIB SUGRA SUSY variations should generically be inconsistent with the SUSY algebra and so would be highly restricted as a consequence.

\section{Discussion}
\label{discussionSec}
We have analyzed the existence of the $AdS_5 \cross S^5$ solution of IIB SFT to all orders in the $\frac{1}{R}$ expansion in sections \ref{susySec} and \ref{obstructionSec}, and found that the solution is possibly only obstructed at orders $n$ with $n = 3 \bmod 4$ and $n = 0 \bmod 4$. Despite not establishing an all-order existence proof, in section \ref{solext} we have found the solution up to third order in the large radius expansion in the flat-vertex frame. Up to second order, we have also explicitly verified that the solution preserves maximal super-isometry in the form of SFT gauge transformations. To determine the solution to higher orders in perturbation theory would require the flat $n$-string bracket for $n\geq 4$ as outlined in appendix \ref{sec:flatvert}, whose explicit construction is somewhat involved and the relevant computations remain a technical challenge. Note that in the pp-wave limit of section \ref{sec:pp} the equations of motion get simplified, which allowed us to find an all-order solution. A similar simplification is not immediate for the full extended system, and consequently showing the all-order maximal super-isometry of the pp-wave solution is left for the future.

We would like to extend the analysis of the axion spectrum in $AdS_5\times S^5$ in section \ref{sec:axion} to massive string modes. This requires restoring the massive components of the string field via (\ref{massivesf}) and analyzing the dispersion relation of the fluctuation modes, following the strategy outlined in \cite{deLacroix:2017lif}. In the pp-wave background, this analysis has been carried out to the second order in \cite{Cho:2018nfn}. It remains to be seen whether the simplicity of the all-order massless string field solution describing the pp-wave background of section \ref{sec:pp} can lead to a first-principle determination of the all-order massive string spectrum in SFT that reproduces the well-known results of \cite{Metsaev:2001bj, Metsaev:2002re}.

Another potential application of the SFT approach is the computation of string amplitudes in $AdS_5\times S^5$. In the flat-vertex frame, a tree-level $n$-point amplitude of closed string modes can in principle be extracted from the string field solution that describes nonlinear fluctuations around the background at the $(n-1)$-th order. The way this works is that the nonlinear fluctuation captures the Green's function with all propagators amputated, except for the one on the outgoing leg. After amputating this propagator, the tree-level amplitude is recovered. It will be interesting to carry out this analysis for the 4-point amplitude in the $\frac{1}{R}$ expansion and compare to the conjectured results of \cite{Alday:2023mvu, Alday:2024yax}.

\section*{Acknowledgements}
JG would like to thank FAPESP grants 2022/04105-0 and 2024/00082-1 for partial financial support. The work of MC is supported by Clay C\'ordova's Sloan Research Fellowship from the Sloan Foundation. We thank the International Centre for Theoretical Physics, Trieste, Italy, for its hospitality during the course of this work. XY thanks the Aspen Center for Physics, which is supported by National Science Foundation grant PHY-2210452. This work was supported by DOE grant DE-SC0007870.

\appendix

\section{Worldsheet conventions} \label{WSconventions}
The worldsheet theory of IIB string theory has 10 free bosons $X^\mu$, 10 free fermions $\psi^\mu, \bar{\psi}^\mu$, the $b,c, \bar{b}, \bar{c}$ fermionic ghosts and the $\beta, \gamma, \bar{\beta}, \bar{\gamma}$ bosonic ghosts. The bosonic ghosts are re-bosonized in terms of the $(\phi,\eta, \xi)$ system as
\begin{equation}
    \beta = \partial \xi e^{-\phi}, \, \, \gamma = \eta e^{\phi}
\end{equation}
and similar for its antiholomorphic counterpart. Here we follow the conventions of \cite{Sen:2024nfd}, which are those of \cite{Sen:2021tpp} apart from a factor of 2 in the PCO. These are obtained from \cite{Polchinski:1998rr} by rescaling
\begin{align}
    & \beta \to -\beta/2, \, \, \gamma \to 2 \gamma,  \, \, \xi \to \xi/2 \, \, \eta \to 2\eta, \, \, \phi \to \phi, \,\, X^\mu \to X^\mu, \, \, \psi^\mu \to -i \psi^\mu/\sqrt{2}.
\end{align}
Note that in these conventions, one also has an additional factor of $\frac{1}{2}$ in the definition of the supercurrent $G_m$ compared to  \cite{Polchinski:1998rr}.

{\allowdisplaybreaks

The basic operator products are
\be
\begin{aligned}
    &\partial X^\mu (z) \partial X^\nu (w) \sim -\frac{\eta^{\mu \nu}}{2} \frac{1}{(z-w)^2}, \\
    &\psi^\mu (z) \psi^\nu (w) \sim -\frac{\eta^{\mu \nu}}{2} \frac{1}{z-w}, \\
    & c(z) b(w) \sim \frac{1}{z-w}, \\
    & \eta(z) \xi(w) \sim \frac{1}{z-w}, \\
    & \partial \phi(z) \partial \phi(w) \sim - \frac{1}{(z-w)^2}, \\
    &e^{i k_1 \cdot X(z)} e^{i k_2 \cdot X(w)} \sim (z-w)^{-\frac{1}{2} k_1 \cdot k_2} e^{i(k_1+k_2) X(w)},\\
&e^{q_1 \phi(z)} e^{q_2 \phi(w)} \sim (z-w)^{-q_1 q_2} e^{(q_1 + q_2)\phi(w)}.
\end{aligned}
\ee
}
Note that using
    $\partial X^\mu(z) f(X(w)) \sim -\frac{1}{2} \frac{1}{z-w} (\partial^\mu f)(X(w))$, we can write that
    \begin{equation}
        \oint \partial X^\mu \to -\frac{1}{2} \partial^\mu
    \end{equation}
    when acting on a function of $X$ in position space. In momentum space $\oint \partial X^\mu \to -\frac{1}{2} i k^\mu$.

Then we have the matter stress-energy tensor $T_m$ and its associated supercurrent $G_m$
\be
\begin{aligned}
    T_m &= - \partial X^\mu \partial X_\mu + \psi_\mu \partial \psi^\mu, \\
    G_m &= - \psi_\mu \partial X^\mu.
\end{aligned}
\ee
Upon rebosonization, these give the BRST charge $Q = Q^L + Q^R$ where
\begin{align}
  Q^L = \oint \dd z \biggr(c T_m - e^{\phi} \eta G_m + b c \partial c + c\big(-\eta \partial \xi - \frac{1}{2} \partial \phi \partial \phi - \partial^2 \phi\big) - \frac{1}{4} b e^{2\phi}\eta \partial \eta\biggr)
\end{align}
and $Q^R$ is its antiholomorphic counterpart. The holomorphic PCO is given by
\begin{equation}
    \mathcal{X}(z) = \acomm{Q}{\xi(z)} =  c \partial \xi + e^\phi G_m - \frac{1}{4} \partial \eta e^{2\phi}b-\frac{1}{4}\partial(\eta e^{2\phi}b).
\end{equation}

There are also the chiral spin fields $S_\alpha$ and the anti-chiral spin fields $S^\beta$ that are assembled into the Dirac spinor
\ie{}
S_A =
\begin{pmatrix}
	S_\alpha \\
	S^\beta
\end{pmatrix}.
\fe
In this basis, the gamma matrices take the off-diagonal form
\ie{}
(\Gamma^\mu)_A^{~B} =
\begin{pmatrix}
0 & (\gamma^\mu)_{\alpha\beta} \\
(\gamma^\mu)^{\alpha\beta} & 0
\end{pmatrix},
\fe
where the matrices $(\gamma^\mu)_{\alpha\beta} = (\gamma^\mu)_{\beta\alpha}$ have real-valued entries, and satisfy $\{\gamma^\mu, \gamma^\nu\} = 2\eta^{\mu\nu}$.  The charge conjugation matrix $C = \Gamma^0$ and its inverse are given by
\ie{}
C_{AB} =
\begin{pmatrix}
0 & \delta_\alpha^{~\beta} \\
- \delta^\alpha_{~\beta} & 0
\end{pmatrix}, \qquad
(C^{-1})^{AB} =
\begin{pmatrix}
0 & -\delta^\alpha_{~\beta} \\
\delta_\alpha^{~\beta} & 0
\end{pmatrix}.
\fe
Similarly, the chirality matrix $\Gamma \equiv \Gamma^0 \cdots \Gamma^9$ takes the form
\ie{}
\Gamma_A^{~B} =
\begin{pmatrix}
-\delta_\alpha^{~\beta} & 0 \\
0 & \delta^\alpha_{~\beta}
\end{pmatrix},
\fe
such that the projectors onto chiral and anti-chiral spinors $P_\pm = \frac12(1\pm \Gamma)$ read
\ie{}
(P_+)_A^{~B} =
\begin{pmatrix}
0 & 0 \\
0 & \delta^\alpha_{~\beta}
\end{pmatrix}, \quad
(P_-)_A^{~B} =
\begin{pmatrix}
\delta_\alpha^{~\beta} & 0 \\
0 & 0
\end{pmatrix}.
\fe
In other words,  and $P_+ S^\alpha = 1$ and $P_- S_\alpha = -1$. It is useful to note that
\ie{}
(\Gamma^\mu C)_{AB} =
\begin{pmatrix}
-(\gamma^\mu)_{\alpha\beta} & 0 \\
0 & (\gamma^\mu)^{\alpha\beta}
\end{pmatrix}, \quad
(C^{-1}\Gamma^\mu )^{AB} =
\begin{pmatrix}
-(\gamma^\mu)^{\alpha\beta} & 0 \\
0 & (\gamma^\mu)_{\alpha\beta}
\end{pmatrix}.
\fe

The basic OPEs for the spin field are
\be
\begin{aligned}
\psi^\mu(z)~ e^{-\frac12\phi}S_A(w) &\sim \frac{i}{2}{(\Gamma^\mu)_A^{\:\:\: B} \over \sqrt{(z-w)}}~ e^{-\frac12 \phi}S_B(w), \\
 e^{-\frac12\phi}S_A(z)~  e^{-\frac12\phi}S_B(w) &\sim  - {C_{AB} \over (z-w)^\frac{3}{2}}~ e^{-\phi}(w) -i{(\Gamma_\mu C)_{AB} \over z-w}~ e^{-\phi}\psi^\mu(w).
\end{aligned}
\ee
The composite operators $e^{-\frac12\phi}S^\alpha$ and $e^{-\frac32\phi}S^\alpha$ have weight $h=1$ and are GSO even and Grassmann odd, as expected for spacetime fermions. We can use the fact that $e^{\pm \phi}$ is GSO odd and Grassmann odd to obtain other such composite operators, e.g. $e^{\frac12\phi}S^\alpha$ and $e^{-\frac52 \phi}S^\alpha$. From the basic OPEs above, we have that
\be
\begin{aligned}
\psi^\mu(z)~e^{-\frac12\phi}S^\alpha(w) &\sim \frac{i}{2}{(\gamma^\mu)^{\alpha\beta} \over \sqrt{z-w} }~e^{-\frac12\phi}S_{\beta}(w) + \ldots, \\
\psi^\mu(z)~e^{-\frac12\phi}S_{\alpha}(w) &\sim  \frac{i}{2}{(\gamma^\mu)_{\alpha\beta} \over \sqrt{z-w} }~e^{-\frac12\phi}S^\beta(w) + \ldots, \\
e^{-\frac12\phi}S^\alpha(z)~e^{-\frac12\phi}S^\beta(w) &\sim  -i{(\gamma_\mu)^{\alpha\beta} \over z-w }~e^{-\phi}\psi^\mu(w) + \ldots, \\
e^{-\frac12\phi}S_\alpha(z)~e^{-\frac12\phi}S_\beta(w) &\sim  i{(\gamma_\mu)_{\alpha\beta} \over z-w }~e^{-\phi}\psi^\mu(w) + \ldots, \\
e^{-\frac32\phi}S_\alpha(z)~e^{-\frac12\phi}S^\beta(w) &\sim -{\delta_\alpha^{~\beta} \over (z-w)^2}~ e^{-2\phi}(w) + \ldots, \\
e^{-\frac32\phi}S^\alpha(z)~e^{-\frac12\phi}S_\beta(w) &\sim {\delta^\alpha_{~\beta} \over (z-w)^2}~ e^{-2\phi}(w) + \ldots.
\label{SOPEs}
\end{aligned}
\ee
Using these, we can construct a plethora of other useful OPEs, e.g.
\be
\begin{aligned}
e^{-\phi}\psi^\mu(z)~ e^{-\frac12\phi} S^\alpha(w) &\sim \frac{i}{2}{(\gamma^\mu)^{\alpha\beta} \over z-w}~ e^{-\frac32\phi} S_{\beta}(w) + \ldots, \\
e^{\phi}\psi^\mu(z)~ e^{-\frac12\phi} S^\alpha(w) &\sim \frac{i}{2}(\gamma^\mu)^{\alpha\beta}  e^{\frac12\phi} S_{\beta}(w) + \ldots, \\
e^{-\phi}\psi^\mu(z)~ e^{\frac12\phi}S_{\alpha}(w) &\sim -\frac{i}{2}(\gamma^\mu)_{\alpha\beta}  e^{-\frac12\phi}S^\beta(w) + \ldots, \\
e^{\phi} \psi^\mu(z) e^{-\frac{3}{2}\phi}S_\alpha (w)&\sim -\frac{i}{2} (z-w)(\gamma^\mu)^\beta_\alpha e^{-\frac{1}{2}\phi}S_\beta(w)+\ldots, \\
e^{\phi} \psi^\mu(z) e^{-\frac{5}{2}\phi}S_\alpha (w)&\sim \frac{i}{2} (\gamma^\mu)^\beta_\alpha (z-w)^2 e^{-\frac{3}{2}\phi}S_\beta(w)+\ldots.
\end{aligned}
\ee
We also have
\be
\begin{aligned}
e^{\frac12\phi}S_{\alpha}(z)~e^{-\frac12\phi} S^\beta(w) &\sim  -{\delta_\alpha^{~\beta} \over z-w} - \frac{1}{2}\delta_\alpha^{~\beta} \partial \phi(w) + (\gamma_{\mu\nu})_\alpha^{~\beta}~\psi^\mu\psi^\nu(w) + \ldots, \\
e^{-\phi}\psi^\mu(z)~e^{-\phi}\psi^\nu(w) &\sim
{\eta^{\mu\nu} \over 2(z-w)^2}~e^{-2\phi}(w) \\
&~~~- {1 \over z-w}~ \bigg[e^{-2\phi}\psi^\mu\psi^\nu(w) + \frac{1}{2}\eta^{\mu\nu} e^{-2\phi}\partial\phi(w) \bigg] + \ldots.
\end{aligned}
\ee

To see that these OPEs are consistent, we can try to compute the correlator
\begin{equation}
    \langle e^{-\phi}\psi^\mu(z_1) e^{-\frac{1}{2} \phi} S_\alpha (z_2) e^{-\frac{1}{2}\phi} S_\beta(z_3)\rangle
\end{equation}
    in various channels. Let us begin with the $z_{23} \to 0$ channel followed by the $z_{13} \to 0$ channel:
\be
\begin{aligned}
    \langle e^{-\phi}\psi^\mu(z_1) e^{-\frac{1}{2} \phi} S_\alpha (z_2) e^{-\frac{1}{2}\phi} S_\beta(z_3)\rangle &\sim \frac{i}{z_{23}} (\gamma_\nu)_{\alpha \beta} \langle  e^{-\phi}\psi^\mu(z_1) e^{-\phi} \psi^\nu(z_3)\rangle \\
    & \sim \frac{i}{2z_{23}z_{13}^2} (\gamma^\mu)_{\alpha \beta}.
\end{aligned}
\ee
And computing it in the channel where $z_{12} \to 0$ followed by $z_{23} \to 0$ gives
\be
\begin{aligned}
    \langle e^{-\phi}\psi^\mu(z_1) e^{-\frac{1}{2} \phi} S_\alpha (z_2) e^{-\frac{1}{2}\phi} S_\beta(z_3)\rangle &\sim \frac{i}{2 z_{12}} (\gamma^\mu)_{\alpha \gamma}\langle  e^{-\frac{3}{2}\phi} S^\gamma(z_2) e^{-\frac{1}{2}\phi} S_\beta(z_3) \rangle \\
    & \sim \frac{i}{2 z_{12} z_{23}^2} (\gamma^\mu)_{\alpha \beta},
\end{aligned}
\ee
obtaining a match. These limits are consistent with the correlator
\begin{equation}
    \langle e^{-\phi}\psi^\mu(z_1) e^{-\frac{1}{2} \phi} S_\alpha (z_2) e^{-\frac{1}{2}\phi} S_\beta(z_3)\rangle = \frac{i}{2} \frac{(\gamma^\mu)_{\alpha \beta}}{z_{12} z_{13} z_{23}}.
\end{equation}
Similarly one can check the consistency of the correlators
\be
\begin{aligned}
&  \langle e^{-\phi}\psi^\mu(z_1) e^{-\frac{1}{2} \phi} S^\alpha (z_2) e^{-\frac{1}{2}\phi} S^\beta(z_3)\rangle = -\frac{i}{2} \frac{(\gamma^\mu)^{\alpha \beta}}{z_{12} z_{13} z_{23}},\\
&\left\langle e^{-\frac12\phi}S_{\alpha_1}(z_1) \cdots e^{-\frac12\phi}S_{\alpha_4}(z_4)\right\rangle \\
&~~~=  -{(\gamma_\mu)_{\alpha_1\alpha_2}(\gamma^\mu)_{\alpha_3\alpha_4} \over 2z_{12} z_{23}z_{24}z_{34}}-{(\gamma_\mu)_{\alpha_1\alpha_3}(\gamma^\mu)_{\alpha_2\alpha_4} \over 2z_{13} z_{23}z_{34}z_{24}}-{(\gamma_\mu)_{\alpha_1\alpha_4}(\gamma^\mu)_{\alpha_2\alpha_3} \over 2z_{14} z_{24}z_{34}z_{23}} ,\\
&\left\langle e^{-\frac12\phi}S^{\alpha_1}(z_1) \cdots e^{-\frac12\phi}S^{\alpha_4}(z_4)\right\rangle \\
&~~~=  -{(\gamma_\mu)^{\alpha_1\alpha_2}(\gamma^\mu)^{\alpha_3\alpha_4} \over 2z_{12} z_{23}z_{24}z_{34}}-{(\gamma_\mu)^{\alpha_1\alpha_3}(\gamma^\mu)^{\alpha_2\alpha_4} \over 2z_{13} z_{23}z_{34}z_{24}}-{(\gamma_\mu)^{\alpha_1\alpha_4}(\gamma^\mu)^{\alpha_2\alpha_3} \over 2z_{14} z_{24}z_{34}z_{23}}.
\end{aligned}
\ee
{\allowdisplaybreaks
Below we list some useful formulas involving the action of $Q$ on level zero states:
\begin{align}
&Q \cdot f(X) ce^{-\phi}\psi^\mu = \frac14 \Box f(X) c \partial c  e^{-\phi}\psi^\mu  - \frac{1}{4} \partial^\mu f(X) c\eta, \\
&Q \cdot f(X) c \partial c e^{-\phi}\psi^\mu  = \frac{1}{4}\partial^\mu f(X) c \partial c \eta,  \\
&Q \cdot f(X) ce^{-\frac12\phi}S^\alpha  = \frac14 \Box f(X) c \partial c  e^{-\frac12\phi}S^\alpha - \frac{i}{4} \slashed{\partial}^{\alpha\beta} f(X) c \eta e^{\frac{1}{2}\phi} S_\beta, \\
&Q \cdot f(X) c \partial c e^{-\frac12\phi}S^\alpha  =  \frac{i}{4}\slashed{\partial}^{\alpha \beta}f(X) c\partial c \eta e^{\frac{1}{2}\phi}S_\beta + \frac{1}{4} f(X) c \eta \partial \eta e^{\frac32\phi} S^\alpha,\\
&Q \cdot f(X) ce^{-\frac32\phi}S_\alpha = \frac14 \Box f(X) c \partial c e^{-\frac32\phi}S_\alpha, \\
&Q \cdot f(X) c \partial ce^{-\frac32\phi}S_\alpha = 0, \\
& Q \cdot f(X) c \eta e^{\frac{1}{2}\phi} S_\alpha = \frac{1}{4} \Box f(X) c\partial c \eta e^{\frac{1}{2}\phi} S_\alpha - \frac{i}{4} (\slashed{\partial} f)_{\alpha \beta} c \eta \partial \eta e^{\frac{3}{2} \phi} S^\beta, \\
& Q \cdot f(X) c \partial c \eta e^{\frac{1}{2}\phi} S_\alpha = \frac{i}{4} (\slashed{\partial} f)_{\alpha \beta} c \partial c \eta \partial \eta e^{\frac{3}{2} \phi} S^\beta,\\
&Q \cdot f(X) c \partial \xi e^{-\frac52\phi}S^\alpha  = \frac14 \Box f(X) c \partial c \partial \xi e^{-\frac52\phi}S^\alpha - \frac{i}{4} \slashed{\partial}^{\alpha \beta}f(X) c e^{-\frac32\phi}S_\beta, \\
&Q \cdot f(X) c \partial c \partial \xi e^{-\frac52\phi}S^\alpha = \frac{i}{4} \slashed{\partial}^{\alpha \beta}f(X) c \partial c e^{-\frac32\phi}S_\beta,\\
&Q \cdot f(X) c e^{-2\phi}\partial\xi = \frac{1}{4} \Box f(X) c \partial c e^{-2\phi}\partial \xi - \frac{1}{2} \partial_\mu f(X) c e^{-\phi} \psi^\mu, \\
&Q \cdot f(X) c \partial c e^{-2\phi}\partial\xi = -\frac{1}{2} f(X) c \eta + \frac{1}{2}\partial_\mu f(X) c \partial c e^{-\phi} \psi^\mu, \\
& Q \cdot f(X) c \eta=  \frac{1}{4}\Box f(X) c\partial c \eta, \\
& Q \cdot f(X) c\partial c \eta = 0.
\label{QAct}
\end{align}
}
{\allowdisplaybreaks
Similarly, using $\mathcal{X}_0 = \oint \frac{\dd z}{z} \mathcal{X}$, we have the picture-raised states:
\begin{align}
\label{PCO1}
&{\cal X}_0 \cdot f(X) c e^{-\phi}\psi^\mu  = c\big(-\frac{1}{2} f(X) \partial X^\mu - \frac{1}{2} \partial_\nu f(X) \psi^\nu \psi^\mu + \frac{1}{4} \partial^\mu f(X) \partial \phi \big) - \frac{1}{4} f(X) \eta e^{\phi}\psi^\mu, \\
\label{PCO2}
&{\cal X}_0 \cdot f(X) c \partial ce^{-\phi}\psi^\mu = c\partial c \big(-\frac{1}{2} f(X) \partial X^\mu - \frac{1}{2}\partial_\nu f(X) \psi^\nu \psi^\mu + \frac{1}{4} \partial^\mu f(X) \partial \phi \big) - \\ \nonumber
& \hspace{3.8cm}\frac{1}{4}f(X)\left(2 c\partial \eta - \partial c \eta + 2 c \eta \partial \phi \right)e^{\phi}\psi^\mu,\\
\label{PCO3}
&{\cal X}_0 \cdot f(X) ce^{-\frac12\phi}S^\alpha =  -\frac{i}{2} f(X) ce^{\frac12\phi} (\slashed{\partial X} S)^\alpha - \frac{1}{4}f(X)(2 \partial \eta + \eta bc + 2 \eta \partial \phi)e^{\frac32\phi}S^\alpha, \\
\label{PCO4}
&{\cal X}_0 \cdot f(X) \eta ce^{-\frac12\phi}S^\alpha=  -f(X) c\partial c e^{-\frac{1}{2}\phi}S^\alpha - \frac{i}{2} f(X) \eta ce^{\frac12\phi} (\slashed{\partial X} S)^\alpha - \frac{1}{2}f(X)\eta \partial \eta e^{\frac32\phi}S^\alpha, \\
\label{PCO5}
&{\cal X}_0 \cdot f(X) ce^{-\frac32\phi}S_\alpha =  -\frac{i}{4} \slashed{\partial}_{\alpha \beta} f(X) c e^{-\frac12\phi} S^\beta,  \\
\label{PCO6}
&{\cal X}_0 \cdot f(X) c \partial ce^{-\frac32\phi}S_\alpha = -\frac{i}{4} \slashed{\partial}_{\alpha \beta} f(X) c \partial ce^{-\frac12\phi} S^\beta -\frac{1}{4} f(X) c \eta e^{\frac12\phi}S_\alpha,  \\
\label{PCO7}
&{\cal X}_0 \cdot f(X) c \partial \xi e^{-\frac12\phi}S^\alpha = 0, \\
\label{PCO8}
&{\cal X}_0 \cdot f(X) c \partial \xi e^{-\frac52 \phi}S^\alpha   = 0,\\
\label{PCO9}
&{\cal X}_0 \cdot f(X) c \partial c \partial \xi e^{-\frac52 \phi}S^\alpha   = \frac{1}{4} f(X)c e^{-\frac12\phi}S^\alpha,\\
\label{PCO10}
& {\cal{X}}_0 \cdot f(X) c \partial e^{-2\phi} e^{i k \cdot X} = \partial_\mu f(X) c e^{-\phi} \psi^\mu e^{i k \cdot X}, \\
\label{PCO11}
& \mathcal{X}_0 \cdot f(X) c e^{-2\phi}\partial X^\mu = \frac{1}{2} f(X)c e^{-\phi}\psi^\mu, \\
\label{PCO12}
& \mathcal{X}_0 \cdot \partial_\rho f(X) c e^{-3\phi}  \partial X^\rho \psi^\nu \partial \xi =  \frac{1}{4} \partial^\nu f(X) c e^{-2\phi} \partial \xi, \\
\label{PCO13}
&  \mathcal{X}_0\cdot \partial_\rho f(X) c \partial c e^{-3 \phi}\partial X^\rho \psi^\nu \partial\xi = \frac{1}{4} \partial^\nu f(X) c \partial c e^{-2\phi}\partial \xi, \\
\label{PCO14}
&  \mathcal{X}_0  \cdot f(X) c e^{-3\phi} \partial \psi^\nu \partial \xi = \frac{1}{4} \partial^\nu f(X) c e^{-2\phi}\partial\xi, \\
\label{PCO15}
&  \mathcal{X}_0  \cdot f(X) c \partial c e^{-3 \phi}\partial \psi^\nu \partial \xi = \frac{1}{4}\partial^\nu f(X) c\partial c e^{-2\phi}\partial\xi, \\
\label{PCO16}
&  \mathcal{X}_0 \cdot  f(X) c e^{-3 \phi} \psi^\nu \partial^2 \xi = 0, \\
\label{PCO17}
&  \mathcal{X}_0 \cdot f(X) c \partial c e^{-3\phi} \psi^\nu \partial^2 \xi =f(X) c e^{-\phi}\psi^\nu, \\
\label{PCO18}
&  \mathcal{X}_0 \cdot f(X) c \partial e^{-3\phi}\psi^\nu \partial\xi = - \frac{3}{4}
\partial^\nu f(X) c e^{-2\phi}\partial \xi, \\
\label{PCO19}
&  \mathcal{X}_0  \cdot f(X) c \partial c \partial e^{-3 \phi} \psi^\nu \partial \xi = -\frac{3}{2} f(X) c e^{-\phi} \psi^\nu - \frac{3}{4} \partial^\nu f(X) c \partial c e^{-2 \phi}\partial \xi, \\
\label{PCO20}
& \mathcal{X}_0 \cdot f(X) c \partial^2 c e^{-3 \phi} \psi^\mu \partial \xi = \frac{1}{2}f(X) c e^{-\phi}\psi ^\mu, \\
\label{PCO21}
& \mathcal{X}_0 \cdot  f(X) c \partial^3 c \partial^2 \xi \partial \xi  e^{-4 \phi} = 6 f(X) c e^{-2 \phi}\partial \xi, \\
\label{PCO22}
& \mathcal{X}_0 \cdot  f(X) c \partial c \partial^3 c \partial^2 \xi \partial \xi  e^{-4 \phi} = 6 f(X) c \partial c e^{-2 \phi} \partial \xi, \\
\label{PCO23}
   & \mathcal{X}_0 \cdot \partial_\nu f(X) c_0^+ c  \partial^2 c \partial^2 \xi \partial \xi \partial X^\nu e^{-4 \phi} = 0.
\end{align}
}

\section{Flat superstring brackets}
\label{sec:flatvert}


As seen in the context of bosonic SFT in \cite{Mazel:2024alu}, one may consider an asymmetric version of the string vertex (\ref{supersftvert}) where the first puncture is treated differently from the rest, yielding a generalized set of string brackets $[\Psi^{\otimes n}]$ that still satisfy the $L_\infty$ relation (\ref{linfty}) and thereby define consistent gauge-invariant string field equation (\ref{generalEOM}). The asymmetric string vertices define a string field frame related to the one considered in section \ref{sec:stringfieldbracket} by a field redefinition.

For instance, we can choose the coordinate map $f_1$ to be the inversion, and $f_2, \cdots, f_{n+1}$ to be simply Euclidean transformations combined with scaling,
\ie\label{flatcoormaps}
f_1(w_1) = r_0 w_1^{-1},~~~~ f_{i+1}(w_{i+1}) = q_i w_{i+1} + z_i,~~~~ i = 1,\cdots, n,
\fe
The coordinate maps (\ref{flatcoormaps}) give rise to the ``flat'' superstring bracket
\ie
[\Psi^{\otimes n}] =  {\cal G} {b_0^- \mathbb{P}^- r_0^{-L_0^+}\over (-2\pi i)^{n-2}} \int_{\Upsilon_{n;1}} e^{\cal B}\prod_{a=1}^{d_o} \big[ {\cal X}(x_a) - d\xi(x_a) \big] \prod_{\bar a=1}^{\bar d_o} \big[ \bar{\cal X}(\bar x_{\bar a}) - d\bar\xi(\bar x_{\bar a})\big] \prod_{i=1}^{n} (|q_i|^{L_0^+}\Psi)(z_i) ,
\fe
where $\mathbb{P}^-$ is the projector onto $L_0^-=0$ states, and ${\cal B}$ is explicitly given by
\ie\label{binsertflat}
{\cal B} = - \sum_{i=1}^n \left[ dz_i b_{-1}^{(z_i)} + {dq_i\over q_i} b_0^{(z_i)} + d\bar z_i \bar b_{-1}^{(\bar z_i)} + {d\bar q_i\over \bar q_i} \bar b_0^{(\bar z_i)} \right],
\fe
where $b_n^{(z_i)}$ stands for $b_n$ acting on the field operator inserted at $z_i$.
The integration contour $\Upsilon_{n;1}$ is a $(2n-4)$-dimensional chain in $\widehat{\cal Q}_{n+1}\to {\cal M}_{0,n+1}$ specified by coordinate maps of the form (\ref{flatcoormaps}) with a symmetric choice of $(q_i, z_i)$ as functions over the moduli domain $\pi(\Upsilon_{n;1})\subset {\cal M}_{0,n+1}$, together with the choice of holomorphic PCO positions $x_1, \cdots, x_{d_o}$ and anti-holomorphic PCO positions $\bar x_1, \cdots, \bar x_{\bar d_o}$ that satisfy the matching conditions
\ie\label{geommassuperflat}
- \partial \Upsilon_{0,n;1} &= \sum_{\A \subset\{1,\cdots,n\}} \varrho_{\A} \left( \widetilde\Upsilon_{0,|\A|;1}\times \widetilde\Upsilon_{0,n-|\A|+1;1}\times \{q:|q|=1\} \right),
\fe
where $\varrho_{\A}$ is a plumbing map defined analogously to (\ref{plumbingmapab}) that takes $(q'_i, z'_i, x'_{a'}, \bar x'_{\bar a'})\in \Gamma_{0,|\A|;1}$ and $(q''_j, z''_j, x''_{a''}, x''_{\bar a''})\in \Gamma_{0,n-|\A|+1;1}$ to
\ie\label{flatmatch}
& z_{\A(i)} = r_0^{-1} q q_1'' z_i' + z_1'',~~~ |q_{\A(i)}| =r_0^{-1} |q_1'' q_i'| ,~~~i = 1,\cdots |\A|,
\\
& z_{\overline\A(j)} = z_{j+1}'',~~~|q_{\overline\A(j)}| = |q_{j+1}''|,~~~j=1,\cdots,n-|\A|,
\\
& x_{a'} =  r_0^{-1}qq_1'' x'_{a'}+z_1'', ~~~ \bar x_{\bar a'} =  r_0^{-1}q q_1'' \bar x'_{\bar a'} + z_1'',~~~x_{a''} = x''_{a''}, ~~~ \bar x_{\bar a''} = \bar x''_{\bar a''}.
\fe
Here $\A(i)$ stands for the $i$-th element of the index set $\A\subset \{1,\cdots, n\}$ with an arbitrarily chosen ordering, and similarly $\overline\A(j)$ stands for the $j$-th element of the complement of $\A$.

Explicitly, we may take the flat 2-string bracket to be
\ie\label{twostringbracksuper}
[\Psi^{\otimes 2}] = {\cal G^\#} [\Psi^{\otimes 2}]_b,
\fe
where the picture-adjusting operator ${\cal G^\#}$ defined as repeatedly acting with ${\cal X}_0$ or $\bar{\cal X}_0$ to raise the picture number to $-1$ in the NS sector and $-{1\over 2}$ in the R sector, and $[\Psi^{\otimes 2}]_b$ is defined as
\ie\label{eqn:2vt}
[\Psi^{\otimes 2}]_b \equiv b_0^-\mathbb{P}^- r_0^{-L_0^+} \left( \Psi(-z_0)  \Psi(z_0) \right).
\fe
The flat 3-string bracket can be constructed as
\ie\label{threestringbracksuper}
[\Psi^{\otimes 3}] = {\cal G^\#}{ b_0^-\mathbb{P}^- r_0^{-L_0^+} \over -2\pi i} \int_{{\cal D}_{0,4}} dt\wedge d\bar t \, {\cal B}_{\bar t} {\cal B}_t \prod_{i=1}^3 (|q_i|^{L_0^+} \Psi)(z_i) \, - 3 {\rm Vert} \big[[\Psi^{\otimes 2}]_b \otimes \Psi\big],
\fe
where $q_i=q_i(t, \bar t)$, $z_i=z_i(t, \bar t)$, ${\cal B}_t$, ${\cal B}_{\bar t}$, and the domain ${\cal D}_{0,4}$ are constructed as in section 4.2 of \cite{Mazel:2024alu}, and the second term on the RHS is a correction due to vertical integration \cite{Sen:2015hia}. In particular, ${\rm Vert}[\Phi\otimes \Psi]=0$ for an ordinary string field $\Phi$ carrying picture number $-1$ or $-{1\over 2}$. When $\Phi$ carries holomorphic picture number $-2$ or $-{3\over 2}$, and anti-holomorphic picture number $-1$ or $-{1\over 2}$, we have
\ie
{\rm Vert}[\Phi\otimes \Psi] &= {\cal G^\#}\Big( \big[ \xi_0 \bar{\cal X}_0 \Phi \otimes \Psi \big]_b - \xi_0 \big[ \bar{\cal X}_0 \Phi\otimes \Psi \big]_b  \Big),
\fe
and when $\Phi=[\Psi^{\otimes 2}]_b$ carries holomorphic and anti-holomorphic picture number $-2$ or $-{3\over 2}$, we may take
\ie
{\rm Vert}[\Phi\otimes \Psi] &= {\cal G^\#}\Big( \big[ \xi_0 \bar{\cal X}_0 \Phi \otimes \Psi \big]_b - \xi_0 \big[ \bar{\cal X}_0 \Phi\otimes \Psi \big]_b + {\cal X}_0 \big[ \bar\xi_0 \Phi\otimes \Psi \big]_b - {\cal X}_0 \bar\xi_0 \big[ \bar{\cal X}_0 \Phi\otimes \Psi \big]_b \Big).
\fe

For later use, we provide more details on the relevant plumbing construction for the four-punctured sphere. Consider two cubic flat vertices defined in (\ref{eqn:2vt}). Each of them is described by a sphere with a global coordinate $z_i$ $(i=1,2)$ with local charts around the three punctures
\be
\begin{aligned}
    z_1&=f_1(w_1)=r_0 w_1^{-1},~z_1=f_2(w_2)=w_2-z_0,~z_1=f_3(w_3)=w_3+z_0,
    \\
    z_2&=f_1(w'_1)=r_0 {w'_1}^{-1},~z_2=f_2(w'_2)=w'_2-z_0,~z_2=f_3(w'_3)=w'_3+z_0.
\end{aligned}
\ee
The plumbing operation of the flat vertices is performed by making the identifications
\begin{equation}\label{eqn:plumbingPara}
    w'_1w_2=q~\text{ and }~w'_1w_3=q',
\end{equation}
where $|q|\leq1$ and $|q'|\leq1$ are the plumbing parameters. We focus on the first plumbing with $q$-parameter as the latter can be obtained by a replacement $z_0\rightarrow -z_0$. In $z_1$-coordinate system, the plumbing brings points $z_2=\pm z_0$ to $z_1=z_0\left(\pm{q\over r_0}-1\right)$. Performing an $SL(2,\mathbb{C})$ transformation, we can bring points $z_1=\infty,z_0\left({q\over r_0}-1\right),z_0\left(-{q\over r_0}-1\right)$ and $z_0$ to $z=\infty,0,t,1$ where the modulus is expressed in terms of the plumbing parameter as $t={2q\over q-2r_0}$. The corresponding propagator region $|q|\leq1$ is given by
\begin{equation}\label{eqn:propRegion}
    \bigg| {1\over t}-{1\over2}\bigg|\geq r_0.
\end{equation}
Similar plumbing constructions in the other channels lead to the determination of the vertex region of \cite{Mazel:2024alu}
\be
{\cal D}_{0,4} =\{t\in \mathbb{C}: \,\, \abs{t-\frac{1}{2}}, \abs{\frac{1}{t}-\frac{1}{2}}, \abs{\frac{1}{1-t}-\frac{1}{2}} < r_0\}
\label{vertexRegionFour}
\ee
mentioned above. The identification of the plumbing region in terms of the parameter $q$ will be necessary when we discuss the linearized equations of motion in section \ref{sec:linEOM}.

To see the dependence of our solution on the parameter $r_0$ entering into the flat bracket, we will find it useful to define the Hata-Zwiebach double bracket \cite{Hata:1993gf, Mazel:2024alu}
\be
[[\Psi^{\otimes n}]]_{r_0} =  {\cal G} {b_0^- \mathbb{P}^- r_0^{-L_0^+}\over (-2\pi i)^{n-2}} \int_{\Upsilon_{n;1}} \mathcal{B}_{r_0} e^{\cal B}\prod_{a=1}^{d_o} \big[ {\cal X}(x_a) - d\xi(x_a) \big] \prod_{\bar a=1}^{\bar d_o} \big[ \bar{\cal X}(\bar x_{\bar a}) - d\bar\xi(\bar x_{\bar a})\big] \prod_{i=1}^{n} (|q_i|^{L_0^+}\Psi)(z_i),
\label{doubledef}
\ee
where $\mathcal{B}_{r_0}$ is a $b$-ghost insertion measuring the dependence of transition maps on $r_0$.
The bracket is defined such that the $r_0$ dependence of $\Psi$ is determined via
\be
\partial_{r_0} \Psi(r_0) = \sum\limits_{n=2}^\infty [[\Psi^{\otimes n}]_{r_0}.
\ee
In this paper, we need only the explicit form of the flat double 2-bracket, which using (\ref{twostringbracksuper}) and (\ref{doubledef}) is given by
\be
[[\Psi^{\otimes 2}]]_{r_0} = b_0^+ [\Psi^{\otimes 2}].
\label{hata2}
\ee


\section{$AdS_5 \times S^5$ preliminaries}
\label{sugraApp}
In this appendix, we summarize some known facts about the $AdS_5 \times S^5$ SUGRA solution and its geometry that will serve as useful checks on our classical solution.

\subsection{Geometry of $AdS_5 \times S^5$}

The $AdS_5 \times S^5$ geometry as embedded in $\mathbb{R}^{2,10}$ can be described by the level set
\ie{}
-(x^{-1})^2 -(x^{0})^2 + (x^1)^2 + \cdots + (x^4)^2 &= -R^2,  \\
+(x^4)^2 + (x^5)^2 + \cdots + (x^{9})^2 + (x^{10})^2 &= +R^2.
\fe
We parametrize the spacetime with projective embedding coordinates $x^\mu = (x^a, x^i)$ with $a = 0,\ldots,4$ and $i = 5,\ldots,9$ such that
\ie
x^{-1} = \pm \sqrt{R^2 + x_a x^a}, \quad x^{10} = \pm \sqrt{R^2 -x_i x^i}.
\fe
The line element in these coordinates reads
\ie
ds^2 &= dx_a dx^a - {(x_a dx^a)^2 \over R^2+ x_b x^b} + dx_i dx^i + {(x_i dx^i)^2 \over R^2-x_j x^j}\\ &= dx_\mu dx^\mu - \frac{1}{R^2}(x_a dx^a)^2 + \frac{1}{R^2}(x_i dx^i)^2 + {\cal O}(R^{-4}).
\fe
where we define lowered/raised indices using $\eta_{ab}$ and $\delta_{ij}$ and their inverses, \textit{not} with the curved metric $G_{\mu\nu}$. Using the line element, we can read off the metric and its inverse
\ie{}
G_{ab} &= \eta_{ab} - {x_a x_b \over R^2} + {\cal O}(R^{-4}), \quad
G^{ab} = \eta^{ab} + {x^a x^b \over R^2} + {\cal O}(R^{-4}), \\
G_{ij} &= \delta_{ij} + {x_i x_j \over R^2} + {\cal O}(R^{-4}), \quad
G^{ij} = \delta^{ij} - {x^j x^j \over R^2} + {\cal O}(R^{-4}).
\fe
The determinant of the metric is
\ie{}
-G = 1 - {x_a x^a - x_i x^i \over R^2} + {\cal O}(R^{-4})
\fe
and we can read off the volume forms
\be
\begin{aligned}
    \omega_{AdS_5} &= \frac{1}{R} \left(1 + \frac{1}{R^2} X_a X^a \right)^{-1/2} \sim \frac{1}{R} \left(1 - \frac{1}{2R^2} X_a X^a + \ldots\right), \\
\omega_{S^5} &= \frac{1}{R} \left(1 - \frac{1}{R^2} X_i X^i \right)^{-1/2} \sim \frac{1}{R} \left(1 + \frac{1}{2R^2} X_i X^i + \ldots\right).
\end{aligned}
\ee
The Christoffel symbols are
\ie{}
\Gamma^c_{ab} = -{\eta_{ab} x^c \over R^2} + {\cal O}(R^{-4}), \quad
\Gamma^k_{ij} = {\delta_{ij} x^k \over R^2} + {\cal O}(R^{-4}).
\fe
The scalar Laplacian is
\ie{}
\label{AdSBox}
\Box_{\text{AdS}} = \partial_\mu \partial^\mu +{1 \over R^2} \left( x^a x^b \partial_a \partial_b + 5 x^a \partial_a - x^i x^j \partial_i \partial_j - 5 x^i \partial_i \right) + {\cal O}(R^{-4}).
\fe
The divergence of a 1-form $F_\mu$ is
\ie{}
G^{\mu\nu}\nabla_\mu F_\nu = \partial^\mu F_\mu + {1 \over R^2} \left(x^a x^b \partial_a F_b + 5 x^a F_a - x^i x^j \partial_i F_j - 5 x^i F_i \right) + {\cal O}(R^{-4}).
\fe
We also note that the Ricci tensor $R_{\mu\nu}$ of the solution is proportional to
\be
R_{\mu \nu} \sim \frac{1}{R^2}
\begin{pmatrix}
\eta_{ab} & 0 \\ 0 & -\delta_{ij}
\end{pmatrix}_{\mu \nu} \equiv \frac{1}{R^2} \widetilde{R}_{\mu \nu}
\label{ricciT}
\ee
and the Riemann tensor $R_{\mu \nu \lambda \sigma}$ has the nonzero components
\be
\begin{aligned}
&R_{abcd} \sim -\frac{1}{R^2} (\eta_{ac} \eta_{bd} - \eta_{ad} \eta_{bc}) \equiv \frac{1}{R^2} \widetilde{R}_{abcd},\\
&R_{ijkl} \sim \frac{1}{R^2} (\eta_{ik} \eta_{jl} - \eta_{il} \eta_{jk}) \equiv \frac{1}{R^2} \widetilde{R}_{ijkl}.
\label{riemannT}
\end{aligned}
\ee

\subsection{The SUGRA solution}
\label{sugraSol}
The $AdS_5 \cross S^5$ background up to third order in the large radius expansion is supported by the 5-form flux
\ie{}
F &= \frac12(\omega_{AdS_5} + \omega_{S^5}) \dd X^0 \wedge \ldots \wedge \dd X^4 \\
&\sim \frac{1}{R}\big(1 - \frac{1}{4R^2}(X_a X^a - X_i X^i) + \ldots\big)\dd X^0 \wedge \ldots \wedge \dd X^4,
\fe
has the metric
\be
G_{\mu \nu} = \eta_{\mu \nu} - \frac{1}{R^2}(\delta^a_\mu \delta^b_\nu X_a X_b - \delta^i_\mu \delta^j_\nu X_i X_j) + \ldots
\ee
and a constant dilaton profile $\Phi$.

\section{Diffeomorphism as a string field gauge transformation}
\label{diffeoApp}
In SFT, the diffeomorphism that shifts the fields in the massless ansatz is implemented as a special gauge transformation, see (\ref{gaugeTransforms}), with the worldsheet parity odd massless part \cite{Mazel:2025fxj}
\be
\lambda = V_\mu(X)\big(c\bar{c} e^{-\phi} \psi^\mu e^{-2\bar{\phi}}\bar{\partial} \bar{\xi} + c\bar{c} e^{-2\phi}\partial \xi e^{-\bar{\phi}} \bar{\psi}^\mu\big),
\ee
where the target space gauge field $V_\mu(X)$ is a diffeomorphism generator. The operator $ce^{-2\phi}\partial\xi$ and its antiholomorphic counterpart are added to maintain the appropriate picture number as they picture raise to the identity.

To leading order in the number of fields $\psi$ (here with ansatz (\ref{masslessansatz})), one has that $\delta_\lambda \psi = Q \lambda + \ldots$ so one simply computes the action of $Q$
\be
\begin{aligned} \label{Qdiff}
	Q\lambda = &\frac{1}{8}(\partial_\mu V_\nu + \partial_\nu V_\mu) V^{\mu \nu}_{NSNS} + \frac{1}{4}\partial^\mu V_\mu D - \frac{1}{2} \Box V_\mu V^\mu_{aux.},
\end{aligned}
\ee
and we readily conclude that, at the linearized level, the components of $\psi$ in (\ref{masslessansatz}) get shifted by
\be
\begin{aligned}
& F_{\alpha \beta} \to F_{\alpha \beta}  + \ldots, \\
&h_{\mu\nu} \to h_{\mu\nu} + \frac{1}{8}\big(\partial_\mu V_\nu +  \partial_\nu V_\mu\big) + \ldots, \\
&\phi \to \phi + \frac{1}{4}\partial^\mu V_\mu+ \ldots, \\
& A_{\mu} \to A_{\mu} - \frac{1}{2} \Box V_\mu + \ldots.
\end{aligned}
\ee
One sees that the string fields $F_{\alpha \beta}(X)$, $h_{\mu\nu}(X)$ transform as a gauge-invariant field strength and a metric respectively. On the other hand, it is clear that the ghost dilaton $\phi$ does not represent the physical dilaton, which is then given by the scalar combination $\Phi \equiv \frac{1}{4}(h-\phi)$ (overall normalization is pinned down by comparison with the Einstein-Hilbert action in string frame, see also \cite{Bergman:1994qq}). This fact is important in section \ref{solext}, when we interpret the super-isometry transformations.

\section{Details on finding the $AdS_5 \cross S^5$ solution to the extended string field equations}
\label{detailsEOM}
In this appendix, we give more details regarding the construction of the $AdS_5 \cross S^5$ solution to the extended system of string field equations. Note that when we compute string brackets below, we do not keep track of the $r_0$-dependence in the flat vertex (\ref{twostringbracksuper}). This dependence drops out since the string fields whose brackets we take are not sufficiently high-degree functions of $X^\mu$ and the end result is kept invariant by $r_0^{-L_0^+}$.

\subsection{First order}

\subsubsection{Brackets for super-isometry}
\label{firstSUSY}
At first order, one needs to compute the string bracket $[\psi^{(1)} \otimes \lambda^{(0)}]'$,
where
\be
\begin{aligned}
    \psi^{(1)} &= F^{(1)}_{\alpha \beta}(X) c\bar{c}e^{-\frac{1}{2}\phi}S^\alpha e^{-\frac{1}{2}\bar{\phi}}\bar{S}^\beta, \\
    \lambda^{(0)} & = \epsilon^{(0)}_{\alpha} c\bar{c}e^{-\frac{1}{2}\phi} S^\alpha e^{-2\bar{\phi}}\bar{\partial}\bar{\xi}.
\end{aligned}
\ee

We first write out the OPE
\be
\begin{aligned}
\psi^{(1)}(-z_0) \lambda^{(0)}(z_0) &=  F^{(1)}_{\alpha \beta}(X) c\bar{c}e^{-\frac{1}{2}\phi}S^\alpha e^{-\frac{1}{2}\bar{\phi}}\bar{S}^\beta(-z_0) \epsilon^{(0)}_{ \gamma} c\bar{c}e^{-\frac{1}{2}\phi}S^\gamma e^{-2\bar{\phi}}\bar{\partial}\bar{\xi}(z_0) + \\
& ~~~~ F^{(1)}_{\alpha \beta}(X) c\bar{c}e^{-\frac{1}{2}\phi}S^\alpha e^{-\frac{1}{2}\bar{\phi}}\bar{S}^\beta(-z_0) \bar{\epsilon}_{0 \gamma} c\bar{c}e^{-2\phi} \partial \xi e^{-\frac{1}{2}\bar{\phi}}\bar{S}^\gamma(z_0)\\
& \sim - (\gamma_\mu)^{\alpha \gamma}F^{(1)}_{\alpha \beta}(X) (4\partial c \bar{\partial}\bar{c} c\bar{c} z_0 \bar{z}_0) \epsilon^{(0)}_{\gamma} \frac{-i}{-2z_0} e^{-\phi}\psi^\mu \frac{1}{-2\bar{z}_0}e^{-\frac{5}{2}\bar{\phi}} \bar{S}^\beta \bar{\partial}\bar{\xi} + \\
& ~~~~\hspace{-0.07cm} - (\gamma_\mu)^{\beta \gamma}F^{(1)}_{\alpha \beta}(X) (4\partial c \bar{\partial}\bar{c} c\bar{c} z_0 \bar{z}_0)\bar{\epsilon}^{(0)}_{\gamma} \frac{1}{-2z_0}e^{-\frac{5}{2}\phi}S^\alpha \partial \xi \frac{-i}{-2\bar{z}_0} e^{-\bar{\phi}} \bar{\psi}^\mu,
\end{aligned}
\ee
where we only kept the massless part. This then gives
\be
\begin{aligned}
    \mathbb{P}[\psi^{(1)} \otimes \lambda^{(0)}] = & -2i (\gamma_\mu)^{\alpha \gamma} F^{(1)}_{\alpha \beta} \epsilon^{(0)}_{\gamma}c_0^+ c\bar{c}e^{-\phi}\psi^\mu e^{-\frac{5}{2}\bar{\phi}}\bar{S}^\beta \bar{\partial}\bar{\xi} \\
    &  -2i (\gamma_\mu)^{\beta \gamma} F^{(1)}_{\alpha \beta} \bar{\epsilon}^{(0)}_{\gamma}c_0^+ c\bar{c} e^{-\frac{5}{2}\phi}S^\alpha \partial\xi e^{-\bar{\phi}}\bar{\psi}^\mu
\end{aligned}
\ee
and picture raising turns this into
\be
\begin{aligned}
    [\psi^{(1)} \otimes \lambda^{(0)}]' = &+\frac{i}{4} (\gamma_\mu)^{\alpha \gamma} F^{(1)}_{\alpha \beta} \epsilon^{(0)}_{\gamma} c e^{-\phi}\psi^\mu \bar{c}e^{-\frac{1}{2}\phi}S^\beta \\
   &-\frac{i}{4} (\gamma_\mu)^{\beta \gamma} F^{(1)}_{\alpha \beta} \bar{\epsilon}^{(0)}_{\gamma} c e^{-\frac{1}{2}\phi}S^\alpha \bar{c}e^{-\bar{\phi}}\bar{\psi}^\mu.
\end{aligned}
\ee

\subsection{Second order}

\subsubsection{Brackets for the equations of motion}

\label{secondEOMdetails}
To get the source term for the second-order equation of motion, we have to compute $[(\psi^{(1)})^{\otimes 2}]'$ where $\psi^{(1)} = F^{(1)}_{\alpha \beta}(X) c\bar{c}e^{-\frac{1}{2}\phi}S^\alpha e^{-\frac{1}{2}\bar{\phi}}\bar{S}^\beta$. To do this, we write down the OPE
\be
\begin{aligned}
    &F^{(1)}_{\alpha \beta}(X) c\bar{c}e^{-\frac{1}{2}\phi}S^\alpha e^{-\frac{1}{2}\bar{\phi}}\bar{S}^\beta(-z_0) F_{1\gamma\delta}(X) c\bar{c}e^{-\frac{1}{2}\phi}S^\gamma e^{-\frac{1}{2}\bar{\phi}}\bar{S}^\delta(z_0)  \sim \\
    & -F^{(1)}_{\alpha \beta}(X(-z_0)) F_{1\gamma\delta}(X(z_0)) (\gamma_\mu)^{\alpha \gamma} (\gamma_\nu)^{\beta \delta} 4 \partial c \bar{\partial}\bar{c} c\bar{c} z_0 \bar{z}_0\frac{-i}{-2z_0}e^{-\phi} \psi^\mu \frac{-i}{-2\bar{z}_0} e^{-\bar{\phi}}\bar{\psi}^\nu
\end{aligned}
\ee
so that
\be
[(\psi^{(1)})^{\otimes 2}]' = 2 \Tr (F^{(1)} \gamma_\mu F^{(1)} \gamma_\nu) c_0^+ c\bar{c}e^{-\phi}\psi^\mu e^{-\bar{\phi}}\bar{\psi}^\nu.
\label{psi1psi1}
\ee

\subsubsection{Brackets for super-isometry}

\label{secondSUSY}
At second order, the nontrivial massless bracket we have to compute is $[\psi^{(2)} \otimes \lambda^{(0)}]'$, where
\begin{align}
    \psi^{(2)} &= 4h^{(2)}_{\mu\nu}(X)c\B{c}e^{-\phi}\psi^{\mu}e^{-\B{\phi}}\B{\psi}^{\nu}+\phi^{(2)}(X)c\B{c}\big(\eta e^{-2\B{\phi}}\B{\partial}\B{\xi}-e^{-2\phi}\partial\xi\B{\eta}\big) + \\ & \nonumber \hspace{0.6 cm}A^{(2)}_\mu(X)c_0^+c\B{c}\big(e^{-\phi}\psi^{\mu}e^{-2\B{\phi}}\B{\partial}\B{\xi}+e^{-2\phi}\partial\xi e^{-\B{\phi}}\B{\psi}^{\mu}\big)
\end{align}
and it will be sufficient to keep the $\epsilon^{(0)}_{\alpha} c\bar{c}e^{-\frac{1}{2}\phi} S^\alpha e^{-2\bar{\phi}}\bar{\partial}\bar{\xi}$ part of $\lambda^{(0)}$.

We begin by computing the ghost dilaton $\phi$ contribution. It is clear that the only level-matched contribution comes from the OPE
\be
\begin{aligned}
    &-\phi^{(2)}(X) c\bar{c} e^{-2\phi}\partial\xi \bar{\eta}(-z_0) \epsilon^{(0)}_{\alpha} c\bar{c}e^{-\frac{1}{2}\phi} S^\alpha e^{-2\bar{\phi}}\bar{\partial}\bar{\xi}(z_0) \sim \\
    & (\phi^{(2)}(X) - \bar{z}_0 \partial_\mu \phi^{(2)}(X) \bar{\partial} X ^\mu) 4\partial c \bar{\partial} \bar{c} c\bar{c} z_0 \bar{z}_0 \epsilon^{(0)}_{\alpha} \frac{1}{-2z_0} \partial \xi e^{-\frac{5}{2}\phi}S^\alpha \frac{1}{(-2\bar{z}_0)^2}(e^{-2\bar{\phi}} + \bar{z}_0 \bar{\partial} e^{-2\bar{\phi}})
\end{aligned}
\ee
so that
\be
\begin{aligned}
&\mathbb{P}[-\phi^{(2)}(X) c\bar{c}e^{-2\phi}\partial\xi \bar{\eta} \otimes \epsilon^{(0)}_{\alpha} c\bar{c}e^{-\frac{1}{2}\phi} S^\alpha e^{-2\bar{\phi}}\bar{\partial}\bar{\xi}] = \\
&\phi^{(2)}(X) \epsilon^{(0)}_{\alpha} c_0^+c\bar{c} \partial \xi e^{-\frac{5}{2}\phi}S^\alpha \bar{\partial} e^{-2\bar{\phi}} - \partial_\mu \phi^{(2)}(X) \epsilon^{(0)}_{\alpha} c_0^+c\bar{c} \partial \xi e^{-\frac{5}{2}\phi}S^\alpha \bar{\partial}X^\mu e^{-2\bar{\phi}}.
\end{aligned}
\ee
So using (\ref{PCO9})-(\ref{PCO11}),
this gets picture-raised by $\mathcal{G}^\#$ to
\be
-\frac{1}{2} \cdot \frac{1}{4} \cdot \big(-1 -\frac{1}{2} \big) \partial_\mu \phi^{(2)}(X) \epsilon^{(0)}_{\alpha}c e^{-\frac{1}{2}\phi}S^\alpha \bar{c} e^{-\bar{\phi}}\bar{\psi}^\mu = \frac{3}{16} \partial_\mu \phi^{(2)}(X) \epsilon^{(0)}_{\alpha}c e^{-\frac{1}{2}\phi}S^\alpha \bar{c} e^{-\bar{\phi}}\bar{\psi}^\mu
\ee

We continue by computing the metric contribution. The relevant OPE is
\be
\begin{aligned}
    &4h^{(2)}_{\mu\nu}(X)c\B{c}e^{-\phi}\psi^{\mu}e^{-\B{\phi}}\B{\psi}^{\nu}(-z_0) \epsilon^{(0)}_{\alpha} c\bar{c}e^{-\frac{1}{2}\phi} S^\alpha e^{-2\bar{\phi}}\bar{\partial}\bar{\xi}(z_0)  \sim \\
    & 4(\gamma^\mu)^{\alpha \beta}\big(h^{(2)}_{\mu\nu} - \bar{z}_0 \partial_\rho h^{(2)}_{\mu \nu} \bar{\partial}X^\rho) 4 \partial c \bar{\partial} \bar{c} c\bar{c} z_0 \bar{z}_0 \frac{i}{2} \frac{1}{-2z_0}\epsilon^{(0)}_{\alpha}e^{-\frac{3}{2}\phi}S_\beta \\& \frac{1}{(-2\bar{z}_0)^2}(e^{-3\bar{\phi}} + \frac{1}{3}\bar{z}_0 \bar{\partial}e^{-3\bar{\phi}} )(\bar{\psi^\nu} - \bar{z}_0 \bar{\partial} \bar{\psi^\nu})(\bar{\partial}\bar{\xi} + \bar{z}_0 \bar{\partial}^2 \bar{\xi})
\end{aligned}
\ee
so that the bracket equals to
\be
\begin{aligned}
    &+2i(\gamma^\mu)^{\alpha \beta} h^{(2)}_{\mu \nu} \epsilon^{(0)}_{\alpha}c_0^+ c\bar{c} e^{-\frac{3}{2}\phi} S_\beta e^{-3\bar{\phi}}\bar{\psi}^\nu \bar{\partial}^2\bar{\xi} +  \frac{2i}{3}(\gamma^\mu)^{\alpha \beta} h^{(2)}_{\mu \nu} \epsilon^{(0)}_{\alpha}c_0^+ c\bar{c}  e^{-\frac{3}{2}\phi} S_\beta \bar{\partial}e^{-3\bar{\phi}}\bar{\psi}^\nu \bar{\partial}\bar{\xi} - \\
    &- 2i(\gamma^\mu)^{\alpha \beta} h^{(2)}_{\mu \nu} \epsilon^{(0)}_{\alpha}c_0^+ c\bar{c} e^{-\frac{3}{2}\phi} S_\beta e^{-3\bar{\phi}}\bar{\partial}\bar{\psi}^\nu \bar{\partial}\bar{\xi}  - 2i(\gamma^\mu)^{\alpha \beta} \partial_\rho h^{(2)}_{\mu \nu} \epsilon^{(0)}_{\alpha}c_0^+ c\bar{c} e^{-\frac{3}{2}\phi} S_\beta e^{-3\bar{\phi}}\bar{\partial}X^\rho\bar{\psi}^\nu \bar{\partial}\bar{\xi},
\end{aligned}
\ee
which using (\ref{PCO5}), (\ref{PCO6}) and (\ref{PCO12})-(\ref{PCO19})
gets picture-raised to
\be
\begin{aligned}
    &+4\cdot\frac{1}{2}\cdot \frac{i}{2} \cdot \big(-\frac{i}{4}\big) \big(1-\frac{1}{3}\cdot\frac{3}{2}\big)(\gamma^\mu)^{\alpha \beta} \epsilon^{(0)}_{\alpha}\slashed{\partial}_{\beta \gamma} h^{(2)}_{\mu\nu} c e^{-\frac{1}{2}\phi} S^\gamma \bar{c}e^{-\bar{\phi}}\bar{\psi}^\nu + \\ &
    + 4\cdot\frac{1}{2} \cdot \frac{i}{2} \cdot \big(-\frac{i}{4}\big) \cdot \big(-\frac{1}{4}-\frac{1}{4}-\frac{1}{3}\cdot\frac{3}{4}\big) (\gamma^\mu)^{\alpha \beta}\epsilon^{(0)}_{\alpha}\partial^\nu \slashed{\partial}_{\beta \gamma} h^{(2)}_{\mu\nu} c e^{-\frac{1}{2}\phi} S^\gamma  \bar{c}\bar{\partial}\bar{c}e^{-2\bar{\phi}}\bar{\partial}\bar{\xi} - \\
    &-4\cdot \frac{1}{2} \cdot \frac{i}{2} \cdot \frac{i}{4} \cdot \big(-\frac{1}{4}-\frac{1}{4}-\frac{1}{3}\cdot\frac{3}{4}\big) (\gamma^\mu)^{\alpha \beta}\epsilon^{(0)}_{\alpha}\partial^\nu \slashed{\partial}_{\beta \gamma} h^{(2)}_{\mu\nu} c \partial c e^{-\frac{1}{2}\phi} S^\gamma \bar{c}e^{-2\bar{\phi}}\bar{\partial}\bar{\xi} - \\ &
    -4\cdot \frac{1}{2} \cdot \frac{i}{2} \cdot \frac{1}{4} \cdot \big(-\frac{1}{4}-\frac{1}{4}-\frac{1}{3}\cdot\frac{3}{4}\big)
    (\gamma^\mu)^{\alpha \beta} \partial^\nu h^{(2)}_{\mu\nu} \epsilon^{(0)}_{\alpha}c\eta e^{\frac{1}{2}\phi}S_\beta \bar{c} e^{-2\bar{\phi}}\bar{\partial}\bar{\xi},\end{aligned}
    \ee
    with an extra minus from commuting $\bar{c}$ through $e^{-\frac{3}{2}\phi}S_\beta$.
So taken together, we get the bracket
\be
\begin{aligned} \label{h2lambda0}
    &[4h_{\mu\nu}(X)c\B{c}e^{-\phi}\psi^{\mu}e^{-\B{\phi}}\B{\psi}^{\nu} \otimes c\B{c} {\epsilon}_{0 \alpha}(X)e^{-\frac{1}{2}\phi}S^{\alpha}e^{-2\B{\phi}}\B{\partial}\B{\xi}]' = \\
    & \frac{1}{8} (\gamma^\mu)^{\alpha \beta} \epsilon^{(0)}_{\alpha}\slashed{\partial}_{\beta \gamma} h^{(2)}_{\mu\nu} c e^{-\frac{1}{2}\phi} S^\gamma \bar{c}e^{-\bar{\phi}}\bar{\psi}^\nu - \frac{3}{8} (\gamma^\mu)^{\alpha \beta} \epsilon^{(0)}_{\alpha}\partial^\nu \slashed{\partial}_{\beta \gamma} h^{(2)}_{\mu\nu} c_0^+ c\bar{c} e^{-\frac{1}{2}\phi} S^\gamma   e^{-2\bar{\phi}}\bar{\partial}\bar{\xi} + \\
    & +\frac{3i}{16} (\gamma^\mu)^{\alpha \beta} \partial^\nu h^{(2)}_{\mu\nu} \epsilon^{(0)}_{\alpha}c\eta e^{\frac{1}{2}\phi}S_\beta \bar{c} e^{-2\bar{\phi}}\bar{\partial}\bar{\xi}
\end{aligned}
\ee

Lastly, we consider the auxiliary field contribution. This gives two terms, the first one out of which is $[A^{(2)}_\mu(X) c_0^+ c\bar{c} e^{-2\phi}\partial\xi e^{-\bar{\phi}} \bar{\psi}^\mu \otimes \epsilon^{(0)}_{\alpha} c\bar{c}e^{-\frac{1}{2}\phi} S^\alpha e^{-2\bar{\phi}}\bar{\partial}\bar{\xi}]'$. We compute the OPE
\be
\begin{aligned}
    &A^{(2)}_\mu(X) c_0^+ c\bar{c} e^{-2\phi}\partial\xi e^{-\bar{\phi}} \bar{\psi}^\mu(-z_0) \epsilon^{(0)}_{\alpha} c\bar{c}e^{-\frac{1}{2}\phi} S^\alpha e^{-2\bar{\phi}}\bar{\partial}\bar{\xi}(z_0)  \sim \\
    & -A^{(2)}_\mu (-2 \partial c \bar{\partial}\bar{c} c\bar{c} \bar{\partial}^2\bar{c} z_0 \bar{z}_0^2) \epsilon^{(0)}_{\alpha}\frac{1}{-2z_0} \partial\xi e^{-\frac{5}{2}\phi}S^\alpha \frac{1}{(-2\bar{z}_0)^2} e^{-3\bar{\phi}}\bar{\psi}^\mu\bar{\partial}\bar{\xi},
\end{aligned}
\ee
where we used
\be
\begin{aligned}
&c_0^+ c\bar{c}(-z_0) c\bar{c}(z_0) \sim -2 \partial c \bar{\partial}\bar{c} c\bar{c} \bar{\partial}^2\bar{c} z_0 \bar{z}_0^2.
\end{aligned}
\ee
This gives
\be
\mathbb{P}[A^{(2)}_\mu(X) c_0^+ c\bar{c} e^{-2\phi}\partial\xi e^{-\bar{\phi}} \bar{\psi}^\mu\otimes \epsilon^{(0)}_{\alpha} c\bar{c}e^{-\frac{1}{2}\phi} S^\alpha e^{-2\bar{\phi}}\bar{\partial}\bar{\xi}] = - \frac{1}{2}A^{(2)}_\mu\epsilon^{(0)}_{\alpha} c_0^+ c\bar{c} \bar{\partial}^2 \bar{c} \partial\xi e^{-\frac{5}{2}\phi}S^\alpha e^{-3\bar{\phi}}\bar{\psi}^\mu \bar{\partial}\bar{\xi}
\ee
and using (\ref{PCO8}), (\ref{PCO9}) and (\ref{PCO20})
this gets picture-raised to
\be
\frac{1}{4} \cdot \frac{1}{4} \cdot \frac{1}{2} \epsilon^{(0)}_{\alpha} A^{(2)}_\mu(X) c e^{-\frac{1}{2}\phi} S^\alpha \bar{c} e^{-\bar{\phi}}\bar{\psi}^\mu = \frac{1}{32} \epsilon^{(0)}_{\alpha} A^{(2)}_\mu(X) c e^{-\frac{1}{2}\phi} S^\alpha \bar{c} e^{-\bar{\phi}}\bar{\psi}^\mu
\ee
so that
\be
[A^{(2)}_\mu(X) c_0^+ c\bar{c} e^{-2\phi}\partial\xi e^{-\bar{\phi}} \bar{\psi}^\mu \otimes \epsilon^{(0)}_{\alpha} c\bar{c}e^{-\frac{1}{2}\phi} S^\alpha e^{-2\bar{\phi}}\bar{\partial}\bar{\xi}]' = \frac{1}{32} \epsilon^{(0)}_{\alpha} A^{(2)}_\mu(X) c e^{-\frac{1}{2}\phi} S^\alpha \bar{c} e^{-\bar{\phi}}\bar{\psi}^\mu.
\ee
The other term coming from the auxiliary field is $[A^{(2)}_\mu(X) c_0^+ c\bar{c}e^{-\phi}\psi^\mu e^{-2\bar{\phi}}\bar{\partial}\bar{\xi} \otimes \epsilon^{(0)}_{\alpha} c\bar{c}e^{-\frac{1}{2}\phi} S^\alpha e^{-2\bar{\phi}}\bar{\partial}\bar{\xi}]'$. We thus compute the OPE
\be
\begin{aligned}
    &A^{(2)}_\mu(X) c_0^+ c\bar{c}e^{-\phi}\psi^\mu e^{-2\bar{\phi}}\bar{\partial}\bar{\xi} (-z_0) \epsilon^{(0)}_{\alpha} c\bar{c}e^{-\frac{1}{2}\phi} S^\alpha e^{-2\bar{\phi}}\bar{\partial}\bar{\xi}(z_0) = \\
    & A^{(2)}_\mu(X(-z_0)) c_0^+ c\bar{c}(-z_0) c\bar{c}(z_0)\epsilon^{(0)}_{\alpha} e^{-\phi}\psi^\mu(-z_0) e^{-\frac{1}{2}\phi}S^\alpha(z_0) e^{-2\bar{\phi}}\bar{\partial}\bar{\xi}(-z_0) e^{-2\bar{\phi}}\bar{\partial}\bar{\xi}(z_0).
\end{aligned}
\ee
It is clear that we need to expand $c_0^+ c\bar{c}(-z_0) c\bar{c}(z_0)$ to third order in $\bar{z}_0$ and $\bar{\partial}\bar{\xi}(-z_0)\bar{\partial}\bar{\xi}(z_0)$ to second-order in $\bar{z}_0$. Note that the first order in the expansion of $e^{-2\bar{\phi}}(-z_0)e^{-2\bar{\phi}}(z_0)$ vanishes by symmetry. We thus write out
\be
\begin{aligned}
    c_0^+ c\bar{c}(-z_0) c\bar{c}(z_0) & \sim -\frac{1}{2}(-2\partial c c z_0)(- 2 \bar{\partial}\bar{c}\bar{\partial}^2\bar{c}\bar{c}\bar{z}_0^2-\frac{2}{3}\bar{\partial}\bar{c}\bar{c}\bar{\partial}^3\bar{c} \bar{z}_0^3), \\
    \bar{\partial}\bar{\xi}(-z_0) \bar{\partial}\bar{\xi}(z_0) &\sim -2\bar{\partial}^2\bar{\xi}\bar{\partial}\bar{\xi} \bar{z}_0,
    \end{aligned}
\ee
so that
\be
\begin{aligned}
&A^{(2)}_\mu(X(-z_0)) c_0^+ c\bar{c}(-z_0) c\bar{c}(z_0)\epsilon^{(0)}_{\alpha} e^{-\phi}\psi^\mu(-z_0) e^{-\frac{1}{2}\phi}S^\alpha(z_0) e^{-2\bar{\phi}}\bar{\partial}\bar{\xi}(-z_0) e^{-2\bar{\phi}}\bar{\partial}\bar{\xi}(z_0) \sim \\
& (\gamma^\mu)^{\alpha\beta}(A^{(2)}_\mu - \partial_\nu A^{(2)}_\mu \bar{\partial}X^\nu \bar{z}_0)\big[-\frac{1}{2}(-2\partial c c z_0)(- 2 \bar{\partial}\bar{c}\bar{\partial}^2\bar{c}\bar{c}\bar{z}_0^2-\frac{2}{3}\bar{\partial}\bar{c}\bar{c}\bar{\partial}^3\bar{c} \bar{z}_0^3)\big]\epsilon^{(0)}_{\alpha} \\
&\frac{i}{2}\frac{1}{-2z_0}e^{-\frac{3}{2}\phi}S_\beta \frac{1}{(-2\bar{z}_0)^4} e^{-4\bar{\phi}}(-2\bar{\partial}^2\bar{\xi}\bar{\partial}\bar{\xi} \bar{z}_0),
\end{aligned}
\ee
which gives the projected bracket
\be
\begin{aligned}
&\mathbb{P}[A^{(2)}_\mu(X) c_0^+ c\bar{c} e^{-2\phi}\partial\xi e^{-\bar{\phi}} \bar{\psi}^\mu\otimes \epsilon^{(0)}_{\alpha} c\bar{c}e^{-\frac{1}{2}\phi} S^\alpha e^{-2\bar{\phi}}\bar{\partial}\bar{\xi}] = \\&\frac{i}{16} \cdot \frac{2}{3}(\gamma^\mu)^{\alpha \beta}A^{(2)}_\mu(X)\epsilon^{(0)}_{\alpha} c_0^+ c e^{-\frac{3}{2}\phi}S_\beta \bar{c}\bar{\partial}^3\bar{c}\bar{\partial}^2\bar{\xi}\bar{\partial}\bar{\xi}e^{-4\bar{\phi}}  - \\ & \frac{2i}{16} (\gamma^\mu)^{\alpha \beta}\bar{\partial}A^{(2)}_\mu(X)\epsilon^{(0)}_{\alpha} c_0^+ c e^{-\frac{3}{2}\phi}S_\beta \bar{c}\bar{\partial}^2\bar{c}\bar{\partial}^2\bar{\xi}\bar{\partial}\bar{\xi}e^{-4\bar{\phi}}
\end{aligned}
\ee
We now use (\ref{PCO5}), (\ref{PCO6}) and (\ref{PCO21})-(\ref{PCO23}), which gives
\be
\begin{aligned}
&+\frac{i}{8} \cdot \frac{1}{2} \cdot \frac{i}{4} \cdot \frac{1}{3} \cdot 6 \epsilon^{(0)}_{\alpha} \slashed{\partial}_{\beta \gamma} \slashed{A}^{(2)\alpha \beta} c e^{-\frac{1}{2}\phi} S^\gamma \bar{c} \bar{\partial} \bar{c} e^{-2\bar{\phi}}\bar{\partial} \bar{\xi} + \\
&+\frac{i}{8} \cdot \frac{1}{2} \cdot \frac{i}{4} \cdot \frac{1}{3} \cdot 6 \epsilon^{(0)}_{\alpha} \slashed{\partial}_{\beta \gamma} \slashed{A}^{(2)\alpha \beta} c \partial c e^{-\frac{1}{2}\phi} S^\gamma \bar{c} e^{-2\bar{\phi}}\bar{\partial} \bar{\xi}  + \\
& +\frac{i}{8} \cdot \frac{1}{2} \cdot \frac{1}{4} \cdot \frac{1}{3} \cdot 6 \epsilon^{(0)}_{\alpha} \slashed{A}^{(2)\alpha \beta} c \eta e^{\frac{1}{2}\phi} S_\beta \bar{c} e^{-2\bar{\phi}}\bar{\partial} \bar{\xi} = \\
& -\frac{2}{32} \epsilon^{(0)}_{\alpha} \slashed{\partial}_{\beta \gamma} \slashed{A}^{(2)\alpha \beta} c_0^+ c e^{-\frac{1}{2}\phi} S^\gamma \bar{c} e^{-2\bar{\phi}}\bar{\partial} \bar{\xi} + \frac{2 i}{64} \epsilon^{(0)}_{\alpha} \slashed{A}^{(2)\alpha \beta} c \eta e^{\frac{1}{2}\phi} S_\beta \bar{c} e^{-2\bar{\phi}}\bar{\partial} \bar{\xi}.
\end{aligned}
\ee

We thus finally obtain the two-bracket
\be
\begin{aligned}
[\psi^{(2)}\otimes \lambda^{(0)}]' =
&+ \big(\frac{3i}{16} \epsilon^{(0)}_{\beta}(\gamma^\mu)^{ \beta\alpha} \partial^\nu h^{(2)}_{\mu\nu} + \frac{2 i}{64} \slashed{A}^{(2)\alpha \beta} \epsilon^{(0)}_{\beta} \big) c \eta e^{\frac{1}{2}\phi} S_\alpha \bar{c} e^{-2\bar{\phi}}\bar{\partial}\bar{\xi} + \\
& + \big( \frac{1}{8} \epsilon^{(0)}_{\beta}(\gamma^\nu)^{\beta \gamma} \slashed{\partial}_{\gamma \alpha} h^{(2)}_{\mu\nu} + \frac{3}{16} \partial_\mu \phi^{(2)} \epsilon^{(0)}_{\alpha} + \frac{1}{32} \epsilon^{(0)}_{\alpha} A^{(2)}_\mu\big) c e^{-\frac{1}{2}\phi} S^\alpha \bar{c} e^{-\bar{\phi}}\bar{\psi}^\mu \\
& -\big(\frac{3}{8}\epsilon^{(0)}_{\beta}(\gamma^\nu)^{\beta \gamma} \partial^\mu \slashed{\partial}_{ \gamma\alpha} h^{(2)}_{\mu\nu} + \frac{2}{32} \epsilon^{(0)}_{ \beta} \slashed{A}^{(2)\beta\gamma}  \slashed{\partial}_{ \gamma \alpha} \big) c_0^+ c e^{-\frac{1}{2}\phi} S^\alpha \bar{c} e^{-2\bar{\phi}}\bar{\partial} \bar{\xi}.
\end{aligned}
\ee

The final bracket that sources the second-order SUSY equations is the three-bracket $[(\psi^{(1)})^{\otimes2}\otimes\lambda^{(0)}]'$. Starting with the vertex region bracket, we consider
\begin{equation}
	[(\psi^{(1)})^{\otimes2}\otimes\lambda^{(0)}] = \mathcal{G}^\# \frac{b_0^-\mathbb{P}^-r^{-L_0^+}}{-2\pi i} \int_{\mathcal{D}_{0,4}}dt \wedge d\bar{t}\, \mathcal{B}_{\bar{t}} \mathcal{B}_t \, \psi^{(1)}(z_1)\psi^{(1)}(z_2)\lambda^{(0)}(z_3).
\end{equation}
The presence of the $b$-ghost insertions implies that we are always dealing with OPEs which include at most two $c$ and $\bar{c}$ ghosts. After acting with $b_0^-$, the massless level projected terms obtained from the OPE are proportional to
\begin{equation}
	\widetilde{R}_{\mu\nu}\epsilon_0\gamma^{\mu}c_0^+c\B{c}e^{-\frac{3}{2}\phi}S\B{\partial}(e^{-\B{\phi}}\B{\psi}^{\nu})e^{-2\B{\phi}}\B{\partial}\B{\xi}\quad\mathrm{and}\quad \widetilde{R}_{\mu\nu}\epsilon_0\gamma^{\mu}c\B{c}\B{\partial}^2\B{c}e^{-\frac{3}{2}\phi}Se^{-3\B{\phi}}\B{\psi}^{\nu}\B{\partial}\B{\xi}.
\end{equation}
It is easy to see that acting with $\mathcal{G}^\#$ on both of these expressions leads to zero since the Ricci tensor profile $\widetilde{R}_{\mu\nu}$ defined in (\ref{ricciT}) is constant. For the same reason, the brackets $\big[\lambda^{(0)}\otimes\frac{b_0^+}{L_0^+}(1-\mathbb{P})[(\psi^{(1)})^{\otimes 2}]\big]$ and $\big[\psi^{(1)}\otimes\frac{b_0^+}{L_0^+}(1-\mathbb{P})[\psi^{(1)}\otimes\lambda^{(0)}]\big]$ associated to the propagator region also vanish. The vertical integration contribution related to this 3-bracket is also zero. To see why, we first note that $\mathrm{Vert}\big[[(\psi^{(1)})^{\otimes 2}]_b\otimes\lambda^{(0)}\big]=0$ because $[(\psi^{(1)})^{\otimes 2}]_b$ carries holomorphic and anti-holomorphic picture number $-1$. Therefore, the non-trivial contributions are possibly related to $\mathrm{Vert}\big[[\psi^{(1)}\otimes\lambda^{(0)}]_b\otimes\psi^{(1)}\big]$. Following the computations laid out in Appendix \ref{firstSUSY}
\begin{equation}
    [\psi^{(1)}\otimes\lambda^{(0)}]_b\sim \big(\epsilon^{(0)}\gamma_{\mu}F^{(1)}\big)c_0^+c\B{c}e^{-\phi}\psi^{\mu}e^{-\frac{5}{2}\B{\phi}}\B{S}\B{\partial}\B{\xi}.
\end{equation}
The computation of the vertical integration contribution then reduces to $\mathcal{G}^\# \bar{\xi}_0\big[\mathcal{X}_0[\psi^{(1)}\otimes\lambda^{(0)}]_b\otimes\psi^{(1)}\big]_b$
since $\bar{\xi}_0[\psi^{(1)}\otimes\lambda^{(0)}]_b=0$. The operations involved in computing $\big[\mathcal{X}_0[\psi^{(1)}\otimes\lambda^{(0)}]_b\otimes\psi^{(1)}\big]_b$ do not introduce $\bar{\eta}$-modes and thus $\bar{\xi}_0$-action also yields zero. We therefore conclude that the vertical integration contribution to the 3-bracket $[(\psi^{(1)})^{\otimes2}\otimes\lambda^{(0)}]'$ is zero.

\subsubsection{Solving the extended system}
\label{integrabilityAppendix}
In this appendix, we fix the background string field in Siegel gauge $A_\mu^{(2)} = 0$, which is subject to the equations of motion (\ref{curvatureEq})-(\ref{auxEOMeq}) and super-isometry consistency conditions (\ref{dilatonEq}), (\ref{integrEq}) that read (after plugging in the first-order solution)

\begin{align}
    \label{actuallyFirstEqq}
    &\Box h^{(2)}_{\mu\nu} = -8 N_{\text{R}}^2 \widetilde{R}_{\mu \nu}, \\
    \label{firstEqq}
    &\Box \phi^{(2)} = 0, \\
    \label{auxilEOM}
    &\partial_\mu \phi^{(2)} - 2\partial^\nu h^{(2)}_{\mu \nu} = 0, \\
    \label{secondEqq}
    & \partial_\mu (\phi^{(2)} - h^{(2)})  = 0, \\
    \label{thirdEqq}
    &\gamma^\nu \slashed{\partial} \partial_\rho h^{(2)}_{ \mu \nu} +  N_{\text{R}}^2 \widetilde{R}_{\mu \nu} \gamma^\nu \gamma_\rho - \mu \leftrightarrow \rho = 0,
\end{align}
where we have used elementary Diracology to write $F^{(1)} \gamma_\mu F^{(1)} \gamma_\rho = - N_{\text{R}}^2\widetilde{R}_{\mu \nu}\gamma^\nu \gamma_\rho$ so that
\be
-\frac{1}{16 N_{\text{R}}^2} \Tr(F^{(1)} \gamma_\mu F^{(1)} \gamma_\nu) = \widetilde{R}_{\mu \nu},
\ee
with $\widetilde{R}_{\mu\nu}$ the rescaled $AdS_5 \cross S^5$ Ricci tensor (\ref{ricciT}).
We take the ansatz (\ref{hAnsatz})
\be
\begin{aligned}
    h^{(2)}_{ \mu \nu} =  \frac{\alpha_1}{4} &\big(\delta^a_\mu \delta^b_\nu X_a X_b - \delta^i_\mu \delta^j_\nu X_i X_j\big) + \frac{\alpha_2}{4} \big(\eta_{ab}\delta_{\mu}^a\delta_{\nu}^bX_cX^c-\delta_{ij}\delta_{\mu}^i\delta_{\nu}^jX_kX^k\big)
\end{aligned}
\ee
and the zero physical dilaton condition (\ref{secondEqq}) then amounts to setting
\be
\phi^{(2)} = \frac{1}{4}(\alpha_1 + 5\alpha_2)(X_a X^a - X_i X^i) ,
\ee
which automatically satisfies the wave equation (\ref{firstEqq}).
Plugging in the ansatz for the second-order solution, the integrability condition (\ref{thirdEqq}) gives
\be
\begin{aligned}
    &\alpha_1(5\delta^a_\mu \eta_{a\rho}+\delta^a_\mu \gamma_\rho \gamma_a - 5 \delta^i_\mu \eta_{i \rho} - \delta^i_\mu \gamma_\rho \gamma_i) + 2\alpha_2(\delta^a_\mu \delta^b_\rho \gamma_a \gamma_b - \delta^i_\mu \delta^j_\rho \gamma_i \gamma_j) + 4 N_{\text{R}}^2 \widetilde{R}_{\mu \nu} \gamma^\nu \gamma_\rho - \mu \leftrightarrow \rho = 0,
\end{aligned}
\ee
which can be rewritten as
\be
\begin{aligned}
	\alpha_1 \widetilde{R}_{\mu \nu} \gamma_\rho \gamma^\nu + 2\alpha_2 \widetilde{R}_{\mu \nu} \gamma^\nu \gamma_\rho + 4 N_{\text{R}}^2  \widetilde{R}_{\mu \nu} \gamma^\nu \gamma_\rho - \mu \leftrightarrow \rho = 0,
\end{aligned}
\ee
which can be commuted to (the anticommutator drops out)
\begin{align}
	(-\alpha_1 + 2\alpha_2 + 4 N_{\text{R}}^2 ) \widetilde{R}_{\mu \nu}  \gamma^\nu \gamma_\rho  - \mu \leftrightarrow \rho = 0.
\end{align}
This is solved by $\alpha_1 - 2\alpha_2 - 4 N_{\text{R}}^2 = 0$. Finally, the auxiliary equation of motion (\ref{auxilEOM}) gives $-5 \alpha_1 + 3\alpha_2 = 0$, which leads to the unique solution $\alpha_1 = -\frac{12}{7} N_{\text{R}}^2, \alpha_2 = -\frac{20}{7} N_{\text{R}}^2$.

Note that the system is overdetermined and so we have to check whether the last unused equation (\ref{actuallyFirstEqq}) is satisfied and a brief calculation shows that it indeed is since
\be
\Box h_{\mu \nu}^{(2)} = \frac{1}{2}(\alpha_1 + 5\alpha_2) \widetilde{R}_{\mu \nu} = -8 N_{\text{R}}^2 \widetilde{R}_{\mu \nu}.
\ee

\subsubsection{Gauge transformation from Siegel gauge}
\label{gaugeTransformAppendix}
As derived in the appendix \ref{diffeoApp}, linearized diffeomorphism with generator $V_\mu$ acts on the string fields of our second-order solution as
\be
\begin{aligned}
    \delta_\lambda h^{(2)}_{\mu\nu}&=\frac{1}{8}\big(\partial_{\mu}V_{\nu}+\partial_{\nu}V_{\mu}\big), \\
    \delta_\lambda\phi^{(2)}&=\frac{1}{4}\partial^{\mu}V_{\mu}, \\
    \delta_\lambda A^{(2)}_\mu&=-\frac{1}{2}\Box V_{\mu}.
\end{aligned}
\ee
As mentioned in section \ref{secondOrdExt}, these gauge transformations can be used to bring the Siegel gauge solution (\ref{ssol})
\be
\begin{aligned}
	h^{(2)}_{\mu \nu} &=-\frac{3}{7} N_{\text{R}}^2 \big(\delta^a_\mu \delta^b_\nu X_a X_b - \delta^i_\mu \delta^j_\nu X_i X_j\big) -\frac{5}{7} N_{\text{R}}^2 \big(\eta_{ab}\delta_{\mu}^a\delta_{\nu}^bX_cX^c-
    \delta_{ij}\delta_{\mu}^i\delta_{\nu}^jX_kX^k\big), \\
		\phi^{(2)} &= - 4 N_{\text{R}}^2X^2, \\
    A^{(2)}_\mu &= 0
\end{aligned}
\ee
 into a form which makes identification with $AdS_5\times S^5$ geometry more manifest at the cost of introducing a non-trivial auxiliary field. We now spell this out in more detail.

 It is easy to see that we should choose a gauge parameter cubic in the coordinates and we take the ansatz
\begin{equation}
    V_{\mu}=\alpha \big(\delta_{\mu}^{a}X_aX_bX^b-\delta_{\mu}^iX_iX_jX^j\big).
\end{equation}
We now tune $\alpha$ so that we eliminate the $-\frac{5}{7} N_{\text{R}}^2 (\eta_{ab}\delta_{\mu}^{a}\delta_{\nu}^{b}X_cX^c-\eta_{ij}\delta_{\mu}^{i}\delta_{\nu}^{j}X_kX^k)$ piece of the Siegel gauge solution. To do this, we compute the gauge transform
\begin{equation}
    \delta_\lambda h^{(2)}_{\mu\nu}(X)=\frac{\alpha}{4}\big(\eta_{ab}\delta_{\mu}^{a}\delta_{\nu}^{b}X_cX^c+2\delta_{\mu}^a\delta_{\nu}^bX_aX_b-\eta_{ij}\delta_{\mu}^{i}\delta_{\nu}^{j}X_kX^k-2\delta_{\mu}^i\delta_{\nu}^jX_iX_j\big)
\end{equation}
and we see that for $\alpha = \frac{20}{7} N_{\text{R}}^2$, we end up with
\begin{equation}
	h^{(2)}_{\mu\nu}(X)= N_{\text{R}}^2\big(\delta_{\mu}^a\delta_{\nu}^bX_aX_b-\delta_{\mu}^i\delta_{\nu}^jX_iX_j\big).
\end{equation}
 It is then easy to see that the auxiliary field acquires a non-trivial profile
\begin{equation}
	A^{(2)}_{\mu}(X)=-20 N_{\text{R}}^2 \big(\delta_{\mu}^aX_a-\delta_{\mu}^ iX_i\big) = -10 N_{\text{R}}^2\partial_\mu X^2.
\end{equation}
Finally, the transformation of the ghost-dilaton component is
\begin{equation}
	\delta_\lambda\phi^{(2)}(X)=5 N_{\text{R}}^2 X^2
\end{equation}
such that in this relaxed Siegel gauge we get
\begin{equation}
	\phi^{(2)}(X)=N_{\text{R}}^2X^2.
\end{equation}
Therefore, we verify that the second-order solution can be brought to the form
\be
\begin{aligned}
	h^{(2)}_{\mu\nu}(X)&=N_{\text{R}}^2\big(\delta_{\mu}^a\delta_{\nu}^bX_aX_b-\delta_{\mu}^i\delta_{\nu}^jX_iX_j\big),\\
	\phi^{(2)}(X)&=N_{\text{R}}^2X^2,\\
	A^{(2)}_{\mu}(X)&=-10 N_{\text{R}}^2\partial_\mu X^2,
\end{aligned}
\ee
which makes the connection with $AdS_5\times S^5$ geometry much more clear. Indeed, after proper normalization, the expression found above for the $h^{(2)}_{\mu\nu}$ directly matches the second-order term in the $\frac{1}{R}$-expansion of the $AdS_5\times S^5$ metric of (\ref{sugraSol}).

\subsection{Third order}

\subsubsection{Brackets for the equations of motion}
\label{AppEOM3rd}
The computation for the 2-bracket $[\psi^{(2)} \otimes \psi^{(1)}]'$ starts by noting that only the metric fluctuation in $\psi^{(2)}$ contributes so that we only need to consider the OPE
\be
\begin{aligned}
&4h^{(2)}_{\mu\nu}c\B{c}e^{-\phi}\psi^{\mu}e^{-\B{\phi}}\B{\psi}^{\nu}(-z_0)F^{(1)}_{\alpha \beta}c\B{c}e^{-\frac{1}{2}\phi}S^{\alpha}e^{-\frac{1}{2}\B{\phi}}\B{S}^{\beta}(z_0)
    \sim \\
    &\frac{1}{z_0\B{z}_0}\times 4\times\frac{1}{4}\times\Big(\frac{i}{2}\Big)^2h^{(2)}_{\mu\nu}(\gamma^{\mu})^{\gamma\alpha}F^{(1)}_{\alpha \beta}(\gamma^{\nu})^{\beta\delta}c\B{c}(-z_0)c\B{c}(z_0)e^{-\frac{3}{2}\phi}S_{\gamma}e^{-\frac{3}{2}\B{\phi}}\B{S}_{\delta}(z_0).
\end{aligned}
\ee
Expansion around $z=0$ contributes with a factor of $-4c\partial c\B{c}\B{\partial}\B{c}$, such that we find
\begin{equation}
    h^{(2)}_{\mu\nu}(\gamma^{\mu})^{\gamma\alpha}F^{(1)}_{\alpha \beta}(\gamma^{\nu})^{\beta\delta}c\partial c\B{c}\B{\partial}\B{c}e^{-\frac{3}{2}\phi}S_{\gamma}e^{-\frac{3}{2}\B{\phi}}\B{S}_{\delta}.
\end{equation}
After $b_0^-$-action, we conclude that
\begin{equation}
    \mathbb{P}[h^{(2)}_{\mu\nu}V_{NSNS}^{\mu\nu}\otimes F^{(1)}_{\alpha \beta} V^{\alpha \beta}_{RR}]=-2h^{(2)}_{\mu\nu}(\gamma^{\mu})^{\gamma\alpha}F^{(1)}_{\alpha \beta}(\gamma^{\nu})^{\beta\delta}c_0^+c\B{c}e^{-\frac{3}{2}\phi}S_{\gamma}e^{-\frac{3}{2}\B{\phi}}\B{S}_{\delta}.
\end{equation}
This expression has to be picture-adjusted. We use the relations (\ref{PCO5}), (\ref{PCO6})
together with their analogous right-moving counterparts to obtain
\begin{equation}
\begin{split}
   \label{thirdSolBracket}
   [\psi^{(2)}\otimes \psi^{(1)}]'=&-\frac{i}{16}\slashed{\partial}_{\lambda\gamma}h^{(2)}_{\mu\nu}(\gamma^{\mu})^{\gamma\alpha}F^{(1)}_{\alpha \beta}(\gamma^{\nu})^{\beta\delta}ce^{-\frac{1}{2}\phi}S^{\lambda}\B{c}\B{\eta}e^{\frac{1}{2}\B{\phi}}\B{S}_{\delta}\\
    &-\frac{i}{16}\slashed{\partial}_{\lambda\delta}h^{(2)}_{\mu\nu}(\gamma^{\mu})^{\gamma\alpha}F^{(1)}_{\alpha \beta}(\gamma^{\nu})^{\beta\delta}c\eta e^{\frac{1}{2}\phi}S_{\gamma}\B{c}e^{-\frac{1}{2}\B{\phi}}\B{S}^{\lambda}\\
    &+\frac{1}{8}\slashed{\partial}_{\lambda\gamma}(\gamma^{\mu})^{\gamma\alpha}F^{(1)}_{\alpha \beta}(\gamma^{\nu})^{\beta\delta}\slashed{\partial}_{\delta\sigma}h^{(2)}_{\mu\nu}c_0^+c\B{c}e^{-\frac{1}{2}\phi}S^{\lambda}e^{-\frac{1}{2}\B{\phi}}\B{S}^{\sigma}.
\end{split}
\end{equation}

The computation of the 3-bracket $[(\psi^{(1)})^{\otimes 3}]'$ is straightforward and yields zero. To see why, take the definition of the 3-bracket
\begin{equation}
	[(\psi^{(1)})^{\otimes 3}] = \mathcal{G}^\#\frac{b_0^{-}\mathbb{P}^{-}r^{-L_0^+}}{-2\pi i}\int_{\mathcal{D}_{0,4}}dt\wedge d\bar{t}\, \mathcal{B}_{\bar{t}} \mathcal{B}_t \, \psi^{(1)}(z_1)\psi^{(1)}(z_2)\psi^{(1)}(z_3).
\end{equation}
The presence of the $b$-ghost insertions implies that we are always dealing with OPEs which include at most two $c$ and $\bar{c}$ ghosts. After acting with $b_0^-$ and projecting into the massless level sector of the state space, the result of the OPEs is proportional to
\begin{equation}
	(\gamma^{\mu}F_1\gamma^{\nu})\widetilde{R}_{\mu\nu}c_0^+c\B{c}e^{-\frac{3}{2}\phi}Se^{-\frac{3}{2}\B{\phi}}\B{S}.
\end{equation}
 Here $\widetilde{R}_{\mu \nu}$ is the Ricci tensor defined in (\ref{ricciT}). As in the supersymmetry discussion, we do not need to actually compute the integral over $\mathcal{D}_{0,4}$ to see that the bracket vanishes, since this is simply a consequence of acting with $\mathcal{G}^\#$ and the profile being constant, which can be easily checked from the general picture-raising equations in Appendix \ref{WSconventions}. The propagator region brackets $\big[\psi^{(1)}\otimes\frac{b_0^+}{L_0^+}(1-\mathbb{P})[(\psi^{(1)})^{\otimes 2}]\big]$ vanish for the same reason. Finally, the contribution due to vertical integration is zero since $[(\psi^{(1)})^{\otimes 2}]_b$ carries holomorphic and anti-holomorphic picture number equal to $-1$.

\section{Details on finding the axion linearized fluctuation}
\label{sec:linEOM}

In this appendix, we give more details regarding the construction of the axion linearized fluctuation. We work with the second-order background solution in the gauge of (\ref{ssol}).

\subsection{Deriving the sources for the second-order axion equation of motion}
\label{sourcesApp}
In this appendix, we derive the source terms for the second-order axion equation of motion (\ref{lineom2}).

\subsubsection{Computation of the axion 2-bracket}
We first compute the massless 2-bracket $[\psi^{(2)} \otimes \phi^{(0)}]'$. We split $\psi^{(2)} = \tilde{\psi}^{(2)} + \hat{\psi}^{(2)}$, where $\hat{\psi}^{(2)} = -4 N_R^2 \ln r_0 \widetilde{R}_{\mu \nu} V_{NSNS}^{\mu \nu}$ is the local coordinate dependent part of the second-order solution (as discussed below (\ref{oos})).
Proceeding exactly as in the calculation of (\ref{thirdSolBracket}), we can then write
\be
\begin{aligned}
	[\hat{\psi}_2 \otimes \phi^{(0)}]'= -\frac{N_R^2}{4} \ln r_0^2 \Biggr[&+(\slashed{\partial}\gamma^{\mu}f^{(0)}\gamma^{\nu}\overset{\leftarrow}{\slashed{\partial}})_{\alpha \beta}\widetilde{R}_{\mu\nu}c_0^+c\B{c}e^{-\frac{1}{2}\phi}S^{\alpha}e^{-\frac{1}{2}\B{\phi}}\B{S}^{\beta}\\
									      &- \frac{i}{2} (\slashed{\partial} \gamma^{\mu} f^{(0)} \gamma^{\nu})_{\alpha}^\beta \widetilde{R}_{\mu\nu}ce^{-\frac{1}{2}\phi}S^{\alpha}\B{c}\B{\eta}e^{\frac{1}{2}\B{\phi}}\B{S}_{\beta}\\
									      &- \frac{i}{2} (\gamma^{\mu} f^{(0)} \gamma^{\nu} \overset{\leftarrow}{\slashed{\partial}})^\alpha_\beta \widetilde{R}_{\mu\nu}c\eta e^{\frac{1}{2}\phi}S_{\alpha}\B{c}e^{-\frac{1}{2}\B{\phi}}\B{S}^{\beta}\Biggr].
\label{hatpsi2}
\end{aligned}
\ee
As far as the bracket with the result of the solution goes, only the $h_{\mu \nu}^{(2)} V_{NSNS}^{\mu \nu}$ part contributes to the 2-bracket. Again, following (\ref{thirdSolBracket}), but including the nontrivial action of $r_0^{-L_0^+}$, we obtain
\begin{equation}
\begin{aligned}
	&[\tilde{\psi}^{(2)} \otimes \phi^{(0)}]'=\frac{N_{\text{R}}^2}{2} \Biggr[(\slashed{\partial} \gamma^{\mu} f^{(0)} \gamma^{\nu}  h^{(2)}_{\mu\nu} \overset{\leftarrow}{\slashed{\partial}})_{\alpha \beta} c_0^+c\B{c}e^{-\frac{1}{2}\phi}S^{\alpha}e^{-\frac{1}{2}\B{\phi}}\B{S}^{\beta} - \\&
-\frac{i}{2}(\slashed{\partial} \gamma^{\mu}f^{(0)}\gamma^{\nu} h^{(2)}_{\mu\nu} )_{\alpha}^{\beta} ce^{-\frac{1}{2}\phi}S^{\alpha}\B{c}\B{\eta}e^{\frac{1}{2}\B{\phi}}\B{S}_{\beta}
-\frac{i}{2}(\gamma^{\mu} f^{(0)} \gamma^{\nu} h^{(2)}_{\mu\nu} \overset{\leftarrow}{\slashed{\partial}} )^\alpha_\beta c\eta e^{\frac{1}{2}\phi}S_{\alpha}\B{c}e^{-\frac{1}{2}\B{\phi}}\B{S}^{\beta}\Biggr] +
								       \\ &
								       \frac{N_{\text{R}}^2}{4} \ln r_0\Biggr[+\Box (\slashed{\partial} \gamma^{\mu} f^{(0)} \gamma^{\nu} h^{(2)}_{\mu\nu} \overset{\leftarrow}{\slashed{\partial}})_{\alpha \beta}c_0^+c\B{c}e^{-\frac{1}{2}\phi}S^{\alpha}e^{-\frac{1}{2}\B{\phi}}\B{S}^{\beta}
	-\frac{i}{2} \Box (\slashed{\partial} \gamma^{\mu} f^{(0)} \gamma^{\nu} h^{(2)}_{\mu\nu})_{\alpha}^\beta ce^{-\frac{1}{2}\phi}S^{\alpha}\B{c}\B{\eta}e^{\frac{1}{2}\B{\phi}}\B{S}_{\beta}\\
									  &-\frac{i}{2}\Box (\gamma^{\mu} f^{(0)} \gamma^{\nu} h^{(2)}_{\mu\nu}\overset{\leftarrow}{\slashed{\partial}})^\alpha_\beta c\eta e^{\frac{1}{2}\phi}S_{\alpha}\B{c}e^{-\frac{1}{2}\B{\phi}}\B{S}^{\beta}\Biggr].
\end{aligned}
\label{tildepsi2}
\end{equation}

\subsubsection{Computation of the axion 3-bracket}
We now compute the massless 3-bracket $[(\psi^{(1)})^{\otimes 2}\otimes \phi^{(0)}]'$
using the definition (\ref{threestringbracksuper}). It is clear that since both $\psi^{(1)}$ and $\phi^{(0)}$ are in the RR sector, there is no contribution from vertical integration. Also, since both the profiles of $\psi^{(1)}$ and $\phi^{(0)}$ are annihilated by $L_0^+ \sim \Box$, the expression (\ref{threestringbracksuper}) simplifies to
\be
[(\psi^{(1)})^{\otimes 2}\otimes \phi^{(0)}] = {\cal G^\#}{ b_0^-\mathbb{P}^-r_0^{-L_0^+} \over -2\pi i} \int_{{\cal D}_{0,4}} dt\wedge d\bar t \, {\cal B}_{\bar t} {\cal B}_t  \, \psi^{(1)}(z_1) \psi^{(1)}(z_2) \phi^{(0)}(z_3)
\ee
and using (\ref{masslessthree}) gives
\be
\begin{aligned}
	[(\psi^{(1)})^{\otimes 2}\otimes \phi^{(0)}]' = &{\cal G^\#}{ b_0^-\mathbb{P}^-r_0^{-L_0^+}  \over -2\pi i} \mathbb{P}\int_{{\cal D}_{0,4}} dt\wedge d\bar t \, {\cal B}_{\bar t} {\cal B}_t  \, \psi^{(1)}(z_1) \psi^{(1)}(z_2) \phi^{(0)}(z_3)- \\
						& 2 \mathbb{P}\left[\psi^{(1)} \otimes \frac{b_0^+}{L_0^+}(1-\mathbb{P})[\psi^{(1)} \otimes \phi^{(0)}]\right] - \mathbb{P}\left[\phi^{(0)} \otimes \frac{b_0^+}{L_0^+}(1-\mathbb{P})[(\psi^{(1)})^{\otimes 2}]\right].
\end{aligned}
\ee
Now, by performing OPE, it is clear that before acting with $\mathcal{G}^\#$, the result is proportional to the string field $\widetilde{R}_{\mu \nu}(\gamma^\mu f^{(0)}\gamma^\nu)_{\alpha \beta} c_0^+ c\bar{c} e^{-\frac{3}{2}\phi} S^\alpha e^{-\frac{3}{2}\bar{\phi}}\bar{S}^\beta$ and we determine its coefficient by computing the overlap with $T_{RR} = T^{\alpha \beta}(X) (V_{RR})_{\alpha \beta}$. The relevant worldsheet correlator is
\begin{equation}\label{eqn:wscorrel4pt}
	\langle T_{RR}(\infty)\psi^{(1)}(0)\psi^{(1)}(x)\phi^{(0)}(1)\rangle=2N_R^2\int d^{10}X \widetilde{R}_{\mu\nu}\Tr\left(f^{(0)}\gamma^\mu \left(T\right)^T\gamma^\nu\right){x+{\bar x}-2\over|1-x|^2|x|^2}.
\end{equation}

Note that there is a logarithmic divergence as $x \to 0$, while it is absent as $x\to1$ and $x\to\infty$. Therefore, among the three propagator contributions, the latter two simply amount to integrating the correlator (\ref{eqn:wscorrel4pt}) over the respective propagator regions in the worldsheet moduli space, while the $x\sim0$ region requires the subtraction of the logarithmic divergence as dictated by the plumbing construction of the corresponding propagator region \cite{Sen:2019jpm}. The $x\sim0$ propagator region is given by $\abs{\frac{1}{x}-\frac{1}{2}}\geq r_0$ as discussed in (\ref{eqn:propRegion}). The rest of the moduli space is covered by the union of the vertex region (\ref{vertexRegionFour}) and two propagator regions around $x\sim1$ and $x\sim\infty$. Taking the union amounts to removing two of the inequalities in (\ref{vertexRegionFour}), leaving us with integration over $\abs{\frac{1}{x}-\frac{1}{2}}< r_0$. Thus, the contribution from this union is
\be
\begin{aligned}
{}&{\cal G}{ b_0^-\mathbb{P}^-r_0^{-L_0^+}  \over -2\pi i} \mathbb{P}\int_{{\cal D}_{0,4}} dt\wedge d\bar t \, {\cal B}_{\bar t} {\cal B}_t  \, \psi^{(1)}(z_1) \psi^{(1)}(z_2) \phi^{(0)}(z_3)-2 \mathbb{P}\left[\psi^{(1)} \otimes \frac{b_0^+}{L_0^+}(1-\mathbb{P})[\psi^{(1)} \otimes \phi^{(0)}]\right]
\\
&=\frac{1}{-2\pi i}2\cdot 2N_R^2 \int\limits_{\abs{\frac{1}{x}-\frac{1}{2}}<r_0} \dd^2x\Biggr(  \frac{x+\bar{x}-2}{\abs{1-x}^2\abs{x}^2}\Biggr) \mathcal{G}^\# \big[ \widetilde{R}_{\mu \nu}(\gamma^\mu f^{(0)}\gamma^\nu)_{\alpha \beta} c_0^+ c\bar{c} e^{-\frac{3}{2}\phi} S^\alpha e^{-\frac{3}{2}\bar{\phi}}\bar{S}^\beta\big]
\\
&=\frac{1}{-2\pi i}32N_R^2 \frac{-i\pi}{2}\big(\ln 4 -\ln(1+4 r_0^2)\big) \mathcal{G}^\# \big[\widetilde{R}_{\mu \nu}(\gamma^\mu f^{(0)}\gamma^\nu)_{\alpha \beta} c_0^+ c\bar{c} e^{-\frac{3}{2}\phi} S^\alpha e^{-\frac{3}{2}\bar{\phi}}\bar{S}^\beta\big],
\end{aligned}
\ee
where one factor of $2$ comes from the overlap with $T_{RR}$ giving rise to an extra $\frac{1}{2}$ in the correlator.

In contrast, the integration over the remaining propagator region $\abs{\frac{1}{x}-\frac{1}{2}}\geq r_0$ should be performed by explicitly subtracting off the logarithmic divergence dictated by the plumbing parameter $q$ introduced in (\ref{eqn:plumbingPara}) which is related to $x$ by $x={2q\over q-2r_0}$. Therefore,
\be
\begin{aligned}
&-\mathbb{P}\left[\phi^{(0)} \otimes \frac{b_0^+}{L_0^+}(1-\mathbb{P})[(\psi^{(1)})^{\otimes 2}]\right] = \\
&\frac{1}{-2\pi i}2\cdot 16N_R^2 \int\limits_{\abs{q}\leq 1}d^2 q \Biggr({{{q\over r_0}\over {q\over r_0}-2}+{{\bar q\over r_0}\over {\bar q \over r_0}-2}-1  \over |q|^2\bigg|2+{q\over r_0}\bigg|^2} +{1\over4 q\bar q}\Biggr) \mathcal{G}^\# \big[\widetilde{R}_{\mu \nu}(\gamma^\mu f^{(0)}\gamma^\nu)_{\alpha \beta} c_0^+ c\bar{c} e^{-\frac{3}{2}\phi} S^\alpha e^{-\frac{3}{2}\bar{\phi}}\bar{S}^\beta\big] = \\
&\frac{1}{-2\pi i}32N_R^2 \frac{-i\pi}{2}\big(\ln(1+4r_0^2)-\ln 4 - \ln r_0^2\big) \mathcal{G}^\# \big[\widetilde{R}_{\mu \nu}(\gamma^\mu f^{(0)}\gamma^\nu)_{\alpha \beta} c_0^+ c\bar{c} e^{-\frac{3}{2}\phi} S^\alpha e^{-\frac{3}{2}\bar{\phi}}\bar{S}^\beta\big],
\end{aligned}
\ee
where we explicitly performed the subtraction coming from $1-\mathbb{P}$.

Adding the above contributions together, we obtain
\be
[(\psi^{(1)})^{\otimes 2}\otimes \phi^{(0)}]' = -8 N_R^2 \ln r_0^2 \, \mathcal{G}^\# \big[\widetilde{R}_{\mu \nu}(\gamma^\mu f^{(0)}\gamma^\nu)_{\alpha \beta} c_0^+ c\bar{c} e^{-\frac{3}{2}\phi} S^\alpha e^{-\frac{3}{2}\bar{\phi}}\bar{S}^\beta\big],
\ee
so that after picture-adjusting
\be
\begin{aligned}
	\label{axion3B}
	[(\psi^{(1)})^{\otimes 2} \otimes \phi^{(0)}]'= \frac{N_R^2}{2} \ln r_0^2 \Biggr[&+(\slashed{\partial} \gamma^{\mu}f^{(0)}\gamma^{\nu}\overset{\leftarrow}{\slashed{\partial}})_{\alpha \beta}\widetilde{R}_{\mu\nu}c_0^+c\B{c}e^{-\frac{1}{2}\phi}S^{\alpha}e^{-\frac{1}{2}\B{\phi}}\B{S}^{\beta}\\
											 &-\frac{i}{2} (\slashed{\partial} \gamma^{\mu} f^{(0)} \gamma^{\nu})_{\alpha}^\beta \widetilde{R}_{\mu\nu}ce^{-\frac{1}{2}\phi}S^{\alpha}\B{c}\B{\eta}e^{\frac{1}{2}\B{\phi}}\B{S}_{\beta}\\
    &-
\frac{i}{2} (\gamma^{\mu} f^{(0)} \gamma^{\nu} \overset{\leftarrow}{\slashed{\partial}})^\alpha_\beta \widetilde{R}_{\mu\nu}c\eta e^{\frac{1}{2}\phi}S_{\alpha} \B{c}e^{-\frac{1}{2}\B{\phi}}\B{S}^{\beta}\Biggr].
\end{aligned}
\ee

\subsection{Explicitly writing the second-order equations of motion for the axion}
\label{axionEOMAppendix}
Using the results of the previous subsections, we can explicitly write down the source term of (\ref{lineom2}). It is clear that in this source, the 3-bracket (\ref{axion3B}) cancels with the 2-bracket (\ref{hatpsi2}) coming from the $r_0$-dependent piece of $\psi^{(2)}$, leaving us with the piece coming from (\ref{tildepsi2})
\be
\begin{aligned}
&-[\psi^{(2)} \otimes \phi^{(0)}]' - \frac{1}{2}[(\psi^{(1)})^{\otimes 2}\otimes \phi^{(0)}]' =  \frac{1}{8} \Biggr[-(\slashed{\partial} \gamma^{\mu} f^{(0)}  \gamma^{\nu}  h^{(2)}_{\mu\nu} \overset{\leftarrow}{\slashed{\partial}})_{\alpha \beta} c_0^+c\B{c}e^{-\frac{1}{2}\phi}S^{\alpha}e^{-\frac{1}{2}\B{\phi}}\B{S}^{\beta} + \\&
+\frac{i}{2}(\slashed{\partial} \gamma^{\mu}f^{(0)}\gamma^{\nu} h^{(2)}_{\mu\nu} )_{\alpha}^{\beta} ce^{-\frac{1}{2}\phi}S^{\alpha}\B{c}\B{\eta}e^{\frac{1}{2}\B{\phi}}\B{S}_{\beta}
+\frac{i}{2}(\gamma^{\mu} f^{(0)} \gamma^{\nu} h^{(2)}_{\mu\nu} \overset{\leftarrow}{\slashed{\partial}} )^\alpha_\beta c\eta e^{\frac{1}{2}\phi}S_{\alpha}\B{c}e^{-\frac{1}{2}\B{\phi}}\B{S}^{\beta}\Biggr] +
								       \\ &
								       \frac{1}{16} \ln r_0\Biggr[-\Box (\slashed{\partial} \gamma^{\mu} f^{(0)} \gamma^{\nu} h^{(2)}_{\mu\nu} \overset{\leftarrow}{\slashed{\partial}})_{\alpha \beta}c_0^+c\B{c}e^{-\frac{1}{2}\phi}S^{\alpha}e^{-\frac{1}{2}\B{\phi}}\B{S}^{\beta}
	+\frac{i}{2} \Box (\slashed{\partial} \gamma^{\mu} f^{(0)} \gamma^{\nu} h^{(2)}_{\mu\nu})_{\alpha}^\beta ce^{-\frac{1}{2}\phi}S^{\alpha}\B{c}\B{\eta}e^{\frac{1}{2}\B{\phi}}\B{S}_{\beta}\\
									  &+\frac{i}{2}\Box (\gamma^{\mu} f^{(0)} \gamma^{\nu} h^{(2)}_{\mu\nu}\overset{\leftarrow}{\slashed{\partial}})^\alpha_\beta c\eta e^{\frac{1}{2}\phi}S_{\alpha}\B{c}e^{-\frac{1}{2}\B{\phi}}\B{S}^{\beta}\Biggr].
\end{aligned}
\ee
From this result, we easily read off the equation of motion (remembering that only one Dirac equation is independent)
\be
f^{(2)}  \overset{\leftarrow}{\slashed{\partial}} =\frac{1}{4}\Big[\slashed{\partial} \gamma^{\mu}f^{(0)} \gamma^{\nu} h^{(2)}_{\mu\nu} + \frac{1}{2} \ln r_0 \big(\gamma^\kappa \gamma^{\mu}\gamma^{\lambda}\gamma^{\nu}\big)\partial_{\kappa}\Box\big(f^{(0)}_\lambda h^{(2)}_{\mu\nu}\big)\Big],
\ee
and we wish to solve it by splitting it into 0-form and 2-form parts.

First, we write
\be
\begin{aligned}
	\gamma^{\mu}f^{(0)} \gamma^{\nu} h^{(2)}_{\mu\nu} &=  N_{\text{R}}^2\big( \gamma^a f^{(0)} \gamma^b X_a X_b - \gamma^i f^{(0)} \gamma^j X_i X_j\big)  \\
									     &=  N_{\text{R}}^2\big( 2 f^{(0)a} \gamma^b X_a X_b -  \gamma^\mu f^{(0)}_{\mu}X_a X^a - 2 f^{(0)i} \gamma^j X_i X_j +  \gamma^\mu f^{(0)}_{ \mu} X_i X^i\big),
\end{aligned}
\ee
so that
\be
\begin{aligned}
	&\frac{1}{4} N_{\text{R}}^2\slashed{\partial} (\gamma^a f^{(0)} \gamma^b X_a X_b - \gamma^i f^{(0)} \gamma^j X_i X_j) = \\
	&\frac{1}{2} N_{\text{R}}^2 \gamma^\mu \partial_\mu \big(f^{(0)a} \gamma^b X_a X_b  - f^{(0)i} \gamma^j X_i X_j -  \frac{1}{2}\gamma^\nu f^{(0)}_{ \nu}(X_a X^a - X_i X^i)\big).
\end{aligned}
\ee
We then use $\gamma^\mu \gamma^\nu = \eta^{\mu \nu} + \gamma^{\mu \nu}$, so that we can write
\be
\begin{aligned}
    &\gamma^\mu \big(f^{(0)a} \gamma^b X_a X_b - f^{(0)i} \gamma^j X_i X_j- \frac{1}{2}\gamma^\nu f^{(0)}_{ \nu}(X_a X^a - X_i X^i)\big) = \\
    &f^{(0)a} X_a X_b \eta^{\mu b} + \gamma^{\mu b} f^{(0)a} X_a X_b
    -f^{(0)i} X_i X_j \eta^{\mu j} - \gamma^{\mu j}f^{(0)i} X_i X_j \\
    & - \frac{1}{2} (\eta^{\mu \nu} f^{(0)}_{ \nu} + \gamma^{\mu \nu} f^{(0)}_{ \nu})(X_a X^a - X_i X^i),
\end{aligned}
\ee
and thus using $\slashed{\partial} f^{(0)} = 0$, we get
\be
\begin{aligned}
    &\gamma^\mu \partial_\mu \big(f^{(0)a} \gamma^b X_a X_b - f^{(0)i} \gamma^j X_i X_j- \frac{1}{2}\gamma^\nu f^{(0)}_{ \nu}(X_a X^a - X_i X^i)\big) = \\
    &5 f^{(0)a} X_a + \partial^b f^{(0)a} X_a X_b - 5 f^i_0 X_i - \partial^j f^i_0 X_i X_j + \\
    &\partial_\mu (\gamma^{\mu b} f^{(0)a} X_a X_b - \gamma^{\mu j}f^{(0)i} X_i X_j -\frac{1}{2}(X_a X^a - X_i X^i)f^{(0)}_{\nu}\gamma^{\mu \nu}).
\end{aligned}
\ee
Thus, we have split the $r_0$-independent source into the 0-form and 2-form parts, giving
\be
\begin{aligned}
	\frac{1}{4}\slashed{\partial} \gamma^{\mu}f^{(0)} \gamma^{\nu} h^{(2)}_{\mu\nu}  = &\frac{1}{2}N_{\text{R}}^2\big(5 f^{(0)a} X_a + \partial^b f^{(0)a} X_a X_b - 5 f^i_0 X_i - \partial^j f^i_0 X_i X_j\big)+ \\
&\frac{1}{2}N_{\text{R}}^2 \partial_\mu (\gamma^{\mu b} f^{(0)a} X_a X_b - \gamma^{\mu j}f^{(0)i} X_i X_j - \frac{1}{2} (X_a X^a - X_i X^i) \gamma^{\mu \nu} f^{(0)}_{ \nu}).
\end{aligned}
\ee

We continue by computing the $r_0$-dependent source by using
\begin{equation}
\big(\gamma^\kappa \gamma^{\mu}\gamma^{\lambda}\gamma^{\nu}\big)h^{(2)}_{\mu\nu}=\Big[2\eta^{\lambda\nu}\big(\eta^{\mu\kappa}-\gamma^{\mu\kappa}\big)-\eta^{\mu\nu}\big(\eta^{\lambda\kappa}-\gamma^{\lambda\kappa}\big)\Big]h^{(2)}_{\mu\nu}
\end{equation}
to write
\be
\begin{aligned}
\big(\gamma^{\kappa}\gamma^{\mu}\gamma^{\lambda}\gamma^{\nu}\big)\partial_{\kappa}\Box\big(f^{(0)}_\lambda h^{(2)}_{\mu\nu}\big) = &\Box (2\partial^\mu f^{(0)\nu} h^{(2)}_{\mu \nu}+2f^{(0)\nu} \partial^\mu h^{(2)}_{\mu \nu} - f^{(0)}_\mu \partial^\mu h^{(2)}) + \\
&\gamma^{\mu \nu} \partial_\mu \Box(2 f^{(0)\rho} h^{(2)}_{\nu \rho} - f^{(0)}_\nu h^{(2)}),
\label{4gamma}
\end{aligned}
\ee
so that in the end, we get the explicit equations of motion
\be
\begin{aligned}
	f^{(2)}  \overset{\leftarrow}{\slashed{\partial}} =  &\frac{1}{2}N_{\text{R}}^2\big(5 f^{(0)a} X_a + \partial^b f^{(0)a} X_a X_b - 5 f^i_0 X_i - \partial^j f^i_0 X_i X_j\big)+ \\
&\frac{1}{2}N_{\text{R}}^2 \partial_\mu (\gamma^{\mu b} f^{(0)a} X_a X_b - \gamma^{\mu j}f^{(0)i} X_i X_j - \frac{1}{2} (X_a X^a - X_i X^i) \gamma^{\mu \nu} f^{(0)}_{ \nu}) + \\
& \frac{1}{8} \ln r_0 \big[\Box (2\partial^\mu f^{(0)\nu} h^{(2)}_{\mu \nu}+2f^{(0)\nu} \partial^\mu h^{(2)}_{\mu \nu} -  f^{(0)}_\mu \partial^\mu h^{(2)}) + \\
&\gamma^{\mu \nu} \partial_\mu \Box(2 f^{(0)\rho} h^{(2)}_{\nu \rho} -  f^{(0)}_\nu h^{(2)})
\big].
\label{axionEOMs}
\end{aligned}
\ee

\printbibliography
\end{document}